%% file: main.tex
\documentclass[fleqn,usenatbib]{mnras}
\hypersetup{breaklinks=true}
\input{header}

\title[MHD log-density intermittency]{The density distribution and physical origins of intermittency in supersonic, highly magnetised turbulence with diverse modes of driving}

\author[Beattie, Mocz, Federrath \& Klessen]{James R. Beattie$^{\orcidicon{0000-0001-9199-7771}\,1,2}$\thanks{E-mail: james.beattie@anu.edu.au}, 
Philip Mocz$^{\orcidicon{0000-0001-6631-2566}\,3\divideontimes,4}$\thanks{E-mail: mocz1@llnl.gov},
Christoph Federrath$^{\orcidicon{0000-0002-0706-2306}\,1,5}$\thanks{E-mail: christoph.federrath@anu.edu.au}
and Ralf S. Klessen$^{\orcidicon{0000-0002-0560-3172}\,6,7}$
\\
$^{1}$Research School of Astronomy and Astrophysics, Australian National University, Canberra, ACT 2611, Australia \\
$^{2}$Department of Astronomy and Astrophysics, University of California, Santa Cruz, 1156 High Street, Santa Cruz, CA 96054\\
$^{3}$Department of Astrophysical Sciences, Princeton University, 4 Ivy Lane, Princeton, NJ 08544, USA \\
$^{4}$Lawrence Livermore National Laboratory, 7000 East Ave, Livermore, CA 94550, USA \\
$^{5}$Australian Research Council Centre of Excellence in All Sky Astrophysics (ASTRO3D), Canberra, ACT 2611, Australia \\
$^{6}$Universit\"at Heidelberg, Zentrum f\"ur Astronomie, Institut f\"ur Theoretische Astrophysik, Albert-Ueberle-Str. 2, 69120 Heidelberg, Germany\\ 
$^{7}$Universit\"at Heidelberg, Interdisziplin\"ares Zentrum f\"ur Wissenschaftliches Rechnen, Im Neuenheimer Feld 205, 69120 Heidelberg, Germany \\
$^{\divideontimes}$Einstein Fellow 
}

\date{Accepted XXX. Received YYY; in original form ZZZ}

\pubyear{2021}

\begin{document}
\label{firstpage}
\pagerange{\pageref{firstpage}--\pageref{lastpage}}
\maketitle

\begin{abstract}
The probability density function (PDF) of the logarithmic density contrast, \mbox{$s=\ln (\rho/\rho_0)$}, with gas density $\rho$ and mean density $\rho_0$, for hydrodynamical supersonic turbulence is well-known to have significant non-Gaussian (intermittent) features that monotonically increase with the turbulent Mach number, $\M$. By studying the mass- and volume-weighted $s$-PDF for an ensemble of 36 sub-to-trans-Alf\'venic mean-field, supersonic, isothermal turbulence simulations with different modes of driving, relevant to molecular gas in the cool interstellar medium, we show that a more intricate picture emerges for the non-Gaussian nature of $s$. Using four independent measures of the non-Gaussian components, we find hydrodynamical-like structure in the highly magnetised plasma for $\mathcal{M} \lesssim 4$. However, for $\mathcal{M} \gtrsim 4$, the non-Gaussian signatures disappear, leaving approximately Gaussian $s$-statistics -- exactly the opposite of hydrodynamical turbulence in the high-$\mathcal{M}$ limit. We also find that the non-Gaussian components of the PDF increase monotonically with more compressive driving modes. To understand the $\mathcal{M} \lesssim 4$ non-Gaussian features we use one-dimensional (1D) pencil beams to explore the dynamics along and across the large-scale magnetic field, $\Bo$. We discuss kinetic, density and magnetic field fluctuations from the pencil beams, and identify physical sources of non-Gaussian components to the PDF as single, strong shocks coupled to fast magnetosonic compressions that form along $\Bo$. We discuss the Gaussianisation of the $\mathcal{M} \gtrsim 4$ $s$-fields through the lens of two phenomenologies: the self-similarity of the $s$-field and homogenisation of the dynamical timescales between the over- and under-dense regions in the compressible gas.
\end{abstract}




\section{Introduction}\label{sec:intro}
    The PDF of the gas density is a valuable tool for understanding the nature of compressible turbulence and star formation in the interstellar medium of galaxies \citep{Vazquez1994,Padoan1997,Passot1998,Federrath2008,Federrath2009,Brunt2010a,Brunt2010b,Krumholz2005,Hennebelle2011,Federrath2012,Burkhart2012,Konstandin2012a,Molina2012,Hopkins2013,Nolan2015,Federrath2015,Squire2017,Pan2019,Mocz2019,Menon2020,Menon2020b,Khuller2021,Sharda2022_driving_mode}. Simple models of molecular clouds (MCs) in the interstellar medium (ISM), which are supersonic and magnetised, and have not yet started to collapse under their own self-gravity have an \textit{approximately} Gaussian volume-weighted $s$-PDF\footnote{Or equivalently, a lognormal $\rho/\rho_0$-PDF.}, where $s\equiv \ln(\rho / \rho_0)$, and $\rho$ is the cloud density, with $\rho_0$ the value of the volume-weighted mean. It follows
    \begin{align}
        p_{\text{N},V}(s;\sigma_{s,V}^2) &= \frac{1}{\sqrt{2\pi\sigma_{s,V}^2}} \exp \left\{ -\frac{(s- s_{0,V})^2}{2\sigma_{s,V}^2} \right\}, \label{eq:Vdis} \\
        s_{0,V} &= - \frac{\sigma_{s,V}^2}{2}, \label{eq:vars_0 relation}\\
        \sigma^2_{s,V} &= f(\M,\Ma,b,\gamma,\Gamma). \label{eq:vars_V}
    \end{align}
    The log-density variance, $\sigma_s^2$, captures the density variations induced by different physical processes in a MC. It is a function of (i) the turbulent Mach number,
    \begin{align} \label{eq:M}
    \M = \sigma_{V}/c_s,
    \end{align}
    where $\sigma_{V}$ is the velocity dispersion on system scale $L$, and $c_s$ is the sound speed \citep{Vazquez1994,Padoan1997,Passot1998,Price2011,Konstandin2012a}, (ii) the Alfv\'en Mach number, 
    \begin{align}\label{eq:Ma}
        \Ma = \sigma_{V} / V_{\rm A} = c_s \M / V_{\rm A},
    \end{align}
    where $V_\text{A} = B / \sqrt{4\pi\rho}$ is the root-mean-squared Alfv\'en wave velocity and $B$ is the magnetic field \citep{Padoan2011,Molina2012,Beattie2021}, (iii) the turbulent driving parameter, $b$\footnote{Note that $b$, the driving parameter, is directly related to the amount of solenoidal and compressive modes being injected into the turbulence via the \textit{source of the turbulence}, the so-called $\zeta$ parameter (not the modes that are then generated in the momentum field). As derived in \citet{Federrath2010_solendoidal_versus_compressive} for forced, hydrodynamical, three-dimensional turbulence an empirical relation is $$b(\zeta) = \frac{1}{3} + \frac{2}{3}\left( \frac{(1-\zeta)^2}{1-2\zeta+3\zeta^2} \right)^3.$$}, which captures the influence of compressive ($\nabla\times\vecB{F}=0$) or solenoidal ($\nabla\cdot\vecB{F}=0$) forcing modes on the density fluctuations, where $\vecB{F}$ is a source of turbulent forcing \citep{Federrath2008,Federrath2010_solendoidal_versus_compressive,Menon2020}, (iv) the adiabatic index $\gamma$ \citep{Nolan2015}, and (v) the polytropic index $\Gamma$ \citep{Federrath2015}. 
    Likewise, the mass-weighted distribution is,
    \begin{align}
        p_{\text{N}, M}(s;\sigma_{s,M}^2) &= \frac{\exp\left\{s\right\}}{\sqrt{2\pi\sigma_{s,M}^2}} \exp \left\{ -\frac{(s- s_{0,M})^2}{2\sigma_{s,M}^2} \right\}, \label{eq:Mdis} \\
        s_{0,M} &= \frac{\sigma_{s,V}^2}{2},\label{eq:mean_M} \\
        \sigma^2_{s,M} &= \sigma^2_{s,V},\label{eq:vars_M}
    \end{align}
    which shows how, in a lognormal density-fluctuation theory, the mass-weighted and volume-weighted distributions are intrinsically linked through the mean (Equation~\ref{eq:mean_M}) and variance (Equation~\ref{eq:vars_M}) of the volume-weighted $s$-PDF \citep{Li2003}. In hydrodynamical, supersonic turbulence it is well-known that there is a significant difference between $\sigma^2_{s,M}$ and $\sigma^2_{s,V}$ (an excess of $\sigma^2_{s,V}$), which in essence is from the emergence of non-Gaussian, intermittent events in the fluid \citep{Passot1998,Kritsuk2007,Federrath2010_solendoidal_versus_compressive,Federrath2013,Hopkins2013,Squire2017,Mocz2019}. 
    
    The $s$-statistics and intermittency of strongly magnetised turbulence, where the flow is globally anisotropic about a coherent, strong mean-field has not been studied in great detail, even though it is indeed a relevant flow regime for either some regions inside of MCs \citep{HuaBai2013,Federrath2016_brick,Hu2019,Heyer2020,Skalidis2020,Hwang2021,Hoang2021,Skaladis2021_subAlf_transition} or even perhaps most MCs \citep{HuaBai2021}. In this study we probe the physics of the $s$-PDFs in isothermal simulations with $2 \lesssim \M \lesssim 20$ and $0.1 \lesssim \Mao \lesssim 2$, where $\Mao = c_s \M \sqrt{4\pi\rho_0} / |\Bo|$, revealing the nature of voids, rarefactions and over-dense structures in the trans- to sub-Alfv\'enic mean-field turbulence, i.e., turbulence where the energy in the mean magnetic field is larger than that of the kinetic turbulent energy, with different types of turbulent driving. We show that the $s$-intermittency for isothermal, compressible, hydrodynamical turbulence is different from turbulence in the highly-magnetised mean-field regime. Specifically, we find that Equation~\ref{eq:vars_M} approximately holds for moderate to high-$\M$ magnetised flows, which is not the case for purely hydrodynamical turbulence due to strong intermittency. We discuss a range of phenomenological models for why this may be the case and, for the first time, identify physical sources of $s$-intermittency in real-space, confirming that an important source of intermittency in these flows are strong shocks that produce intermittent, deep, volume-poor rarefactions in the gas density along the coherent magnetic field.
    
    This study is organised as follows. First, in \S\ref{sec:lognormal} we revisit the lognormal model for the density fluctuations. In \S\ref{sec:intermittent} we extend the discussion to $s$-fluctuation models that include intermittency effects. In \S\ref{sec:numerics} we outline the supersonic MHD turbulence simulations that we use to explore the logarithmic density fluctuations. In \S\ref{sec:densityPDF} we analyse the morphology of the full three-dimensional (3D) volume-weighted and mass-weighted $s$-PDFs from the simulation data, including fitting lognormal and non-lognormal models for the volume-weighted PDFs. Next, in \S\ref{sec:intermittency_results} we compute four independent measures of $s$-intermittency, and show how the highly-magnetised regime has a more complicated intermittency structure than the hydrodynamical regime. In \S\ref{sec:driving_parameter} we repeat our analysis for different mixtures of solenoidal and compressive modes in the turbulence driving function. In \S\ref{sec:physical_intermittency} we use one-dimensional (1D) probes of the low-$\M$ turbulence to reveal the parallel and perpendicular to mean magnetic field $s$, velocity, and magnetic field dynamics, and use this 1D analysis to identify the rarest (intermittent) events in the turbulence, in real-space. In \S\ref{sec:high_mach_gauss} we discuss the lack of intermittency in high-$\M$, magnetised turbulence. Finally in \S\ref{sec:conclusion} we summarise the key results of this study.
    
    \section{The lognormal density-PDF}\label{sec:lognormal}
    \subsection{The lognormal model}
    Lognormal models for the PDF of turbulent $\delta = \rho / \rho_0$ fluctuations originate from \citet{Vazquez1994}. They consider $\delta$ in the self-similar, hierarchical structure of the cool, isothermal ISM, where global pressure is negligible $(\M \gg 1)$ and self-gravity is yet to dominate the dynamics of the region. The lognormal PDF is motivated by assuming that for time, $t_n$, the density can be expressed as a multiplicative interaction through independent jumps $\delta$ in the Eulerian frame of the gas,
    \begin{align}
        \frac{\rho(t_n)}{\rho_0} = \delta_n \delta_{n-1} \hdots \delta_1 \delta_0 \frac{\rho(t_0)}{\rho_0} = \left(\prod^n_{i=0} \delta_i\right) \frac{\rho(t_0)}{\rho_0},
    \end{align}
    where $\rho(t_0)/\rho_0 = 1$ is the initial density in units of the mean. This means that under the log-transformation, fluctuations become additive,
    \begin{align}
        s(t_n) = \ln\frac{\rho(t_n)}{\rho_0} = \sum^n_{i=0} \ln\delta_i = \sum^n_{i=0} s_i,
    \end{align}    
    If the density fluctuations are
    \begin{enumerate}
        \item generated by the same underlying probability distribution, and
        \item are independent from one another (i.e., not temporally correlated),
    \end{enumerate}
    then the central limit theorem states that the distribution of the logarithmic density fluctuations should tend towards a Gaussian distribution as the number of fluctuations increase. This framework lets us understand the nature of a single (Eulerian) density fluctuation changing in time. However, \citet{Vazquez1994} further argued that since the hydrodynamical equations are self-similar in space (i.e., invariant to arbitrary length scaling) the fluctuations should be lognormal on all scales. With these assumptions, the lognormal distribution should aptly describe the density fluctuations on any length scale of the turbulence. 
    
    \subsection{Key issues with the lognormal model}\label{sec:problems_with_normal}
    \subsubsection{Mass conservation}
    \citet{Hopkins2013} articulates that having a lognormal distribution on all scales in the flow violates mass conservation. We show this by considering the $\rho/\rho_0$-PDF, $p_L(\delta)$, on the system scale, $L$. This distribution is the convolution of all PDFs from scales below the system scale, $L/\Gamma$, $L/\Gamma^2$, $\hdots$, $L/\Gamma^n$, $\hdots$, where $\Gamma \gtrsim 1$, in the turbulence, which is simply an application of the Law of Total Probability\footnote{This is simply because $(\rho/\rho_0)_{L/\Gamma^n}$ on each scale is a random variable composed of the (volume-weighted) linear some of densities on smaller scales, $(\rho/\rho_0)_{L/\Gamma^n} \propto (\rho/\rho_0)_{L/\Gamma^{n+1}}+(\rho/\rho_0)_{L/\Gamma^{n+2}}+\hdots$ This, by definition, leads to the PDF on each scale being an infinite convolution of PDFs from scales below it.} \citep{Casting1996}. Writing this formerly, 
    \begin{align} \label{eq:conv_pdf}
        p_{L}(\delta) &= p_{L/\Gamma}(\delta) \otimes \hdots \otimes  p_{L/\Gamma^n}(\delta) \otimes  \hdots =  \bigotimes_{n=1}^{\infty} p_{L/\Gamma^n}(\delta),
    \end{align}
    where $\otimes$ is the convolution operator, $f(x) \otimes g(x) = \int^{\infty}_{-\infty}\d{t}\,f(x-t)g(t)$. Now we assume that all of the PDFs, $p_{L/\Gamma}(\delta)\hdots p_{L/\Gamma^n}(\delta)\hdots$, on the RHS of Equation~\ref{eq:conv_pdf} are lognormal distributions and inquire about the LHS of the equation. This is a well-known and historical problem in statistics, motivated by problems from a broad range of disciplines, from telecommunications to the biosciences \citep[see for example,][]{Wu2005,Lo2012,Hcine2015}. The key conclusion one makes is that the LHS of Equation~\ref{eq:conv_pdf} cannot possibly be a lognormal distribution. In fact, because no scale is special in the turbulence, and instead of picking $p_{L}(\delta)$, we pick any arbitrary $p_{L/\Gamma^n}(\delta)$, then it is clear that only one of the infinite $p_{L/\Gamma^n}(\delta)$ can possibly be lognormal. This is directly related to mass conservation. It follows by assuming that all $p_{L/\Gamma^n}(\delta)$ are lognormal, then the total mass is, 
    \begin{align}
        M_1 = \int_{0}^{\infty} \d{\delta} \, \delta p_{L}(\delta),
    \end{align}
    which ought to equal 
    \begin{align}
    M_2 = \int_{0}^{\infty} \d{\delta} \, \delta \bigotimes_{n=1}^{\infty} p_{L/\Gamma^n}(\delta). 
    \end{align}
    However, because $\bigotimes_{n=1}^{\infty} p_{L/\Gamma^n}(\delta)$ does not converge to a lognormal distribution, and because we also have assumed that \emph{all} PDFs are, including $p_{L}(\delta)$, then $M_1 \neq M_2$, and hence a lognormal model on all scales violates mass conservation.
    
    \subsubsection{The sonic scale}
    This is not the only reason why the lognormal distribution, which describes a scale-free random process, cannot be strictly ``the right model''\footnote{Note that it can, however, be very close to a lognormal model, and for practical purposes this might suffice, but theoretically the $s$-PDF can not be \emph{exactly} a Gaussian model.} for the density fluctuations in supersonic turbulence. Recent numerical experiments by \citet{Federrath2021} reveal that supersonic turbulence is not scale-free, and has a characteristic length, the sonic scale, $\ell_s$, for which
    \begin{align}
        \left(\left\langle |\vecB{v}(\vecB{r}) - \vecB{v}(\vecB{r} + \ell_s)|^2 \right\rangle_{\vecB{r}}/2\right)^{1/2} = c_s,       
    \end{align}
    where $\vecB{v}(\vecB{r})$ is the velocity field of the turbulence. Furthermore, $c_s$ is the sound speed and $\left\langle \hdots \right\rangle_{\vecB{r}}$ is the ensemble average over positions, $\vecB{r}$ where the turbulence above $\ell_s$ behaves like \citet{Burgers1948} turbulence (the turbulence of interacting sawtooth waves), and the scales below \citet{Kolmogorov1941} turbulence (the turbulence of inertially interacting eddies)\footnote{Note that \citet{Federrath2021} showed that this was the case only for second-order structure functions, and that there was some deviation from perfect \citet{Kolmogorov1941} turbulence on small-scales, which they attributed to intermittency effects in the subsonic cascade.}. This means that one can not model the small-scale turbulence and scale the results to the system-scale, as in \citet{Vazquez1994}. In fact, through direct measurements of the $s$-PDF below and above $\ell_s$ \citep{Federrath2021} show that on scales below $\ell_s$ the $s$-PDF becomes peaked and kurtotic, and on scales above, negatively skewed. It is clear that the $s$-PDF is non-Gaussian, even on the subsonic scales of the turbulence.
    
    \section{Beyond lognormality}\label{sec:intermittent}
    In both ISM observations and in high-resolution numerical studies of supersonic turbulence, we find that the $s$-PDF deviates away from being a perfect Gaussian \citep{Kritsuk2007,Federrath2008,Federrath2010_solendoidal_versus_compressive,Price2011,Konstandin2012,Federrath2013,Hopkins2013,Pan2019,Menon2020,Sharda2022_driving_mode}. The reason for these deviations (ignoring gravity and power-law tails due to gravitationally bound, collapsing, dense structures; see \citealt{Klessen2000b,Federrath2008Physica,Kritsuk2011,Federrath2013,Girichidis2014,Mocz2017,Burkhart2018,Khuller2021}, or due to non-isothermal effects; see \citealt{Passot1998,Nolan2015,Federrath2015}) are due to intermittency, in reference to intermittent, i.e., rare events that are non-Gaussian in nature (i.e., fluctuations in space or time that are not described by Gaussian statistics). More quantitatively, \citet{Federrath2010_solendoidal_versus_compressive} described three different realisations that intermittency manifests itself in turbulent flows:
    \begin{enumerate}
        \item non-Gaussian wings / non-Gaussian higher-order moments of PDFs for turbulent quantities, e.g., $\rho$, $\vecB{v}$, $\vecB{B}$, their derivatives, and combinations of the quantities \citep{Burkhart2009,Hopkins2013,Mocz2019,Seta2020,Mohapatra2020,Beattie2020c};
        \item anomalous scaling of the higher-order structure functions of the velocity field \citep{She1994,Casting1996,Kowal2007,Schmidt2008,Konstandin2012,Hopkins2013};
        \item structures with intense vorticity, $\nabla \times \vecB{v}$, and energy dissipation, such as strong shocks and filamentary structures, $\nabla \cdot \vecB{v} < 0$, and rarefied regions $\nabla \cdot \vecB{v} >0$  \citep{Kritsuk2007,Federrath2013,Squire2017,Park2019,Mocz2019,Beattie2020,Yoffe2021}.
    \end{enumerate}
    In supersonic turbulence, let alone magnetised supersonic turbulence, the intermittent structures described above dominate the flow, for example through populations of shocks. This has led to large efforts to understand intermittency, which is still one of the most fundamental problems in turbulence, and to characterise it in the ISM \citep{Falgarone1995,Falgarone2009,Blant2009,Falgarone2015}. It means to properly model the density field one must include intermittency in the recipe. We now discuss two models that are based upon the life and happenings of shocked regions in the turbulence, and give some insight into the processes underpinning intermittency in the turbulent density field.

    \begin{figure*}
        \centering
        \includegraphics[width=\linewidth]{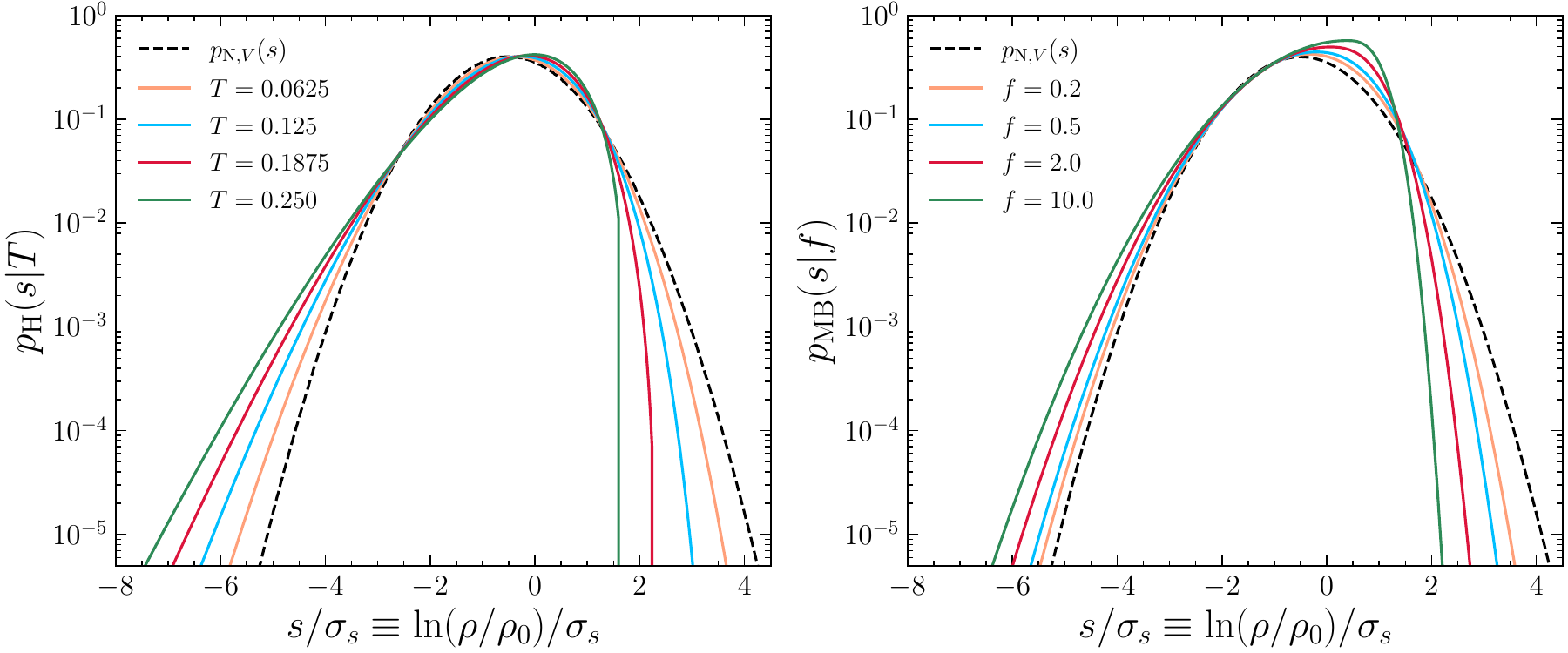}
        \caption{\textbf{Left:} The \citetalias{Hopkins2013} $s$-PDF model shown for different intermittency parameters: $0\, (p_{\text{N}, V}) \leq T \leq 0.25$. This parameter range spans from no intermittency (lognormal, shown with black, dashed line style, $T=0$) to strong intermittency ($T=0.25$). We plot in the variable $s/\sigma_s$ to draw attention to just the non-Gaussian morphology of the distribution. \textbf{Right:} The same as left plot but for the \citetalias{Mocz2019} $s$-PDF model, plotted for different $f$ values, corresponding to changing the dynamical timescale of over-dense and under-dense regions in the turbulence. $f = 0 \iff p_{\text{N}, V}$ is the lognormal model, and $f = 10.0$ is the extreme case where over-dense regions exist on 16$\times$ shorter timescales than under-dense regions.}
        \label{fig:theory_pdfs}
    \end{figure*}

    \subsection{Compound log-Poisson model}
    \citet{Hopkins2013} (hereafter called \citetalias{Hopkins2013}) considers the \citet{Casting1996} model to describe a general steady-state result from a class of multiplicative random relaxation processes that produce random multiplicative changes to the density. The essence of the \citetalias{Hopkins2013} model is that an exponential PDF describes the continuous distribution jumps\footnote{This is best explained in \S3.1 of \citet{Squire2017}. In essence, if we consider multiplicative interactions of shocks (in log space) then the statistics of the density and volume jumps of the fluctuations ought to follow discrete, log-Poisson statistics, since each event is discrete. However, the idea underlying compound log-Poisson statistics is that the jump size itself is a continuous random variable, parameterised by the ``intermittency parameter" $T$.} in $s$ between two neighbouring scales (i.e., $\left\{L/\Gamma^n,L/\Gamma^{n+1}\right\},\,\forall n > 1$) in the turbulence, the $(s_{L/\Gamma^n}-s_{L/\Gamma^{n+1}})$-PDF. Then, using an infinite convolution series, as in Equation~\ref{eq:conv_pdf}, of $(s_{L/\Gamma^n}-s_{L/\Gamma^{n+1}})$-PDFs from each $L/\Gamma^{n}$, the $s$-PDF is derived to be: 
    \begin{align}
        p_{\rm{H}}(s;\sigma_{s,V}^2,T) &= I_1(2\sqrt{\lambda u})  \sqrt{\frac{\lambda}{T^2u}} \exp\left\{ -(\lambda + u) \right\}, \label{eq:Hopkins1} \\
        u &\equiv \frac{\lambda}{1 + T} - \frac{s}{T}, \, u \geq 0, \\
        s_{0,V} & = -\frac{\sigma_{s,V}^2}{2}(1+T)^{-1}, \\
        \sigma_{s,V}^2 & =  2\lambda T^2, \label{eq:Hopkins2}
    \end{align}
    where $I_1(x)$ is the modified Bessel function of the first kind, $\lambda$ is associated with the probability of having a density jump (in a Poisson manner) between two arbitrary, neighbouring scales in the turbulence, $L/\Gamma^n$, $L/\Gamma^{n+1}$, and $T$ is the magnitude of the mean logarithmic density jump between these scales,
    \begin{align}\label{eq:T_average_jump}
        T = \Exp{s_{L/\Gamma^n}-s_{L/\Gamma^{n+1}}}_{L/\Gamma^n} = \Exp{\ln\frac{(\rho/\rho_0)_{L/\Gamma^n}}{(\rho/\rho_0)_{L/\Gamma^{n+1}}}}_{L/\Gamma^n}.
    \end{align}
    Hence $T$ can be thought of as an mean $s$-fluctuation between neighbouring scales, averaged over all scales in the turbulence \citep{Squire2017}. Large values (large in this context is not much greater than 0, noting that $T$ has scales similar to that of $s$, as shown in left panel of Figure~\ref{fig:theory_pdfs}) of $T$ implies that the $\Exp{s_{L/\Gamma^n}-s_{L/\Gamma^{n+1}}}_{L/\Gamma^n}$ is significant, due to, for example, low-volume filling, high-mass over-dense regions. We show illustrations of the \citetalias{Hopkins2013} PDF in the left panel of Figure~\ref{fig:theory_pdfs} for different $T$ values, showing how larger values of $T$ increase the negative skewness of the PDF. For small values of $T$, there are only negligible density jumps (on average) between any two neighbouring scales. This means as $T \rightarrow 0$, the $s$-field becomes perfectly self-similar, $\Exp{s_{L/\Gamma^n}-s_{L/\Gamma^{n+1}}}_{L/\Gamma^n} = 0$, and can be approximated by Gaussian-like statistics. Hence $T$ can also be thought of as the average deviation away from perfect self-similarity in the $s$-field, across all scales. Yet another way of interpreting $T$ is by rewriting Equation~\ref{eq:Hopkins2} in terms of the mass-weighted variance, 
    \begin{align}\label{eq:T_sigma_var}
        \left(\frac{\sigma_{s,V}}{\sigma_{s,M}}\right)^2 &= \left( 1 + T \right)^3, \\
        T &= \left(\frac{\sigma_{s,V}}{\sigma_{s,M}}\right)^{2/3} - 1.
    \end{align}
    As we mentioned in \S\ref{sec:intro}, for a lognormal density-fluctuation theory $\sigma_{s,V}^2=\sigma_{s,M}^2$, corresponding to $T = 0$ in the above expression. Therefore non-zero values of $T$ lead a modified relationship between the $\sigma_{s,V}^2$ and $\sigma_{s,M}^2$, which defines a non-lognormal map between the mass-weighted and volume weighted distributions. We will discuss this in more detail in \S\ref{sec:intermittent}. 

    The \citetalias{Hopkins2013} and \citet{Casting1996} models were motivated for isotropic, homogeneous turbulence, and hence may not be phenomenologically appropriate for MHD. Still we find that the PDF, at least empirically, captures the morphology of the $s$ fluctuations well over a wide range of $\M$ and for $\Mao \lesssim 2$. In this study we explore how well this PDF can be extrapolated to the highly-magnetised, highly-supersonic regime and using the $T$ statistic, we explore how the intermittency of $s$ behaves in the presence of strong magnetic fields and high-$\M$, which was a parameter range not explored in the original \citetalias{Hopkins2013} analysis (large $s$ variance, strong $\Bo$-field).
    
    \subsection{Langevin model}
    Recent work by \citet{Mocz2019}, herein \citetalias{Mocz2019}, model the density-PDF of hydrodynamical density fluctuations using a Markov process framework. \citetalias{Mocz2019} construct a Langevin model,
    \begin{align}\label{eq:MoczPDF}
        s(t + \d{t}) &= s(t) + A(s)\d{t} + \mathcal{N}(0,1)\sqrt{D(s)\d{t}},\\
        A(s) &= -\frac{s-s_{0,V}}{\tau_{A}}\left[1 + H(s-s_{0,V})\frac{3f}{2} \right], \\
        D(s) &= \frac{2\sigma_{s,V}^2}{\tau_{D}},
    \end{align}
    where $A(s)$ is the deterministic or advective term in the model, $D(s)$ is the stochastic or diffusive term, and $\mathcal{N}(0,1)$ is a standard normal distribution. The stochastic term, $D(s)$, contains the turbulent fluctuations $\sigma_s^2$, which change on a dynamical timescale
    \begin{align} \label{eq:fluc_timescale}
        \tau_{D} = \ell_{0}/(c_s\M), 
    \end{align}
    where $\ell_{0}$ is the turbulent driving scale. The deterministic term, $A(s)$, encodes how logarithmic density dynamically fluctuates about $s_0$ (as a mean-reverting random walk) on timescales 
    \begin{align}\label{eq:advect_timescale}
        \tau_{A} = \frac{\tau_{A,0}}{[1 + H(s-s_{0,V})\frac{3f}{2}]},
    \end{align}
    where $H(s-s_{0,V})$ is the Heaviside function, $\tau_{A,0}$ is the characteristic timescale for the fluctuations and $f$ is a constant that encodes how high-density structures, such as the density contrast caused by a shock, live on shorter timescales than the rest of the density fluctuations in the fluid \citep{Robertson2018,Scannapieco2018}. For example, for $s > s_{0,V}$, $\tau_{A}$ is reduced by $1 + 3f/2$, and hence $f$ becomes the fitting parameter for how much shorter the dynamical timescales are for the shocked density structures compared to rarefied and mean-density regions in the turbulence. \citetalias{Mocz2019} found $f \approx 0.15$ using 1D simulations over a large range of $\M$, which means shocked regions operate on dynamical timescales $\approx 20\%$ shorter than the low-density regions. In this study, we test if $f$ is constant or rather varies over a large range of $\M$ and $\Mao$, extending the analysis of \citetalias{Mocz2019}.
    
    \begin{figure*}
        \centering
        \includegraphics[width=\linewidth]{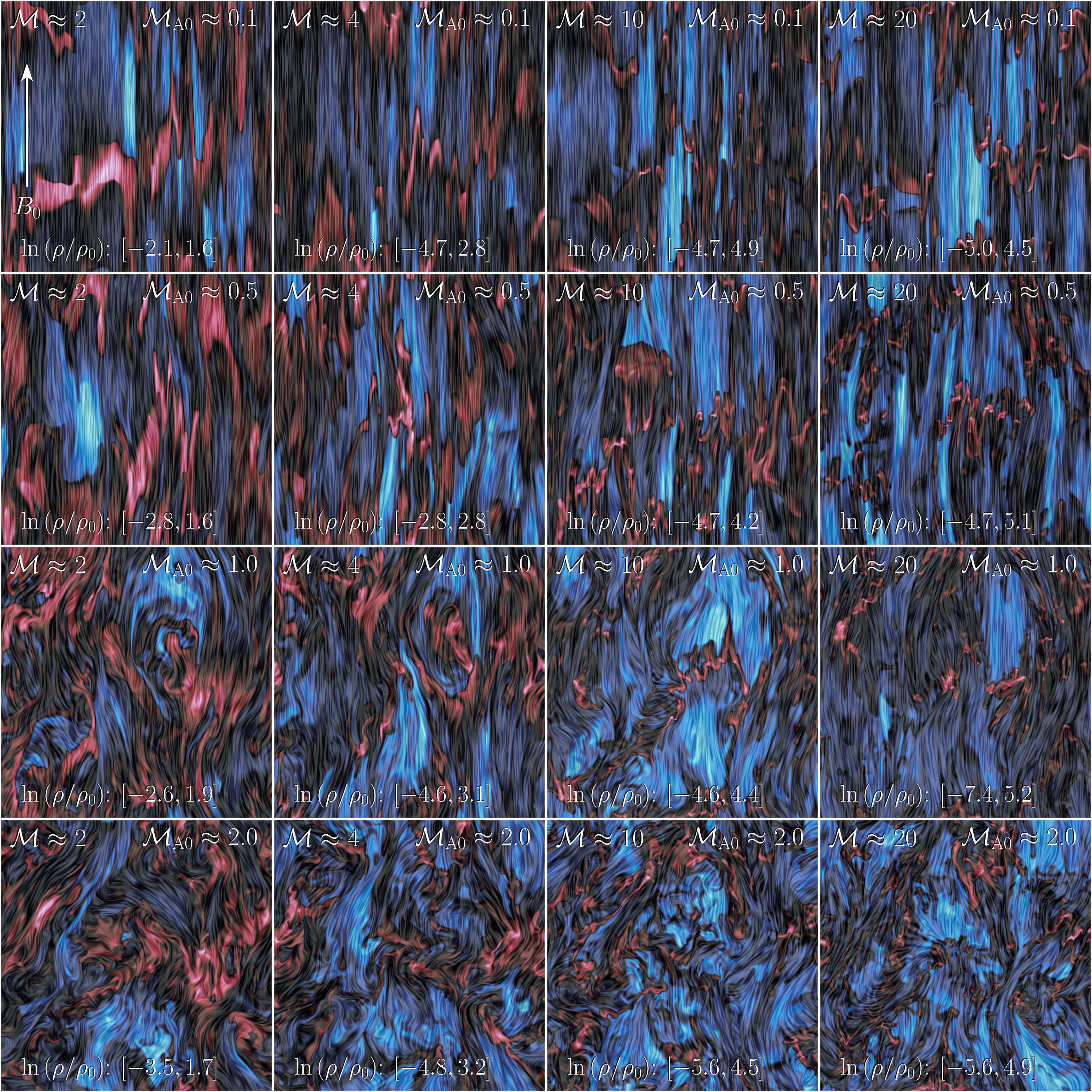}
        \caption{A 2D slice through the 3D logarithmic density field in the $x = L/2$ plane parallel to the mean magnetic field (the direction of $\Bo$ is indicated in the top-left panel). Red indicates the over-densities $(\rho/\rho_0 > 1)$ and blue the under-densities $(\rho/\rho_0 < 1)$; the density range is indicated in each panel. The plots are organised such that the weakest turbulence (smallest root-mean-squared velocities) is on the left, $\M \approx 2$, and strongest on the right, $\M \approx20$, and the strongest mean magnetic field is on the top, $\Mao\approx0.1$ and weakest on the bottom, $\Mao\approx2.0$. Overlaid on each plot is an iterative line integral convolution through the corresponding 2D slice of the magnetic field, revealing the $\vecB{B}$-field topology. Qualitatively, the top panels show density structures that are highly-anisotropic and stretched along $\Bo$ whilst in the bottom panels the structures become more isotropic as the total $\vecB{B}$ becomes dominated by the turbulent component of the field which is, among other things, advected with the supersonic, turbulent velocities.}
        \label{fig:16_panel}
    \end{figure*} 
    
    The PDF of \citetalias{Mocz2019}'s Langevin model, $p_{\rm{MB}}$, defines a solution to the steady-state (time-independent) Fokker-Planck equation,
    \begin{align}
        \frac{\partial }{\partial s}\left[A(s)p_{\rm{MB}}(s)\right] = \frac{1}{2} \frac{\partial^2 }{\partial s^2} \left[D(s)p_{\rm{MB}}(s)\right],
    \end{align}
    which has a solution of the form,
    \begin{align} \label{eq:mocz_pdf_model}
        p_{\rm{MB}}(s) &\propto \exp\left\{ -\frac{(s-s_{0,V})^2[1 + f(s - s_{0,V})H(s-s_{0,V})]}{2\sigma_{s,V}^2 (\tau_{A}/\tau_{D})} \right\},
    \end{align}
    which we show for a number of different $f$ parameters in the right panel of Figure~\ref{fig:theory_pdfs}, where $f=0$ corresponds to no difference between the high- and low-density fluctuation timescales, i.e., a Gaussian PDF, up to $f=10$, which corresponds to high-density structures that operate on timescales 10 times shorter than the low-density regions. It is obvious that the value of $f$ encodes a $3^{\rm rd}$ moment, skewness (discussed more in \S\ref{sec:intermittency_results}), into the PDF through the $\mathcal{O}([s-s_0]^3)$ term in the exponential. This provides a similar morphology to the \citetalias{Hopkins2013} model but, with a softer high-density tail (exponential rather than Bessel function).
    
    We will use the \citetalias{Mocz2019} and \citetalias{Hopkins2013} non-lognormal models as building blocks to interpret results in \S\ref{sec:densityPDF}. To summarise, \citetalias{Mocz2019}'s Langevin model describes skewness in terms of the difference in dynamical timescales between over-dense and under-dense regions in the turbulence. In contrast, the \citetalias{Hopkins2013} model describes a continuous log-Poisson model process, which quantifies the intermittency with the parameter $T$, which is phenomenologically associated with the deviation from perfect self-similarity of the $s$-field, and can alternatively be seen as a modification to the mass-weighted and volume-weighted variance relation. Finally, we note that both of the models we discuss were constructed to address the hydrodynamical phenomenology of intermittency. We believe both models, which in essence rely upon the statistics and timescales of shocked density regions, are able to provide some insight into general supersonic turbulent flows, because large populations of shocked regions form in the density field, with or without a magnetic field present \citep{Lehmann2016,Park2019,Beattie2020,Beattie2020c,Beattie2021}. Before discussing sub-Alfv\'enic $s$-PDF data, we turn to the details of the numerical experiments that we use in our study.
    
\section{Supersonic turbulence Simulations}\label{sec:numerics}

    \begin{figure*}
        \centering
        \includegraphics[width=\linewidth]{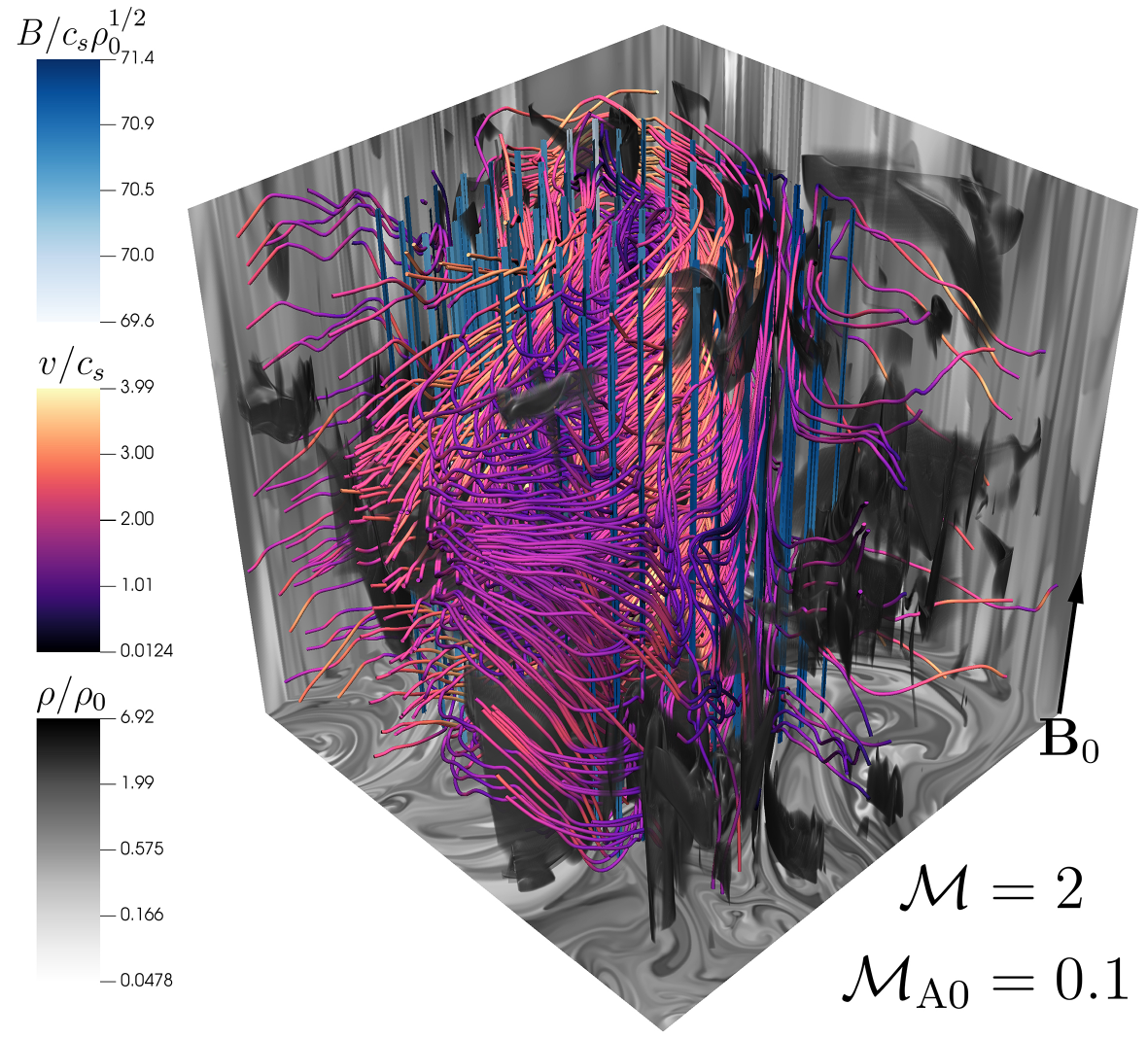}
        \caption{An example of the \texttt{M2MA01} simulation (Table~\ref{tb:simtab}) shown at $t/\tau = 6$ with a section of the magnetic field (blue) and velocity streamlines (magma) surrounded by a box of $\ln(\rho/\rho_0)$ slices in each direction, and with volume-rendered over-dense regions $(\rho/\rho_0 \gg 1)$, shown in black. The velocity structure reveals a deformed vortex twisting around the magnetic field. Typical of sub-Alfv\'enic mean-field turbulence, the magnetic field is extremely ordered, showing no turbulent component at all via the streamline visualisation. This is because all of the energy is contained in the $\Bo$ (direction indicated at the bottom-right of the plot) and in the sub-Alfv\'enic mean-field regime the turbulent velocity fluctuations are too weak to significantly bend the field \citep{Beattie2022_energy_balance,Beattie2022_va_fluctuations,Sampson2022_SCR_diffusion}. However, because $\vecB{B}$ is not fixed at the boundary, the turbulent motions can advect the magnetic fields in the $\vecB{r}\perp\Bo$ plane, establishing supersonic vortical motions with dynamical timescales $\tau = \ell_{0}/(c_s\M)$, where $\ell_{0}$ is the energy injection or driving scale.}
        \label{fig:3d_M2MA01}
    \end{figure*}

    \begin{figure}
        \centering
        \includegraphics[width=\linewidth]{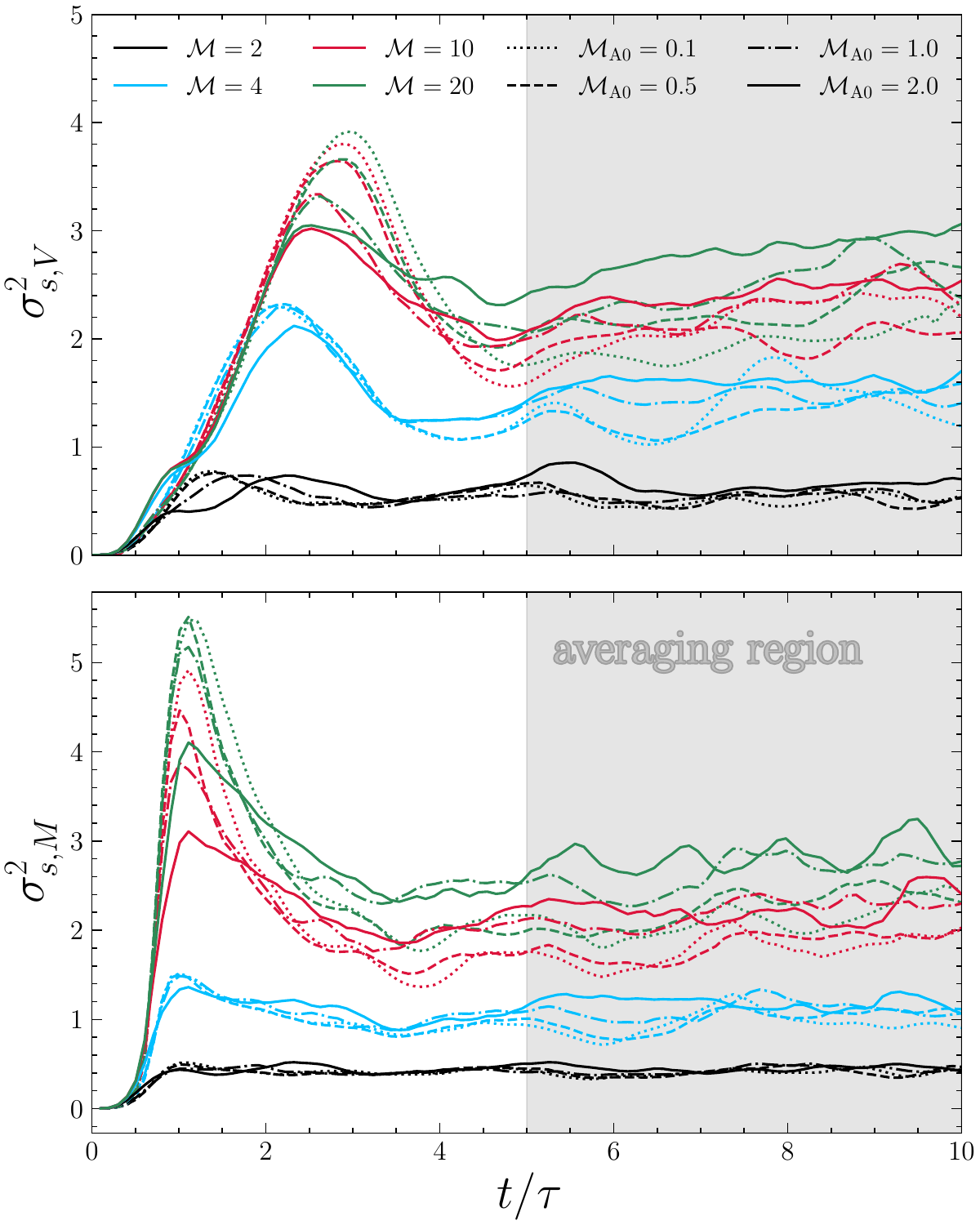}
        \caption{The volume-weighted (top) and mass-weighted (bottom) variances of the logarithmic density field for each of the simulations in Table~\ref{tb:simtab}, as a function of eddy turnover time, $t/\tau$, coloured by $\M$ and with different line styles for $\Mao$. We find that almost all simulations have become approximately stationary between $5 \leq t/\tau \leq 10$, which is the interval where we average over to develop all of the statistics in our study. $\M$ (changing colour) plays the largest role in setting the value of the variance, and the magnetic field (changing line style) only plays a significant role for the high-$\M$ simulations, because it preferentially suppresses small-scale fluctuations that are introduced in high-$\M$ flows \citep{Beattie2021}, coincident with when the turbulence is becoming more globally isotropic.}
        \label{fig:variances_in_time}
    \end{figure}
    
    \input{table.tex}

    \subsection{Ideal, isothermal (M)HD model}
    In this study we analyse the $s$-PDFs using high-resolution, 3D turbulent, ideal, isothermal magnetohydrodynamical (MHD) models,
    \begin{align}
        \frac{\partial \rho}{\partial t} + \nabla\cdot(\rho \vecB{v}) &= 0 \label{eq:continuity}, \\
        \rho \left( \frac{\partial}{\partial t} + \vecB{v}\cdot\nabla \right) \vecB{v} &= \frac{(\vecB{B} \cdot \nabla)\vecB{B}}{4\pi} - \nabla \left(c_s^2 \rho + \frac{\vecB{B}\cdot\vecB{B}}{8\pi}\right) + \rho \vecB{F},\label{eq:momentum} \\
        \frac{\partial \vecB{B}}{\partial t} &= \nabla \times (\vecB{v} \times \vecB{B}),\label{eq:induction}\\
        \nabla \cdot \vecB{B} &= 0, \label{eq:div0}
    \end{align}
    where $\vecB{v}$ is the fluid velocity, $\rho$ the density, $\vecB{B}$ the magnetic field, $c_s$ the sound speed, and $\vecB{F}$ the turbulent acceleration field. To solve the equations we use a modified version of \textsc{flash} based on version~4.0.1 \citep{Fryxell2000,Dubey2008} in a periodic box with dimensions $\V=L^3$, on a uniform grid with resolution $288^3-576^3$, using the multi-wave, approximate Riemann solver framework described in \citet{Bouchut2010} and implemented in \textsc{flash} in \citet{Waagan2011}. We utilise up to approximately 6000 compute cores in parallel, for roughly one million compute hours for the most supersonic and magnetised simulations, running at almost $99\%$ efficiency on the Gadi supercomputer hosted by the National Computing Infrastructure Australia. For details on the performance of the modified \textsc{flash} code we refer the readers to \citet{Federrath2021} and for more details about the current simulations \citet{Beattie2020}, \citet{Beattie2020c}, and \citet{Beattie2021}. With a grid resolution of $\sim 288^3$ many previous studies have found that the 1-point density statistics are well converged \citep[e.g.][]{Kowal2007,Kitsionas2009,Federrath2010_solendoidal_versus_compressive,Price2010,Kritsuk2011,Federrath2013,Mohapatra2021,Beattie2022_va_fluctuations}, hence using a $\gtrsim 288^3$ computational grid is an appropriate resolution for discussing converged $s$ statistics. 
    
    \subsection{Turbulent driving}
    In order to drive turbulent motions we use the methods and code described in \citet{Federrath2010_solendoidal_versus_compressive} and \citet{Federrath2022_turbulence_driving_module}.
    The turbulence-generating acceleration field $\vecB{F}$ follows an Ornstein-Uhlenbeck process that satisfies the stochastic differential equation,
    \begin{align} \label{eq:OU}
        \d{\hat{\vecB{F}}}(\vecB{k},t) = F_0(\vecB{k}) \mathbf{\mathbb{P}}(\vecB{k}) \d{\vecB{W}}(t) - \hat{\vecB{F}}(\vecB{k},t)\frac{\d{t}}{\tau},
    \end{align}
    where $\hat{\vecB{F}}(\vecB{k},t)$ is the Fourier transform of $\vecB{F}$, with correlation time $\tau$, such that $\hat{\vecB{F}}(\vecB{k},t) \sim F_0(\vecB{k}) \exp\left\{ - t/ \tau \right\}$ and $\tau = \ell_0/\sigma_V = L/(2 c_s \M)$ where $\ell_0 = L/2$ is the energy injection scale. Every $\tau$ the driving field loses an $e$-fold of its previous structure. By controlling $\tau$ and $F_0(\vecB{k})$ we are able to set $2 \lesssim \M \lesssim 20$, encapsulating the $\M$ values of supersonic molecular gas clouds in the interstellar medium \citep[e.g.,][]{Schneider2013,Federrath2016_brick,Orkisz2017,Beattie2019b}. $\d{\vecB{W}}(t)$ is a Wiener process, which draws delta-correlated random Gaussian increments from $\mathcal{N}(0,\d{t})$, a mean-zero Gaussian distribution with variance $\d{t}$, which is then projected onto $\vecB{F}$ isotropically in $k$-space with amplitude $F_0(\vecB{k})$. A filter is chosen such that the driving spectrum is concentrated at $|\vecB{k}L/2\pi|=2$ and falls off to zero with a parabolic spectrum between $1 \leq |\vecB{k}L/2\pi| \leq 3$. The projection is performed using the projection tensor
    \begin{align}\label{eq:zeta_equation}
    \mathbb{P}_{ij} = \overbrace{\zeta \left(\delta_{ij} + \frac{k_ik_j}{|k|^2} \right)}^{\rm{solenoidal\; modes}} + \underbrace{(1 -\zeta)\frac{k_ik_j}{|k|^2}}_{\rm{compressive \; modes}},
    \end{align}
    where $\delta_{ij}$ is the Kronecker delta tensor. We control the contribution from each of the driving modes, indicated with the annotations for the two terms in the projection tensor, through the $\zeta$ parameter. For $\zeta = 1$ we obtain purely solenoidal driving $(\nabla \cdot \vecB{F}=0)$, and $\zeta = 0$ produces purely compressive driving  $(\nabla \times \vecB{F}=0)$ \citep[see][for a detailed discussion of the driving]{Federrath2008,Federrath2009,Federrath2010_solendoidal_versus_compressive,Federrath2022_turbulence_driving_module}. Our main, high-resolution experiments are run with $\zeta = 0.5$ to mimic a turbulent source that is mixed with both solenoidal and compressive modes. However, to explore the dependence of the $s$-intermittency on $\zeta$, we run sub-Alfv\'enic $\Mao = 0.1$ and trans-Alfv\'enic $\Mao = 1$ experiments with either $\M = 2$ or $\M = 10$, using $\zeta = \left\{ 0.0, 0.25, 0.5, 0.75, 1.0 \right\}$ turbulent forcing. 
    
    \subsection{Initial conditions and processing}
    The initial velocity field is set to $\vecB{v}(x,y,z,t=0)=(0,0,0)$, with units $c_s=1$, and the density field $\rho(x,y,z,t=0)=\rho_0$, with units $\rho_0=1$. The magnetic field is composed out of a fluctuating, $\delta\vecB{B}(t)$, and a large-scale field, $\Bo$, as $\vecB{B}(t) = B_0 \vecB{\hat{z}} + \delta\vecB{B}(t)$. The $\vecB{B}$-field is initialised with $\Bo$ threaded through $\hat{\vecB{z}}$ of the simulations. This means $\Exp{B_z}_{\V}=B_0$, hence $\partial_t B_0 = \partial_{x_i} B_0 = 0$, $x_i \in \left\{x,y,z \right\}$. The other components satisfy $\Exp{B_x}_{\V}=\Exp{B_y}_{\V}=0$. Because of the periodic boundary conditions and through magnetic flux-conservation, $\Exp{B_z}_{\V}=B_0$ and $\Exp{B_x}_{\V}=\Exp{B_y}_{\V}=0$, $\forall\, t/\tau$. The higher-order moments of the magnetic field evolve self-consistently with the MHD equation with $\Exp{(\delta \vecB{B})^2}^{1/2} \propto \M\Mao$ \citep{Federrath2016c,Beattie2020c,Skalidis2020,Beattie2022_energy_balance}. This corresponds to a systems where $\Bo$ evolves on much larger timescales than $\delta\vecB{B}(t)$. This could be, for example, a cool H$_2$ region that is pierced by a galactic scale mean-field where the large-scale field evolves on galactic timescales, $\sim\mathcal{O}(\rm Gyr)$, and the fluctuating field, akin to the turbulence in the region, on much shorter timescales, $\sim\mathcal{O}(\rm Myr)$. Hence, the mean, coherent field is frozen in magnitude in the turbulence. Even though the gradients of the large-scale field are zero, field lines, which are dominated by $\Bo$ in the sub-to-trans-Alfv\'enic regime \citep{Beattie2022_va_fluctuations}, are able to ``walk" (via turbulent advection) \citep[akin to magnetic field line wander, ][]{Howes2017} in the plane $\perp \Bo$. $B_0$ is set by using the definition of the Alfv\'en velocity and $\M$, $B_0 = 2c_s\sqrt{\pi\rho_0}\M/\Mao$, where $\Mao$ is the desired Alfv\'enic Mach number of $\Bo$. For the magnetised simulations we fix this value between $0.1 \lesssim \Mao \lesssim 2.0$, ensuring that the large-scale field is sufficiently strong compared to the turbulence \citep{Beattie2020c,Beattie2020}. Based on energy balance arguments, in the highly-magnetised, supersonic regime, $\Exp{\delta\vecB{B}^2}_{\mathcal{V}}^{1/2}/|\vecB{B}_{0}| = \mathcal{M}_{A0}^2/2, \, \Mao < 2$ \citep{Beattie2020c,Beattie2022_energy_balance}. For the hydrodynamical simulations we set $\vecB{B} = 0$, which implies $\Mao = \infty$. We use these simulations to compare with the MHD case throughout the study, but they are not the focus of the study. 
    
    We run the simulations from $t/\tau = 0-10$. We construct the $s$-field and bin the data into volume-weighted and mass-weighted distributions. We also extract the volume- and mass-weighted variance and volume-weighted skewness of the $s$-field data. We show the time evolution of the volume- and mass-weighted variance in Figure~\ref{fig:variances_in_time} across the entire run time of the simulations, $0 \leq t/\tau \leq 10$. All results in this study, including $s$-PDF fits, will be based on time-averages across 51~realisations, within $5 \leq t/\tau \leq 10$ to gather data only when the turbulence is in a statistically stationary state, unless explicitly indicated otherwise. We show this region with the grey band in Figure~\ref{fig:variances_in_time}. Compared to hydrodynamical turbulence, which takes roughly $2\tau$ to become stationary \citep{Federrath2009,Price2010} strong ($\Mao \lesssim 2$) large-scale field MHD turbulence takes longer to reach a statistically stationary state, $\sim 5\tau$. Next we discuss the results from the volume-weighted and mass-weighted $s$-PDFs. 
    
    \begin{figure*}
        \centering
        \includegraphics[width=\linewidth]{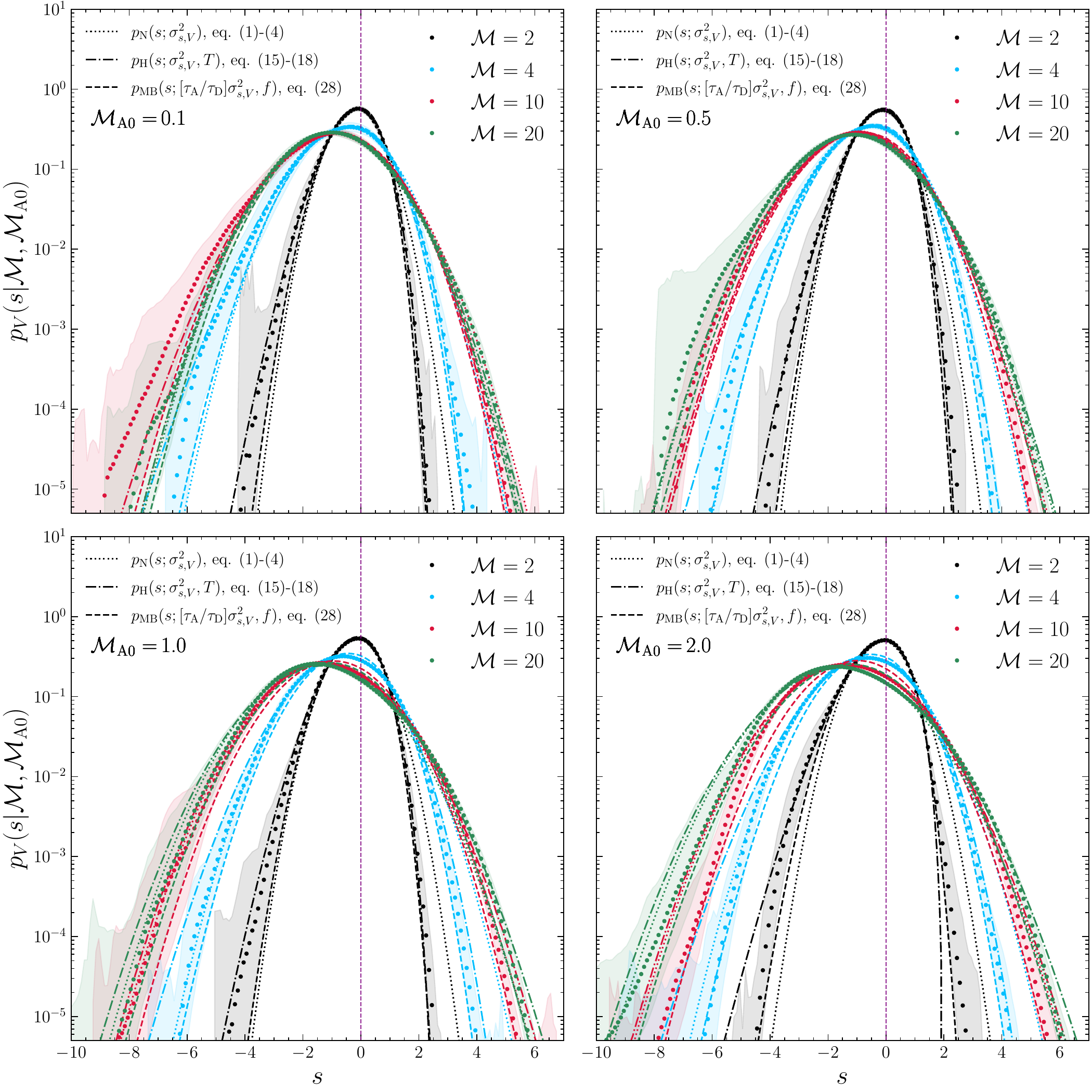}
        \caption{The volume-weighted $s$-PDFs for each of the MHD simulations, shown with $1\sigma$ fluctuations from averaging over 51 time realisations between $5 \leq t/\tau \leq 10$. The \citetalias{Hopkins2013} fit, $p_{\text{H}}(s; \sigma_{s,V}^2,T)$, Equations~\ref{eq:Hopkins1}-\ref{eq:Hopkins2}, is shown with dashed-dots, the \citetalias{Mocz2019} fit, $p_{\text{MB}}(s; [\tau_{\rm A}/\tau_{\rm D}]\sigma_{s,V}^2,f)$, Equation~\ref{eq:mocz_pdf_model}, with dashes, and the Gaussian fit, $p_{\rm N}(s;\sigma_{s,V}^2)$, Equations~\ref{eq:Vdis}-\ref{eq:vars_0 relation}, is shown with dots. Overall, the $p_{\text{H}}(s; \sigma_s^2,T)$ and $p_{\text{MB}}(s; [\tau_{\rm A}/\tau_{\rm D}]\sigma_{s,V}^2,f)$ fits work best for low-$\M$ simulations, where there is an extended low-$s$ tail, and all fits work equally well at high-$\M$. There are significant temporal fluctuations in the low-$s$ tail, which indicates the volume of the rarefaction waves and voids is a highly-volatile quantity, compared to the volume of the highest density structures. We show the $\rho/\rho_0 = 1$ line in purple, which marks the transition between over- and under-densities.}
        \label{fig:3DPDF_v}
    \end{figure*}    
    
    \begin{figure*}
            \centering
            \includegraphics[width=\linewidth]{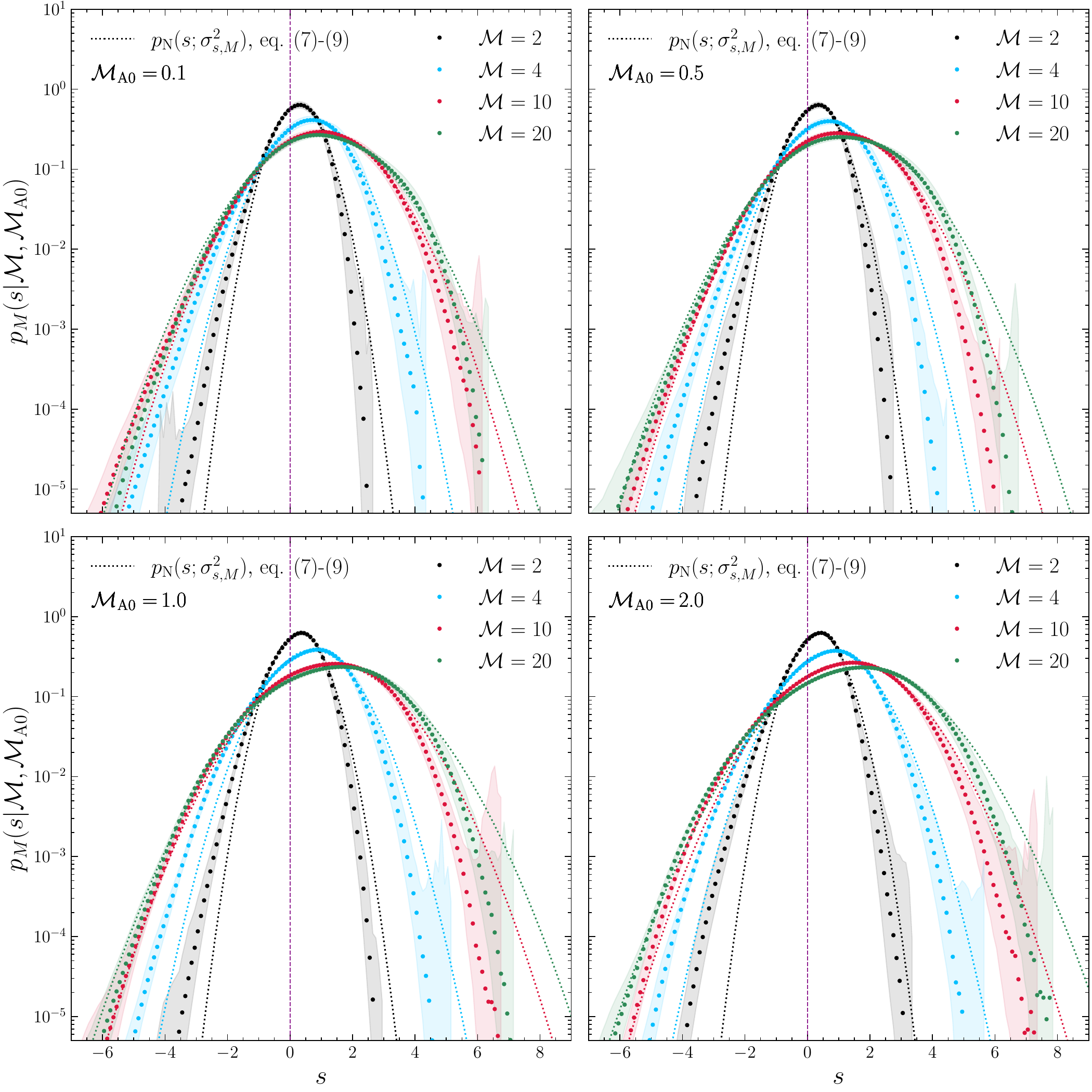}
            \caption{The same as Figure~\ref{fig:3DPDF_v} but for the mass-weighted $s$-PDFs. The Gaussian fit, $p_{\text{N}, M}(s;\sigma_{s,M}^2)$, Equations~\ref{eq:Mdis}-\ref{eq:vars_M}, is shown with dots. Most of the lognormal PDFs overestimate the high-$s$ tail, and underestimate the low-$s$ tail. The high-$s$ tail has significant temporal fluctuations, revealing that the high-mass, high-density structures: filaments, over-dense sheets and shocked regions in the turbulence, fluctuate significantly in mass. Conversely, the lowest density structures, relative to the high-density tail, do not fluctuate significantly.}
            \label{fig:3DPDF_M}
    \end{figure*}

\section{Morphology of the 3D density-PDFs $(\zeta = 0.5)$}\label{sec:densityPDF}
    In \S\ref{sec:intermittent} we discussed two non-Gaussian volume-weighted $s$-PDF models that have been used to describe logarithmic density fluctuations in supersonic turbulence. In Figure~\ref{fig:3DPDF_v} we plot the volume-weighted $s$-PDFs for each of the magnetised simulations from Table~\ref{tb:simtab}. We fit a \citetalias{Hopkins2013} model, $p_{\text{H}}(s; \sigma_s^2,T)$, Equations~\ref{eq:Hopkins1}-\ref{eq:Hopkins2}, \citetalias{Mocz2019} model, $p_{\text{MB}}(s; \sigma_s^2,T)$, Equations~\ref{eq:mocz_pdf_model} and a Gaussian model, $p_{\text{N}, V}(s;\sigma_s^2)$, Equations~\ref{eq:Vdis}-\ref{eq:vars_V} to the $s$-PDFs. The \citetalias{Hopkins2013} and \citetalias{Mocz2019} fits are qualitatively much better than the Gaussian model for the low-$\M$ simulations, which show strong intermittent behaviour, for all $\Mao$. This demonstrates the robustness of both models for describing the $s$-fluctuations, with or without a magnetic field. We draw purple dashed lines where $\ln\rho/\rho_0 = 0$. This shows that the low-$\M$ intermittency is not necessarily from an excess of volume-weighted high-density structures, but from an excess of low-density rarefactions and density voids. As we discussed in \S\ref{sec:densityPDF}, for $\M\gtrsim4$ the fits become, qualitatively, equally as good as the Gaussian model. For the mass-weighted $s$-PDFs, shown in Figure~\ref{fig:3DPDF_M}, we fit a Gaussian model, $p_{\text{N}, M}(s;\sigma_s^2)$, Equations~\ref{eq:Mdis}-\ref{eq:vars_M}. In the next subsections we split the discussion of the volume- and mass-weighted PDFs and their fits into the low- and high-density tail.
    
    \subsection{High-density tail}\label{sec:high_dens_tail}
    The high-density tail of the PDFs in Figure~\ref{fig:3DPDF_v} and \ref{fig:3DPDF_M} trace the shocks, filaments, sheets and other over-dense structures in the supersonic turbulence \citep{Robertson2018,Federrath2021}. These structures make-up a small fraction of the total volume, and most of which is filled with rarefactions and voids that are found in the low-density tail between $1/\M \leq \rho/\rho_0 \leq 1$. However, most of the mass is found between $1 \leq \rho/\rho_0 \leq \M$, so the high-density tail contains most of the mass, but little of the volume \citep{Robertson2018}. 
    
    Because the high-density tail is set by the shocked gas, it is no surprise that \citet{Pan2019} found the tail is amplified by the $\nabla\cdot\vecB{v}$ term in the continuity equation and suppressed by the $\nabla P = c_s^2\nabla \rho$ term in the momentum equation for hydrodynamical turbulence. Since $\nabla\cdot\vecB{v} \sim v/L = c_s \M / L$ for a fixed $L$ the high-density tail ought to increases systematically with $\M$. However, we find that it is asymptotic at high-$\M$, which \citet{Beattie2021} attributes to the total volume of the shocks forming along $\Bo$ reducing until they no longer have a significant contribution to the spread of the PDF. \citet{Beattie2021} suggests that at this point the largest contribution is from weakly compressible MHD shocks that form from field line compressions (fast magnetosonic shocks), which set the maximum limit of the spread of the PDF, along with the type of turbulent driving.
    
    We find that the Gaussian model (shown with dotted markers) systematically overestimates the PDF in the high-density tail for the $\M = 2-4$ simulations, regardless of $\Mao$. This is likely from a mixture of the strong magnetic pressure, $\nabla B^2/(8\pi)$, and the magnetic tension, $(\vecB{B}\cdot\nabla)\vecB{B}$. Both the pressure and tension terms scale with $\mathcal{M}_{\rm A0}^{-2}$, and when $\Mao$ is sufficiently small, the pressure acts to homogenise the density, and the tension acts to smooth out any curvature in the magnetic field, which in turn reduces any large contrasts in the density through $B \propto \rho$ flux-freezing \citep{Landau1959,Mocz2018,Yuen2020}. The overall result is that the magnetic field limits both high- and low-density fluctuations \citep{Nordlund1999,Molina2012,Hennebelle2013,Mocz2018,Beattie2021}. 
    
    In this $|\vecB{B}_{0}| \gg |\delta\vecB{B}|$ regime, the strongest shocks are only able to form along the mean magnetic field \citep{Beattie2020,Beattie2020c,Beattie2021}. This results in less available volume where strong shocks are able to form and therefore fewer over-densities. Because there are fewer shocked regions there is less dense material that is able to contribute to the high-density tail of the PDFs. We explore this in more detail in \S\ref{sec:physical_intermittency} using one-dimensional pencil beams that probe the flow. However, as $\M$ increases to $\M\gtrsim 10$, the $s$-field on large scales becomes more isotropic, allowing for more space where shocks can form, filling up again the high-density tail of the PDF which is then fit well by the Gaussian model. Shock frequencies have previously been found to increases with $\M$ and decrease with $|\vecB{B}_0|$ \citep{Park2019}. \citet{Beattie2020} observed large-scale mixing and isotropisation in the density field by studying the 2D power spectra, which were isotropic on large scales, and then anisotropic on the scales of the individual shocks for $\M\gtrsim 10$. 
    
    In the mass-weighted PDFs, shown in Figure~\ref{fig:3DPDF_M}, we find that the Gaussian model performs best for the high-density tails at low-$\M$, low-$\Mao$ and worst in the high-$\M$, trans-Alfv\'enic flows. Similar to the volume-weighted PDF, the Gaussian model over-predicts the amount of high-density structures in the turbulence. Unlike the volume-weighted PDF, the $1\sigma$ temporal fluctuations (indicated by using transparent bands) from the mass-weighted PDF are strongest in the high-density tail. By tracking and analysing individual over-dense regions \citet{Robertson2018} showed that the temporal fluctuations come from the transient nature of the shock life-cycle, where shocks form, accumulate, lose mass and are torn apart within a fraction of $t_{\rm cross} = L/c_s$. These are found to a much lesser degree in the volume-weighted PDFs, which demonstrates how the highest density structures in the turbulence must dynamically fluctuate through time, in mass, but be relatively stable in volume. The temporal fluctuations in the high-density bins increase with $\M$, supporting the ideas of \citet{Padoan1997}, \citet{Robertson2018} and \citet{Mocz2018}, which state that the internal structure and geometrical properties of \textit{in situ} shocks is set by the properties of the turbulence.
     
    \subsection{Low-density tail}
    The low-density tail captures the density rarefaction waves and voids in the turbulence \citep{Kritsuk2007,Federrath2010_solendoidal_versus_compressive}. These are coupled to the shocks because as the shocks compress the fluid they also evacuate large regions in the turbulence. These regions occupy the largest volumes in the flow, but have the smallest mass \citep{Robertson2018}. For the volume-weighted PDF the low-density temporal fluctuations are significant, shown by the large $1\sigma$ values in the low-density tails in Figure~\ref{fig:3DPDF_v}. This can be interpreted to mean that the volumes of the low-density regions fluctuate throughout the life-cycle of the under-densities and hence the rarefactions are themselves strong sources of intermittency \citep{Passot1998,Kritsuk2007,Federrath2010_solendoidal_versus_compressive}.
    
    For the mass-weighted PDFs in Figure~\ref{fig:3DPDF_M}, the temporal fluctuations are small in the low-density tail, hence, we find an interesting and understandable symmetry between the low-density and high-density regions. The voids and rarefactions in the turbulence fluctuate in volume, and much less so in mass, and the high-density regions fluctuate in mass, and much less so in volume. This is one of the key results from our analysis of the time-averaged PDFs, unrelated to the Gaussianity. 
    
    Similarly to what we found previously for the high-density tail in \S\ref{sec:high_dens_tail}, as $\M$ increases the Gaussian model fits the $s$-PDF better. At low-$\M$, and for most $\Mao$, the low-density tail under-predicts the amount of rarefactions and voids in the turbulence. We have until now restricted most of our attention to the Gaussian fit. In the next section we consider the more general \citetalias{Hopkins2013} and \citetalias{Mocz2019} fits, which have associated intermittency parameters, $T$, that encodes the deviation away from perfect self-similarity through different scales in the $s$-field, and $f$, that captures the difference between the dynamical times of shocked regions and rarefactions, respectively.
    
    \begin{figure}
        \centering
        \includegraphics[width=\linewidth]{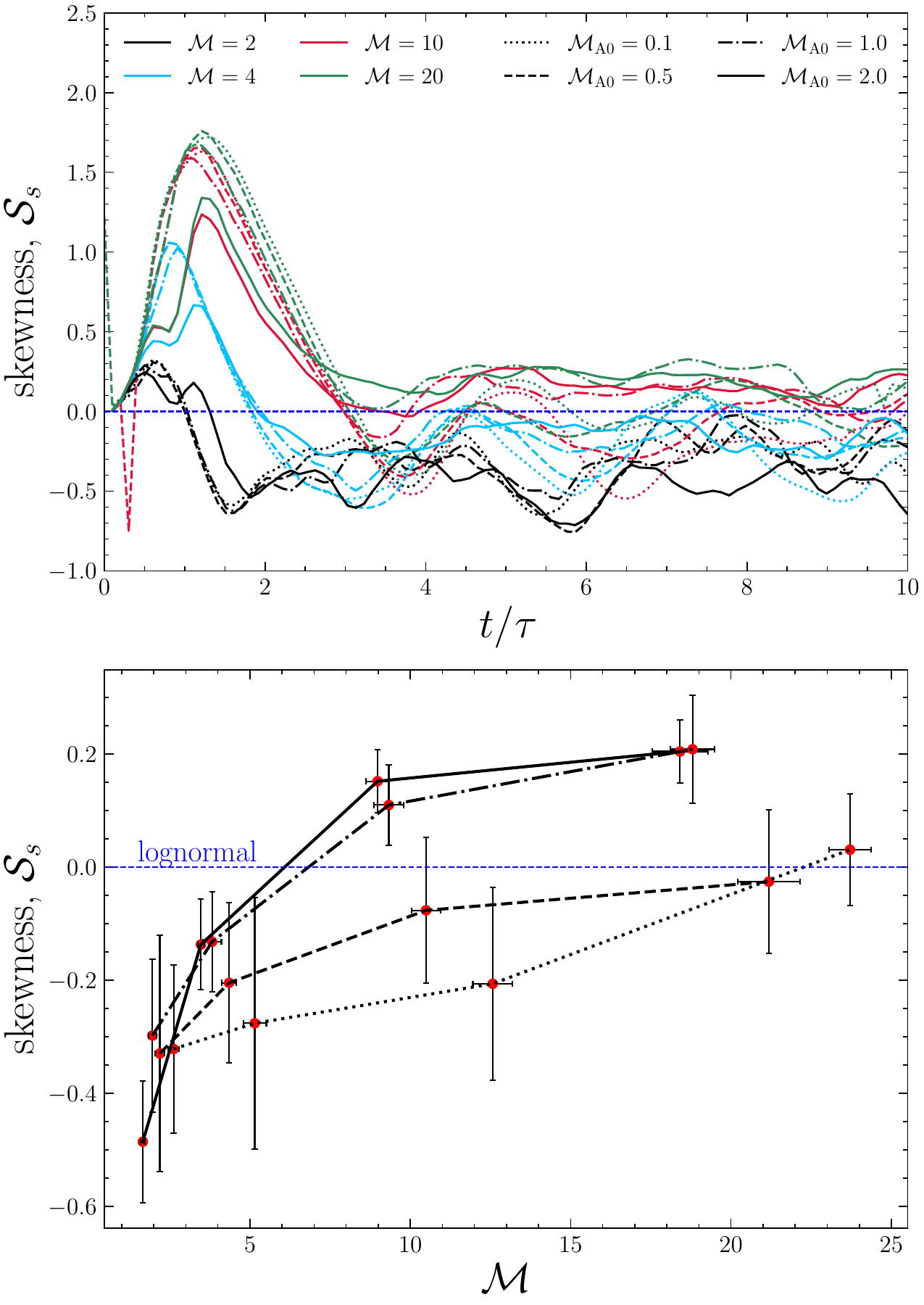}
        \caption{\textbf{Top}: The volume-weighted skewness of the logarithmic density, $\mathcal{S}_{s}$ as a function of turbulent eddy turnover time between $0 \leq t/\tau \leq 10$. As $\M$ increases (black to green curves) the skewness tends towards zero, corresponding to a more Gaussian $s$ (lognormal $\rho/\rho_0$) field. \textbf{Bottom}: $\mathcal{S}_{s}$ as a function of $\M$, averaged over $5 \leq t/\tau \leq 10$, with different line styles for $\Mao$, as indicated in the legend from the top panel.}
        \label{fig:skewness_vol}
    \end{figure}

    \begin{figure*}
        \centering
        \includegraphics[width=\linewidth]{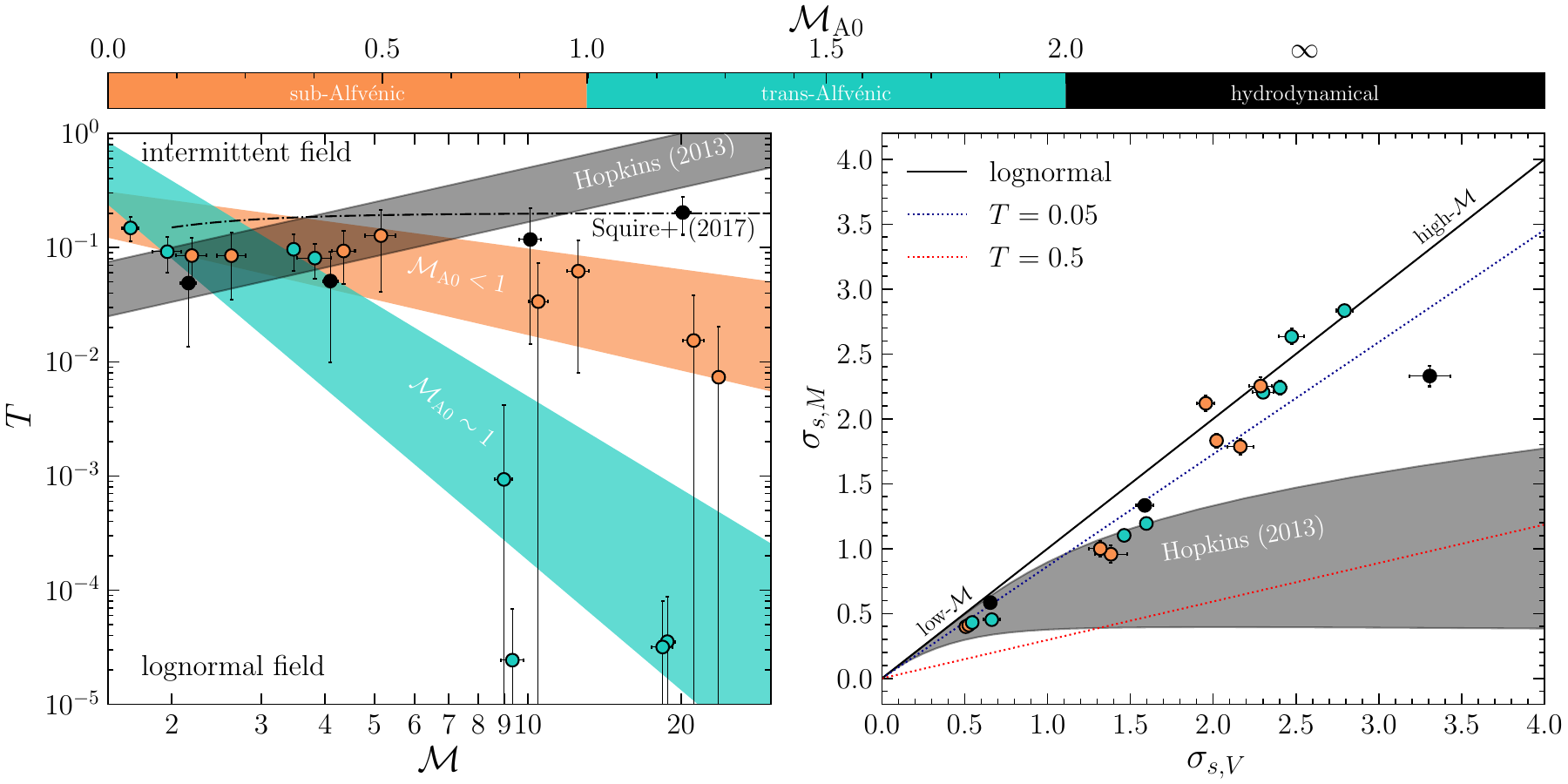}
        \caption{\textbf{Left:} The \citetalias{Hopkins2013} intermittency parameter, $T$, as a function of $\M$, coloured by $\Mao$, split into three categories: sub-Alfv\'enic ($\Mao < 1.0$, orange), trans-Alfv\'enic ($1 \lesssim \Mao \lesssim 2$, aqua) and hydrodynamical ($\Mao = \infty$, black) with the \citetalias{Hopkins2013} hydrodynamical fit shown with the grey band for different driving parameters, $b = 0.3 - 1$. The \citet{Squire2017} fit is shown with the black dot-dashed line, which is based on the shape and density jump of hydrodynamical shocks. A fit to the sub-Alfv\'enic models, $T(\M) = 0.27^{0.39}_{0.19}\M^{-0.81^{-0.60}_{-1.03}}$, is shown with the orange band and trans-Alfv\'enic, $T(\M) = 1.66^{2.51}_{1.10}\M^{-3.24^{-2.70}_{-3.78}}$, with the aqua band. The bands encapsulate the parameter uncertainties for each of the fits. \textbf{Right:} The mass-weighted $s$ dispersion as a function of volume-weighted $s$ dispersion, coloured the same as the left plot. The lognormal model defines the 1:1 line shown in black. The $T = 0.05$ and $T = 0.5$ lines show the cases for weak intermittency and strong intermittency, respectively, from the \citetalias{Hopkins2013} model. We show a band that encapsulated the \citetalias{Hopkins2013} fit with the grey band. In both plots, the $\M=2-4$ simulations show intermittency comparable to the \citetalias{Hopkins2013} fits. However, for $\M \gtrsim 4$ the logarithmic density fields become significantly more Gaussian, with shrinking $T$ and dispersion relation closer to the 1:1 lognormal line, illustrating a key difference between magnetohydrodynamical and hydrodynamical $s$ statistics.}
        \label{fig:variance_compare}
    \end{figure*}
    
\section{Intermittency in the \lowercase{s}-PDF}\label{sec:intermittency_results}
    We qualitatively observed in the volume- and mass-weighted logarithmic density-PDFs that the moderate- to high-$\M$ simulations are fit well by a Gaussian distribution, and the low-$\M$ simulations show significant deviation from the Gaussian fit in both the low- and high-density tail. Now we aim to quantify these non-Gaussian features, focusing primarily on the volume-weighted PDF where the intermittency parameters computed from the  \citetalias{Hopkins2013} and \citetalias{Mocz2019} models are valid.
    
    \subsection{Skewness}
    We quantify the intermittency of the volume-weighted $s$-PDFs in four ways. First, we calculate the $3^{\rm rd}$ central moment, the skewness of $s$. It is defined as
    \begin{align}\label{eq:skewness}
        \mathcal{S}_{s} = \frac{\Exp{\left(s - \Exp{s}_{\V}\right)^3}_{\V}}{\sigma_{s,V}^3}.
    \end{align}
    This gives $\mathcal{S}_{s} = 0$ for the Gaussian distribution, $\mathcal{S}_{s} > 0$ for a distribution with an elongated tail towards the high-density tail and $\mathcal{S}_{s} < 0$ for an elongated low-density tail. Because we compute the central moment the magnitude of $\mathcal{S}_{s}$ is weighted with respect to the volume-weighted variance, $\sigma_s^2$. Hence for $|\mathcal{S}_{s}|\geq 1$ the skewness is just as, or more important for determining the morphology of the $s$-PDF than $\sigma_s$.
    
    We plot the temporal evolution of the skewness for the volume-weighted distribution in the top panel of Figure~\ref{fig:skewness_vol}, coloured by different $\M$ and illustrated with different line styles for different $\Mao$, using the same style as Figure~\ref{fig:variances_in_time}. $\mathcal{S}_{s}$ becomes stationary for $t \gtrsim 5\tau$, which we demonstrate in the top panel. Before $t = 5\tau$, $\mathcal{S}_{s}$ the skewness increases in magnitude with $\M$, which is the opposite trend we found in averaged PDFs. We show the $\mathcal{S}_{s}$ averaged over $5\leq t/\tau \leq 10$ in the bottom panel of Figure~\ref{fig:skewness_vol}. Tracing the different line styles through the plot, which correspond to different $\Mao$, we can immediately see that the absolute value of the skewness decreases and tends towards $\mathcal{S}_{s} \approx 0$ as $\M \gg 1$ for the sub-Alfv\'enic simulations, and $\mathcal{S}_{s} \approx 0.2$ for the trans-Alfv\'enic simulations. For low-$\M$ $\mathcal{S}_{s}$ varies between $-0.3$ and $-0.5$, regardless of $\Mao$, meaning that the peak of the PDF is shifted towards the higher densities, with an elongated tail into the low-densities, consistent with our qualitative findings in \S\ref{sec:densityPDF}, and the type of intermittency described in the \citetalias{Hopkins2013} and \citetalias{Mocz2019} models. The trans-Alfv\'enic simulations show some deviation to higher values than $\mathcal{S}_{s} = 0$ but because $|\mathcal{S}_{s}|$ decreases with $\M$ all of the simulations share the same trend towards more Gaussian statistics. Since we find $|\mathcal{S}_{s}| < 1$, $\mathcal{S}_{s}$ is never more important than $\sigma_{s}^2$ for describing the morphology of the distribution in these strong mean-field, supersonic flows, i.e. a lognormal $\rho/\rho_0$ model is not strictly a bad empirical approximation for the $s$-PDF. However, because any non-zero $\mathcal{S}_{s}$ is a result of asymmetry in the PDF, models that compute astrophysical properties, such as star-formation rate or efficiency from the high-density tail \citep{Krumholz2005,Hennebelle2011,Padoan2011,Federrath2012,Federrath2013b,Burkhart2018}, which is significantly overestimated by the lognormal model (Figures~\ref{fig:3DPDF_v}~and~\ref{fig:3DPDF_M}) may overestimate the amount of shocked and over-dense gas, especially for lower-$\M$, magnetised MCs.
    
    Our $\mathcal{S}_{s}$ results may seem to contradict previous findings of \citet{Kowal2007} and \citet{Burkhart2009}, but they analysed the higher-order statistics of not the logarithmic densities, but the linear densities, $\rho/\rho_0$. The higher-order statistics of $\rho/\rho_0$ and $s$ need not be alike; and in fact we have shown here they are significantly different. Our findings are consistent with those of \citet{Molina2012}, who found that magnetised gas densities are much more lognormal than in hydrodynamical turbulence, but did not quantify it in detail.
    
    \begin{figure}
        \centering
        \includegraphics[width=\linewidth]{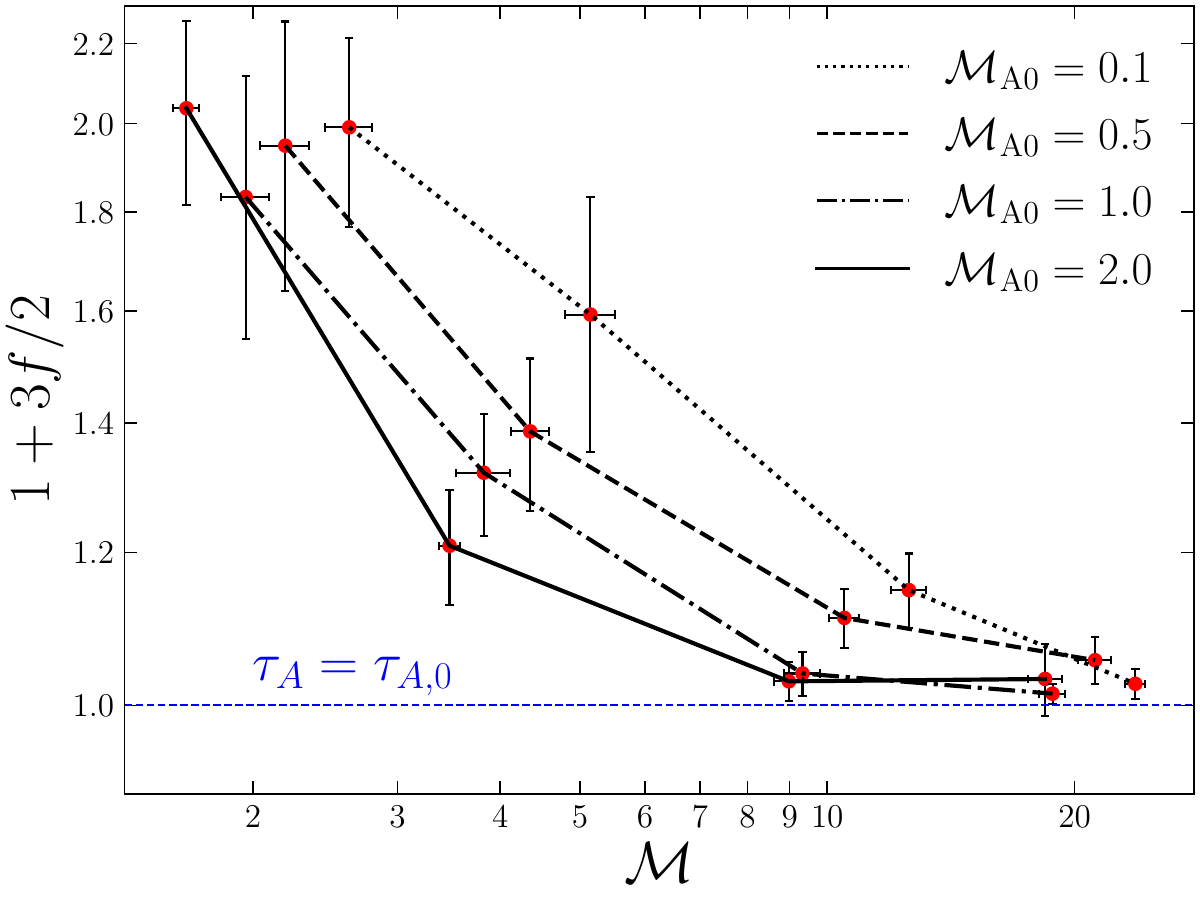}
        \caption{The \citetalias{Mocz2019} intermittency parameter, $f$, written in terms of the reduction factor between the dynamical timescales for the rarefactions and the shocked gas density, $1+3f/2$, as shown in Equation~\ref{eq:advect_timescale}, as a function of $\M$. At low-$\M$ we find that the timescales of shocked gas evolve approximately twice as fast as low-density gas, but as $\M$ increases the timescales homogenise, resulting in more Gaussian $s$-PDFs.}
        \label{fig:f_mach}
    \end{figure}       
    
    \subsection{\citetalias{Hopkins2013} intermittency parameter}
    To understand the transition from the low-$\M$ intermittency to the high-$\M$ Gaussianity we see both in the skewness statistics and directly from the PDFs we plot $T$ as a function of $\M$, coloured by $\Mao$, in the right panel of Figure~\ref{fig:variance_compare}. We overlay an empirical model for the hydrodynamical turbulence that \citetalias{Hopkins2013} fit to the simulation data in grey, $T(\M)\ \approx 0.05\,b\M$. The width of the band demonstrates the impact of the turbulent driving parameter on the fit (where $b$ falls in the range $b = 0.3 - 1$, corresponding to solenoidal to compressive turbulence, respectively; see \citealt{Federrath2010_solendoidal_versus_compressive}). We also show the \citet{Squire2017} model for $T$ with the black dot-dashed line, which is based upon the volume and mass conservation, the size and density contrast of hydrodynamical shocks, $T(\M) = \kappa(1 - \M^{-2})$, where $\kappa = 0.2$ is associated with the shock width being a fraction of $\ell_{\rm s}$ in the turbulence \citep{Federrath2016,Xu2019,Federrath2021}. We also fit our own empirical power-law models of the form $T(\M) \propto \M^{\alpha}$ to the sub-Alfv\'enic and trans-Alfv\'enic data. The bands are associated with the uncertainty in the power-law parameter estimates.
 
    Consistent with our qualitative observations of the PDFs, we find that regardless of $\Mao$, the $\M\lesssim 4$ simulations have \citetalias{Hopkins2013}-like intermittency, shown by how the low-$\M$ $T$ values cluster around the grey model band. We will explore the origin of these intermittent structures in the next section, but for now we just state that the logarithmic density must have some non-Gaussian structures, which lead to a similar value of $T$ as hydrodynamical simulations (shown in black) of the same $\M$. For all of the MHD simulations $T$ peaks in the low-$\M$ and then decreases towards $T = 0$, a Gaussian field, which is the opposite trend found in the hydrodynamical simulations. Note that this does not imply that either \citetalias{Hopkins2013} or \citet{Squire2017} are incorrect, rather that the conclusions made in those studies should not be applied to highly-supersonic, magnetised turbulent flows, assuming that the \citetalias{Hopkins2013} PDF can be used to interpret our MHD data. The sub-Alfv\'enic $T$ values (orange) are fit by a power law, $T(\M) = 0.27^{0.39}_{0.19}\M^{-0.81^{-0.60}_{-1.03}}$ and the trans-Alfv\'enic (aqua) by $T(\M) = 1.66^{2.51}_{1.10}\M^{-3.24^{-2.70}_{-3.78}}$. Hence, the sub-Alfv\'enic simulations seem to be more intermittent than the trans-sonic $\Mao$ simulations, consistent with what we found in the skewness. We interpret this with the \citetalias{Hopkins2013} phenomenology as follows: the average jump in density between neighbouring scales in the turbulence, $T$ (Equation~\ref{eq:T_average_jump}) scales with $\sim \M^{-1}$ in the sub-Alfv\'enic simulations, and $\sim \M^{-3}$ in the trans-Alfv\'enic simulations. We discuss this in much more detail in \S\ref{sec:high_mach_gauss}.
    
    One of the key motivations for \citetalias{Hopkins2013} was to address the discrepancy between the mass-weighted and volume-weighted variances (Equation~\ref{eq:T_sigma_var}), which, as we highlighted in \S\ref{sec:intro}, should be identically equal, $\sigma^2_{s,V} = \sigma^2_{s,M}$, if a lognormal theory describes the density fluctuations. As another independent measure of the intermittency we therefore plot $\sigma_{s,M}$ as a function of $\sigma_{s,V}$ in the left panel of Figure~\ref{fig:variance_compare}. We plot a grey band that encapsulates the empirical relation that \citetalias{Hopkins2013} fit to the hydrodynamical data, the weak ($T = 0.05$, blue, dotted) and strong ($T = 0.5$, red, dotted) intermittency isocontours and the $\sigma_{s,V} = \sigma_{s,M}$ line in black. We find that the low-$\M$ density structures follow more closely the \citetalias{Hopkins2013} intermittency relation, but as $\M$ increases the density field becomes lognormal, tending towards the one-to-one lognormal line with a small amount of scatter, consistent with our previous measurements. 

    \begin{figure*}
        \centering
        \includegraphics[width=\linewidth]{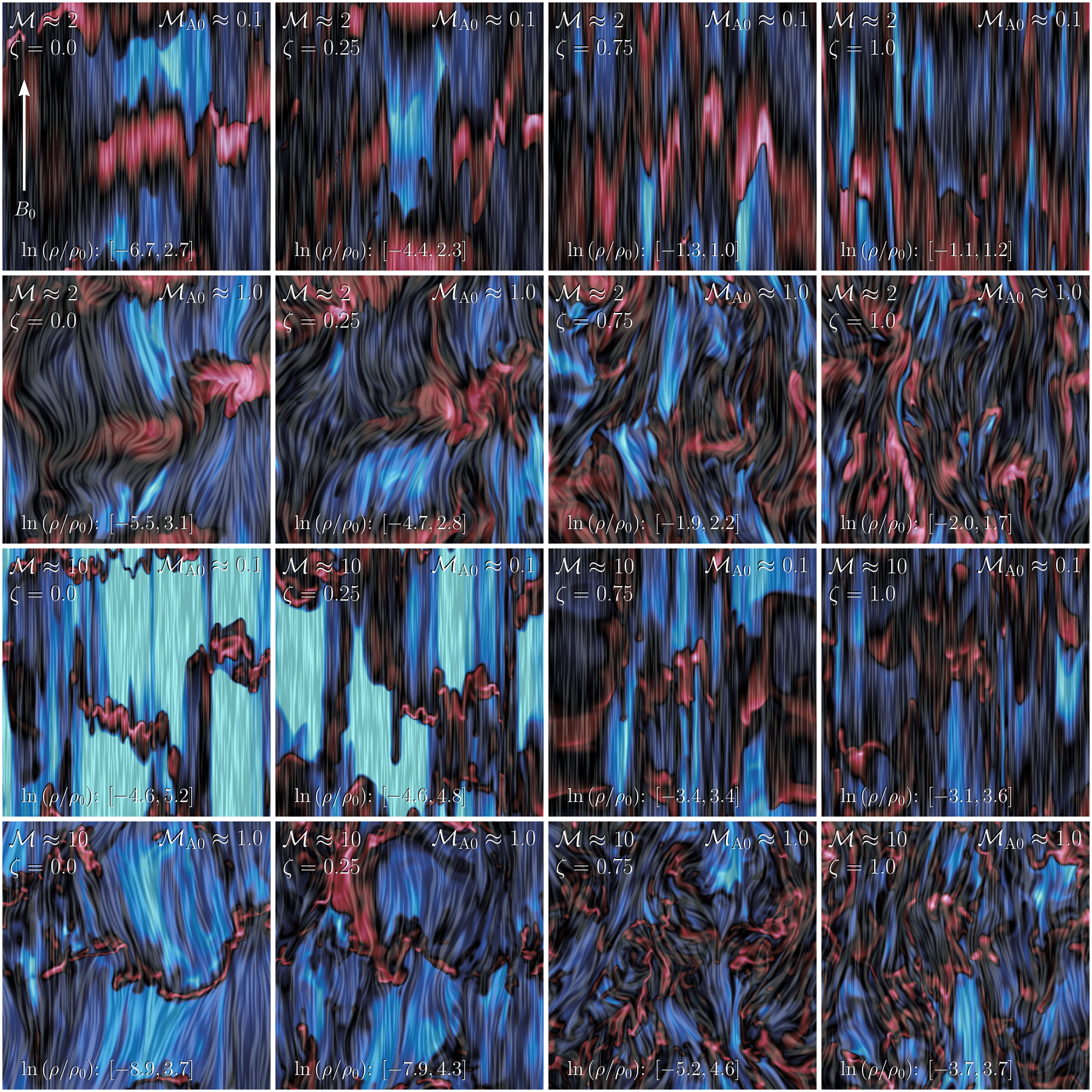}
        \caption{The same as Figure~\ref{fig:16_panel}, but shown for four unique plasma parameters in each row, and different forcing regimes in each column. The first two rows are $\M=2$, varying between $\Mao = 0.1$ in the first row to $\Mao = 1.0$ in the second row. The third and fourths rows are the same as top two rows, but for $\M = 10$. In the first column we show simulations with $\zeta=0.0$ (labelled below $\M$) driving, i.e., the turbulence is excited with purely compressive turbulent modes (see Equation~\ref{eq:zeta_equation}). Likewise, for the last column the turbulence is excited with purely solenoidal modes $\zeta = 1.0$, and the rows in between show incrementally increasing solenoidal fractions in the driving. Qualitatively, the $\zeta=0.0$ plasmas support over-densities packed into small volumes that span across length scales perpendicular to $\Bo$, however as $\zeta \rightarrow 1$, the over-densities become less extreme, and fill more volume in the plasma.}
        \label{fig:16_panel_driving}
    \end{figure*}

    \begin{figure*}
        \centering
        \includegraphics[width=\linewidth]{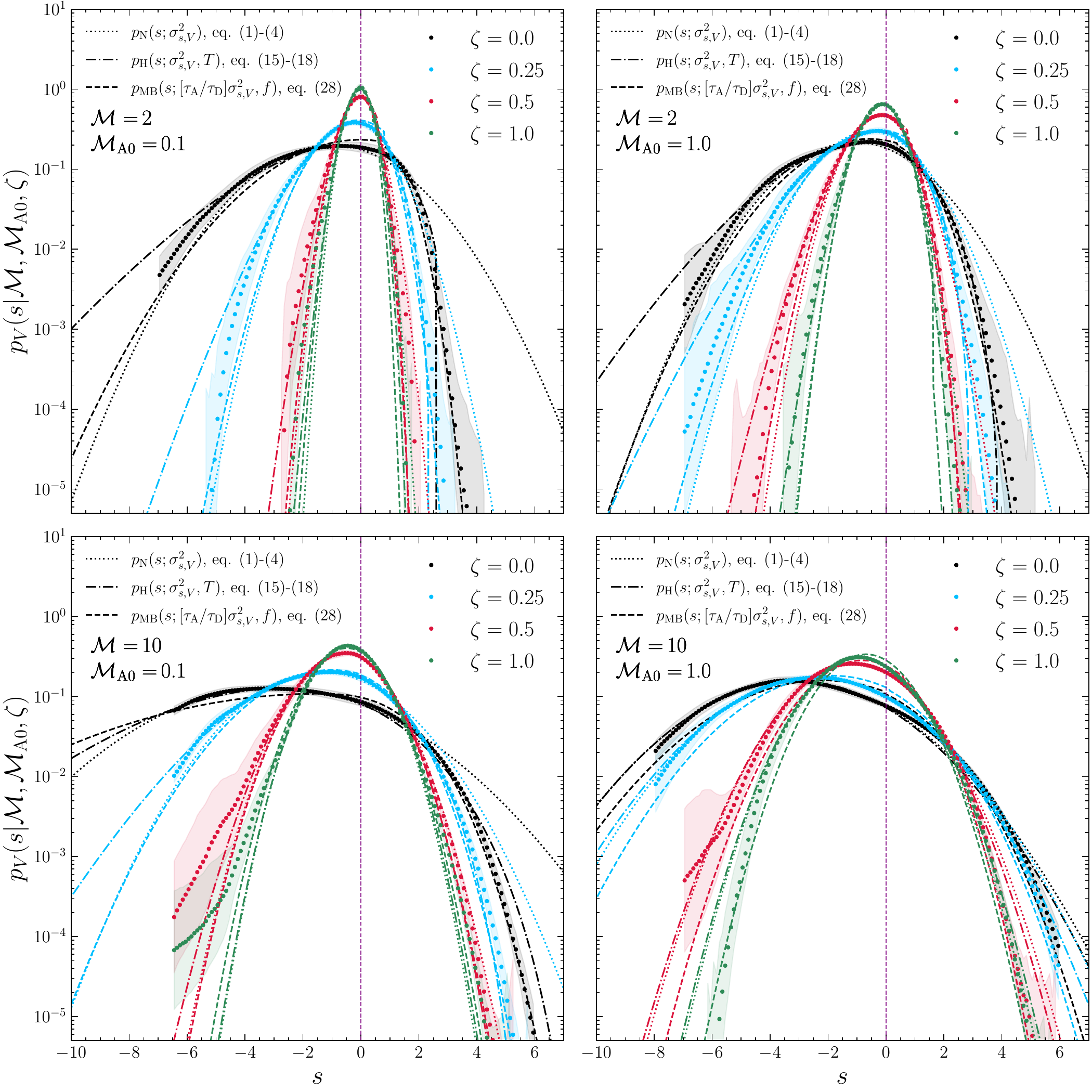}
        \caption{The same as Figure~\ref{fig:3DPDF_v}, but instead of evolving $\M$ in each panel, we change the turbulence forcing parameter $\zeta$ (right-most labels) and fix $\M$ and $\Mao$ (left-most labels). Hence, each panel corresponds to the time-averaged $s$ data from each row in Figure~\ref{fig:16_panel_driving}. As is qualitatively evident in Figure~\ref{fig:16_panel_driving}, $\zeta=0$ simulations are highly intermittent in $s$, and show extreme deviations from lognormality (dotted line) in the $s>0$ tails, whereas the $\zeta > 0.5$ experiments tend to be closer to lognormal.}
        \label{fig:driving_parameter_pdfs}
    \end{figure*}
    
    \begin{figure}
        \centering
        \includegraphics[width=\linewidth]{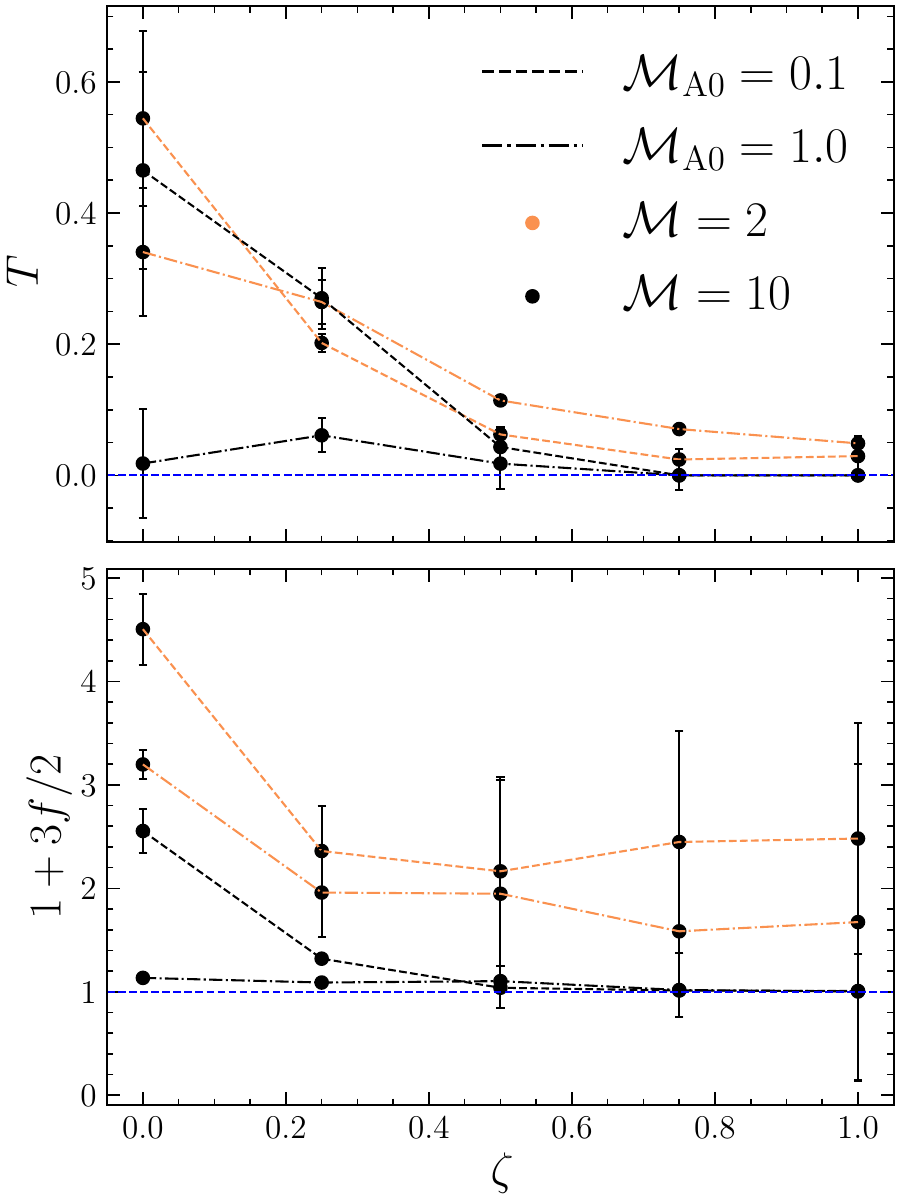}
        \caption{\citetalias{Hopkins2013} (top) and \citetalias{Mocz2019} (bottom) intermittency parameter as a function of $\zeta$ for the $\M=2$ (orange) and $\M=10$ (black) simulations with $\Mao = 0.1$ (dashed) and $\Mao = 1.0$ (dot-dashed). Blue horizontal lines show Gaussian $s$-statistics.}
        \label{fig:driving_parameter_intermittent}
    \end{figure}    
    
    \begin{figure*}
        \centering
        \includegraphics[width=\linewidth]{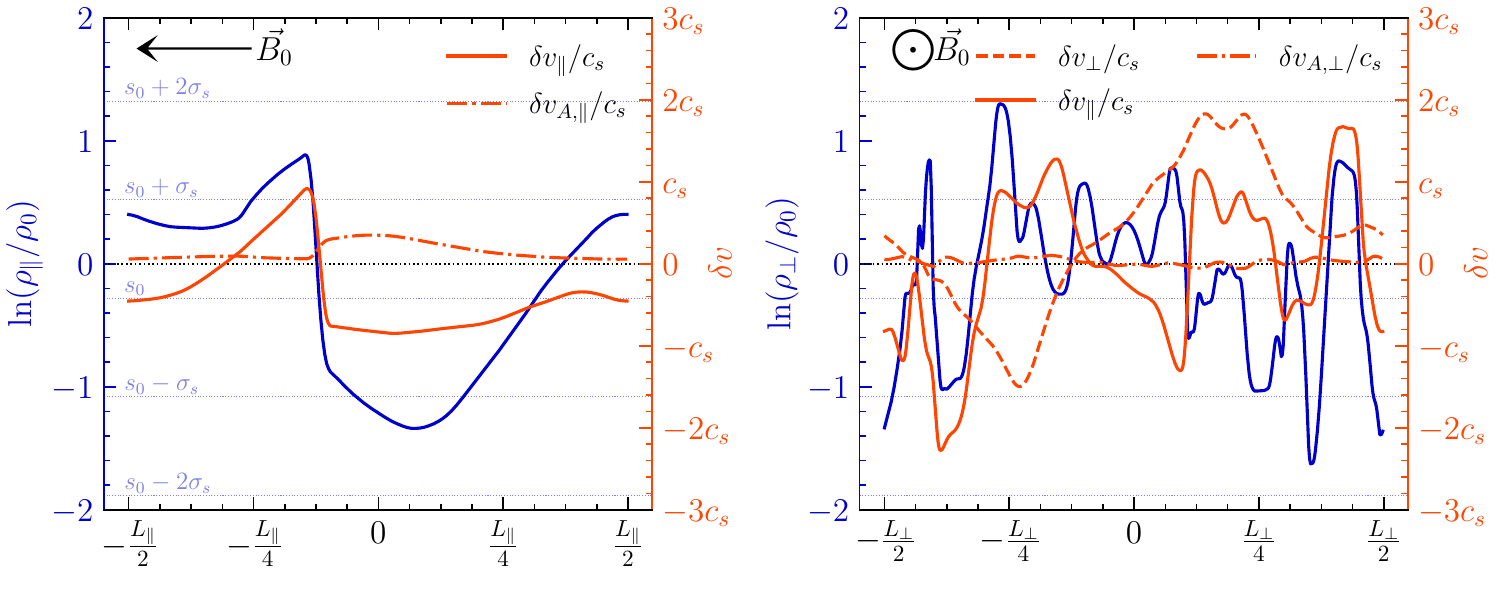}
        \caption{An example of profiles from two typical pencil beams along (left) and across (right) $\Bo$ in the \texttt{M2MA01} simulation. The plot is organised such that the logarithmic density is the blue solid line, corresponding to the left, blue axis, the kinetic velocities are either solid or dashed red lines, corresponding to the right red axis, in units of $c_s$, and Alfv\'en velocity $\delta v_{A} = |\delta \vecB{B}|/\sqrt{4\pi\rho}$, is a dot-dashed red line, also sharing the red axis with the kinetic velocity. Integer values of the global $\sigma_s$ for the $s$ field are shown as horizontal lines across the plots, indicating how extreme the $s$-fluctuations are. The $\Bo$ direction is indicated in the top-left of each panel. \textbf{Left:} The parallel pencil beam reveals large-scale, over-densities with an exponential tail, caused by converging flows (in the grid frame) along $\Bo$, consistent with type~C filament formation \citep{Abe2020}. The structure of the over-density is qualitatively similar to atmospheric shocks studied in \citet{Robertson2018} and \citet{Mocz2018}. A sawtooth shockwave travels down $\Bo$ and is coupled to a volume-filling rarefaction wave, travelling in the opposite direction. At the shockwave interface fast magnetosonic (travelling at $\sim c_s$ with the mean-field subtracted) compression waves are excited that also travel down $\Bo$. \textbf{Right:} The perpendicular pencil beam reveals high-frequency logarithmic density features that are correlated with velocity streams along the field, $\delta v_{\parallel}/c_s$. Large-scale vortical structures are seen in the $\delta v_\perp/c_s$ profile, with two vortices developing on $[-L_{\perp}/2, L_{\perp}/2]$. The shear Alfv\'en wave fluctuations, $\delta v_{\rm A,\perp}/c_s$, are extremely small-scale compared to the kinetic fluctuations.}
        \label{fig:shocks}
    \end{figure*} 
    
    \begin{figure}
        \centering
        \includegraphics[width=\linewidth]{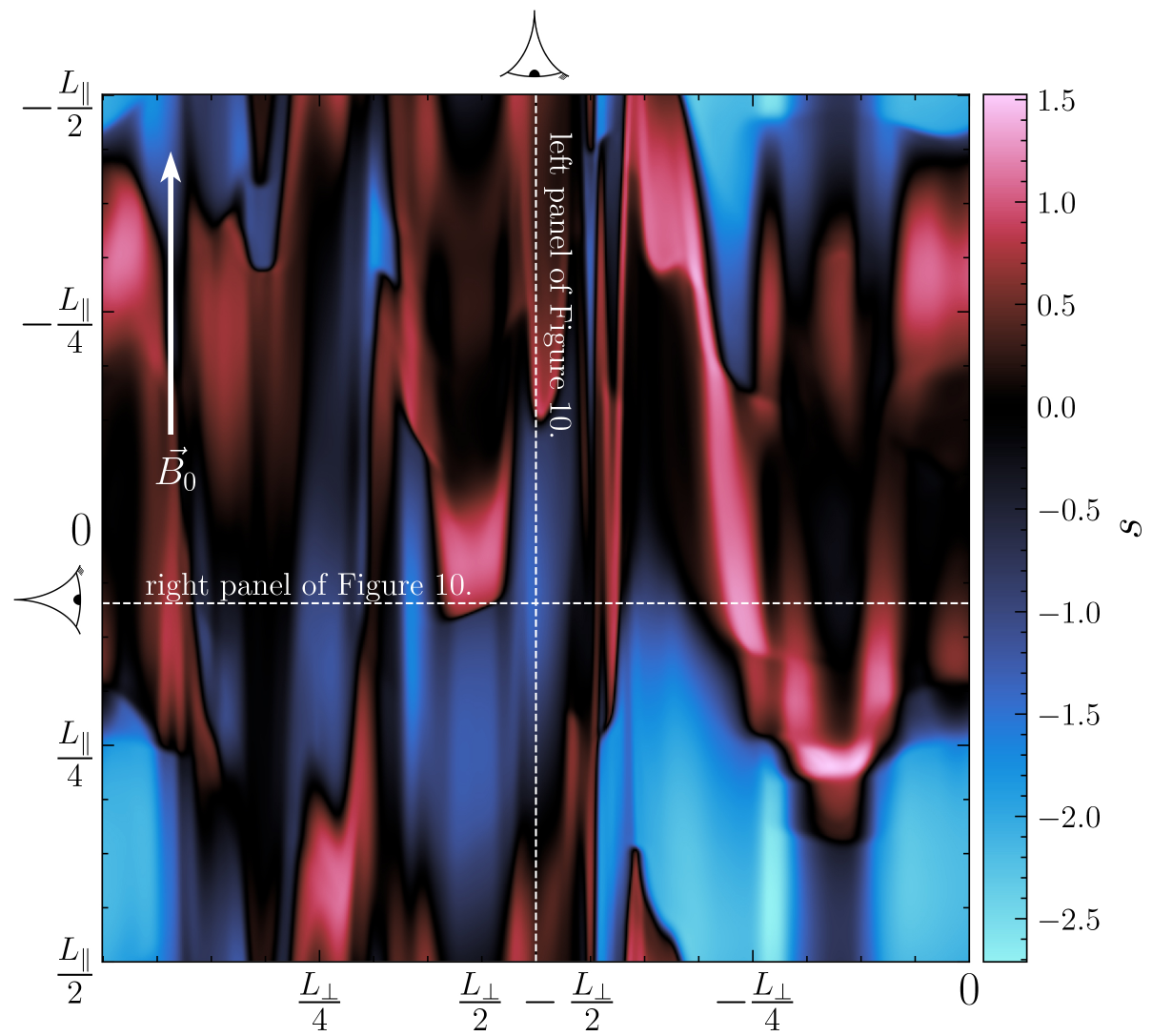}
        \caption{Positions of the 1D pencil beams shown in a 2D slice of the logarithmic density, which are randomly chosen and represent a typical, average pencil beam through the turbulence. The along $\Bo$-field pencil beam (left panel of Figure~\ref{fig:shocks}) is shown with the vertical line and the across $\Bo$-field pencil beam (right panel of Figure~\ref{fig:shocks}) is shown with the horizontal line. The along $\Bo$-field pencil beam has been shifted into the centre of the 2D map for ease of viewing.}
        \label{fig:2D_pencil_beam_positions}
    \end{figure}

    \subsection{\citetalias{Mocz2019} intermittency parameter}
    Using the \citetalias{Mocz2019} fits we extract the intermittency parameter, $f$ (Equation~\ref{eq:advect_timescale}), and plot $1+ 3f/2$ as a function of $\M$, with different line styles for $\Mao$ in Figure~\ref{fig:f_mach}. We use $1+ 3f/2$ because this is the reduction factor between the dynamical timescales for $s > s_0$, $\tau_{\rm A}$ and $s \leq s_0$, $\tau_{\rm A,0}$ structures in the MHD turbulence. The horizontal blue line shows the case where both the under-density and over-density dynamical timescale are the same, i.e. when $f=0$. At low-$\M$, $\tau_{\rm A} \approx \tau_{\rm A,0}/2$, hence the shocked structures are operating on timescales twice as fast as the voids and rarefactions. This results in the most non-Gaussian $s$-PDFs at low-$\M$, similar to what we found using the skewness and $T$ parameter. Consistent with our previous measurements, as $\M$ increases the field becomes more Gaussian, which can be interpreted as the two dynamical timescales becoming equal. The sub-Alfv\'enic simulations take the longest to reach Gaussian-like statistics, and the trans-Alfv\'enic, the fastest, the same as what was found in the $T(\M)$ plots in Figure~\ref{fig:variance_compare}. Again, we discuss and compare these results with the other statistics in more detail in \S\ref{sec:high_mach_gauss}.

    \subsection{Summary of \S\ref{sec:intermittency_results}}
    To summarise this section, we compute the logarithmic density intermittency with four independent measures: (i) the skewness, $\mathcal{S}_s$ (Equation~\ref{eq:skewness}), (ii) the \citetalias{Hopkins2013} $T$ parameter (Equation~\ref{eq:T_average_jump}), (iii) the $\sigma_{s,M}-\sigma_{s,V}$ relation (Equation~\ref{eq:T_sigma_var}) and (iv) the \citetalias{Mocz2019} $f$ parameter (Equation~\ref{eq:advect_timescale}). All statistics point to the conclusion that $s$-intermittency in strong mean-field MHD turbulence is significantly different, and in fact opposite (decreases with $\M$) compared to purely hydrodynamical turbulence (increases with $\M$), which is a key result of this study. However, at low-$\M$ ($\M \lesssim 4$) there considerable \citetalias{Hopkins2013}-like intermittency present in the $s$-field, i.e., $T$ values that are comparable to $T$ values computed for hydrodynamical turbulence. We make the same plots in this section, for each of the intermittency parameters, but as function of $\Mao$ in Appendix~\ref{app:intermittent_Mao_plots}. Until now we have considered the case where there is an equal amount of energy $(\zeta = 0.5)$ in the solenoidal and compressive modes of the source that drives the turbulence, $\vecB{F}$, Equation~\ref{eq:momentum} (the simulations in \textit{Main Simulations}, in Table~\ref{tb:simtab}). However, this need not be the case, and in reality the turbulence may be driven by a diversity of fractions \citep[see Figure~7 in][]{Sharda2022_driving_mode}. Next, we therefore explore the effects of that the energy fraction in the driving modes has on the PDF and intermittency statistics. 
    
\section{The effect of varying the turbulence driving modes}\label{sec:driving_parameter}
    It is a well-established fact that the large- and small-scale fluctuations of the turbulence change in the presence of different ratios of solenoidal and compressive modes $\zeta$ in $\vecB{F}$ (from Equation~\ref{eq:momentum}), the source of the turbulent driving (see top two panels in Figure~14 of \citealt{Federrath2010_solendoidal_versus_compressive} for changes in velocity structure functions in hydrodynamic turbulence and Figure~15 for power spectra). The most drastic effect is perhaps in the $\rho$ (or $s$) statistics, where compressive ($\zeta = 0$) turbulence facilitates the growth of low-volume, high-mass filamentary structures, while solenoidal ($\zeta = 0$) turbulence produces more homogeneous structures. Based on our detailed discussion in \S\ref{sec:intermittent}, this has implications for the higher-order statistics of the $s$-field for these types of magnetised plasmas. We therefore now explore how changing the ratio between solenoidal and compressive modes in the turbulence driving effects the $s$-statistics, using the simulations from \textit{Driving Parameter Simulations} in Table~\ref{tb:simtab}. We perform the same analysis as we did on the $\zeta=0.5$ simulations, starting qualitatively on field visualisations, then turning our attention to the $s$-PDFs, and finally the \citetalias{Hopkins2013} and \citetalias{Mocz2019} intermittency parameters. 
    
    To first develop a qualitative understanding of what varying $\zeta$ does to the $\Mao\lesssim1$ plasma we make the same plot as in Figure~\ref{fig:16_panel}, but showing simulations with different realisations of $\zeta$ (labelled left, under $\M$), Figure~\ref{fig:16_panel_driving}. Note, that similarly to Figure~\ref{fig:16_panel}, each panel has its own independent colour scale, with the minimum and maximum of $s$ indicated in the bottom-left of each panel. In general, for low-$\M$ the structure of the turbulence does not change significantly (i.e., the organisation of $\rho/\rho_0>1$ and $\rho/\rho_0 < 1$ structures in terms of the occupied volume), but the strength of both the voids and over-densities increases as $\zeta\rightarrow0$. This corresponds to increasing the $\sigma_s^2$, but not increasing the intermittency parameters, $T$ or $f$. In contrast, in the lower two rows, when $\M \approx 10$ is significantly high, as $\zeta\rightarrow0$ the $\rho/\rho_0>1$ structures occupy smaller and smaller volumes, as the $\rho/\rho_0<1$ structures occupy more -- exactly the process that facilitates less self-similarity in the $s$-field and increases $T$. We also note that the $\zeta \lesssim 0.25$ simulations exhibit highly-oriented $\rho/\rho_0>1$ structures, consistent with the results in \citet[][e.g., see Figure~2 which demonstrates that the alignment of high-density material is enhanced between the $\zeta=0$ and $\zeta=1$ simulations]{Kortgen2020_alignment_with_driving}. 
    
    We show the time-averaged $s$-PDFs and model fits for each of the $(\M,\Mao,\zeta)$ simulations in Figure~\ref{fig:driving_parameter_pdfs}, in a similar fashion as Figure~\ref{fig:3DPDF_v} (same linestyles for model fits), but now for fixed $(\M,\Mao)$ in each panel, and using the colours to indicate different $\zeta$. We omit the $\zeta=0.75$ $s$-PDFs to avoid cluttering the plots -- the $s$-PDFs all exhibit the same morphology at high-$\zeta$ therefore this is no loss to the reader. There are many similarities between Figure~\ref{fig:3DPDF_v} and Figure~\ref{fig:driving_parameter_pdfs}, so we will focus on the extra information provided by evolving $\zeta$ in each panel and not the same details that were previously discussed.
    
    Firstly, the non-Gaussian features in the $s$-PDFs are amplified as $\zeta\rightarrow0$. This has a significant impact for applications of the $s$-PDF. By comparing the $s>0$ data with the Gaussian fit (dotted line) we find that there are vast (many orders of magnitude in base $e$) deviations from Gaussian $s$-statistics when $\zeta\lesssim0.25$ (light blue and black). This means that \citet{Federrath2012}-esque star formation rate models that integrate the $\rho/\rho_0 > 1$ tail to predict the star formation rate must account for intermittency when $\Mao \lesssim 1$ and the turbulence is being generated by a compressive driving source. In contrast to the purely Gaussian model, both \citetalias{Hopkins2013} and \citetalias{Mocz2019} models perform reasonably well when fitting to the $s>0$ tail ($s$ greater than the purple vertical line), regardless of $\zeta$. Even when the $s$-statistics become very strongly non-Gaussian (see top-left $\M=2$, $\Mao=0.1$ panel for the most extreme case) and the non-Gaussian $s$-PDF models begin to show signs of breaking -- that is, the \citetalias{Hopkins2013} model develops the $s>0$ Bessel function truncation, and the \citetalias{Mocz2019} model develops a $s>0$ humped structure\footnote{We showed that these may develop in the non-Gaussian models when they have extreme values of the $T$ and $f$ parameters in Figure~\ref{fig:theory_pdfs}.} they provide a much better description of the data than the Gaussian model.
    
    We discussed previously in \S\ref{sec:intermittency_results} that the $\M = 10,\,\Mao=1.0$ model hosts the most Gaussian statistics (see Figure~\ref{fig:variance_compare}), which is obviously true based on the morphology of the PDFs. In contrast to the $\M\approx 2$ plasmas, in this regime the Gaussian $s$ model provides a reasonable approximation for the PDF, even at low $\zeta$, maintaining the opposite trend that we find in our MHD turbulence simulations, compared to hydrodynamical turbulence. Now let us quantify the intermittency using the $T$ and $f$ parameters estimated from the PDFs.
    
    We plot the time-averaged intermittency parameters as a function of $\zeta$ in Figure~\ref{fig:driving_parameter_intermittent}, with black ($\M=2$) and orange ($\M=10$) colouring for $\M$ and dashed ($\Mao$=0.1) and dot-dashed ($\Mao$=1.0) linestyles for $\Mao$. In general, as we expect from both the $s$-PDFs and the slice visualisations, as $\zeta\rightarrow0$ the $s$-field becomes more intermittent. The top, $T$ panel suggests this statement is correct for all but the largest $\Mao$ and $\M$ simulation, consistent with what we discussed in the previous paragraph and the morphology shown in Figure~\ref{fig:driving_parameter_pdfs}. A similar picture emerges from the $1+3f/2$  (bottom) panel, where the timescales between the under-and-over-dense gas vary by factors of up to $\approx 5$ in the most extreme case ($\M=2$,$\Mao=0.1$,$\zeta=0.0$). To summarise, unless the turbulence is very strong, or the magnetic field weak, the $s$-fields become monotonically more non-Gaussian as $\zeta\rightarrow0$. 
    
    We have now described the $s$-intermittency in terms of global statistics of the plasma for a range of $\M$, $\Mao$ and $\zeta$, noting that the most extreme intermittency (largest deviations from lognormal $\rho/\rho_0$ statistics) can be found in low-$\M$, low-$\Mao$ plasmas. Therefore, next we turn our attention to identifying what physics leads to this low-$\M$ intermittency by directly identifying the intermittent structures in real-space, following the philosophy from \citet{Imara2021} -- even theoretical astrophysics is, at heart, an observational endeavour that relies upon us being able to visualise the structures we are interested in. 
    
\section{The physics of low-$\M$ intermittency}\label{sec:physical_intermittency}
    In this section we return to the $\zeta=0.5$, high-resolution simulations (\textit{Main Simulations} in Table~\ref{tb:simtab}). \citet{Beattie2020} showed that the density structures in $\M \approx 2 - 4$, sub- to trans-Alfv\'enic turbulence are anisotropic on all length scales. On average, equi-power surfaces in the power spectrum reveal that they are stretched along the mean magnetic field\footnote{Note that on average for $\M \gtrsim 4$ turbulence the density anisotropy is across and not along the mean magnetic field due to formation of perpendicular, high-density filaments; see Figures~2 and 3 in \citet{Beattie2020}.}. For this reason, this regime can naturally be decomposed into two domains: across ($\ell_{\perp}$) and along ($\ell_{\parallel}$) $\Bo$, which \textit{on average} defines a symmetrical axis ($\delta v_x \sim \delta v_y$, $\delta B_x \sim \delta B_y$) in the turbulence\footnote{From time-averaging experiments we find the symmetry is an average property of the turbulence and events, for example, strongly interacting vortices, can lead to velocity streams in preferential directions across the field, ``breaking" the symmetry around $\Bo$.}. Since the \texttt{M2MA01} simulation is the most anisotropic \citep{Beattie2020,Beattie2020c}, this is where such a simple domain decomposition will work the best. Also, since the \texttt{M2MA01} is significantly intermittent (of order the equivalent hydrodynamical simulation), we focus our analysis in this section primarily on the local dynamics and intermittency of this simulation. 
    
    To analyse the real-space intermittent structure we take pencil beams (1D slices) in the $\ell_{\perp}$ and $\ell_{\parallel}$ directions to examine examples of the frozen-in-time local dynamics along each direction. In Figure~\ref{fig:shocks} we show pencil beams for the logarithmic density (blue), kinetic turbulent velocity, $\delta v$, (red, solid, dashed) and Alfv\'en  turbulent velocity, $\delta v_{\rm A} = |\delta\vecB{B}|/\sqrt{4\pi\rho}$ (red, dot-dashed) profiles. To make sure the beams we plot are representative of the average, typical dynamics we check the other $1024$ possible pencil beams and find that these represent that average dynamics in the \texttt{M2MA01} simulation\footnote{See Figure~\ref{fig:3d_M2MA01}.} faithfully (i.e., they are not outliers or peculiar; see the extra 98 random pencil beams in Appendix~\ref{app:random_pencils}). The left panel illustrates a beam along $\Bo$ (magnetic field direction is indicated in the top, left) and the right panel, across. To help the reader contextualise the pencil beams, in Figure~\ref{fig:2D_pencil_beam_positions} we indicate the positions of each of the pencil beams in the frame of a 2D logarithmic density slice. The vertical dashed line shows the position in the left panel of Figure~\ref{fig:shocks} and horizontal in the right panel. 
    
    \subsection{Along $\Bo$-field pencil beam}
    In the left panel we see an \textit{in situ} sawtooth shock coupled to an over-density forming perpendicular to $\Bo$ (solid, blue) from converging kinetic velocities (solid, red line) along $\Bo$, in the simulation frame. Converging flows are a popular hypothesis for filament formation in molecular clouds \citep[e.g.][]{Chen2020,Bonne2020}, and in particular for sub-Alfv\'enic, compressible gases \citep{Padoan1999,Chen2014,Abe2020}. However, the interplay between the magnetic field and gravity orienting the filaments with respect to the magnetic field is still debated in the literature \citep{Planck2016a,Soler2017,Tritsis2018b,Mocz2018,Heyer2020,Pillai2020,Barreto-Mota2021,Girichidis2021}. We show here that perpendicular filaments caused by converging flows along $\Bo$ can be seeded without the presence of gravity, such that dense filaments are primarily created (or at least seeded) by turbulent compression \citep{Federrath2016}. The formation scenario is as follows: the strong $\Bo$-field acts by constraining the turbulence along $\Bo$ (suppressing diagonal velocity modes), compressible modes, which are injected isotropically from our turbulent driving source, are self-organised along $\Bo$, and the gas density along the magnetic field becomes shocked, forming over-dense regions across the $\Bo$. \citet{Beattie2020c} found that the self-organisation of the compressible modes in the velocity was a natural repercussion of the strong $\Bo$, where in the limit of $|\vecB{B}_0|\gg|\delta\vecB{B}|$ the total velocity divergence reduces to just the along-the-field component,
    \begin{align}\label{eq:induction2.1}
        |\nabla \cdot \vecB{v}| = |\partial_{\parallel} \vecB{v}|,
    \end{align}
    where $\partial_{\parallel}$ is the derivative along $\Bo$, and there is only negligible velocity divergence across the field, $\nabla_{\perp} \cdot \vecB{v}_{\perp} \approx 0$, and the compressive motions are along the $\ell_{\parallel}$ beam. It is also worth noting how similar this phenomenology is to the model for weakly-compressible MHD turbulence proposed by \citet{Bhattacharjee1998}.
    
    We highlight and emphasise here that \textit{in situ} turbulent over-densities \textit{are not morphologically the same as simple shock tube models}, which is shown both by the over-density (solid, blue line) and the shock (solid, red line) in the left panel of Figure~\ref{fig:shocks}, which have exponential tails, and sawtooth structure, respectively. This demonstrates that \textit{in situ} over-densities have internal structures (e.g., scale heights), i.e., are not just discontinuities, and that in the strong $\Bo$ regime \citet{Burgers1948}-like turbulence develops along $\Bo$. The over-density we highlight is qualitatively similar to the hydrodynamical atmospheric shocks studied in \citet{Robertson2018} and extended to MHD in \citet{Mocz2018}. \citet{Robertson2018} explains the exponential tail by balancing the ram pressure, which is $\propto \rho v^2$, with the ambient pressure gradient $\nabla P$. We also note how the over-density is approximately consistent with the simple hydrodynamical model proposed by \citet{Padoan2011} for the thickness of the post-shock layers. Like the \citet{Robertson2018} model, \citet{Padoan2011} uses a pressure balance to derive a characteristic width of a shock. For the sawtooth shock we find, $\ln\rho_{\parallel}/\rho_0 = \ln\M^2 \sim 1.4$, and shock width $L_{\parallel}/\M^2 = L_{\parallel}/4$ ($\M=2$ for the \texttt{M2MA01} simulation), which defines a rectangle that captures the basic geometry of the over-density captured in the pencil beam. Many models for the logarithmic density variance \citep[e.g.][]{Molina2012,Federrath2015a,Nolan2015,Beattie2021} rely upon this model for relating the shock jump relations to the variance. \citet{Beattie2021} in particular modelled the variance along $\Bo$ with hydrodynamical shock jump and width conditions, which is supported here.
    
    At the sawtooth shock front (at roughly $-L_{\parallel}/8$ in the left panel of Figure~\ref{fig:shocks}) some fast magnetosonic\footnote{Note here that the MHD waves speeds in the parallel direction are $v = \delta v_{A} + v_{A0}$, but we show the mean-subtracted speeds, $\delta v_{A}$ in Figure~\ref{fig:shocks}. This means that the waves we find in the parallel direction travel at (group or phase) velocities $v \approx \sqrt{c_s^2 + v_{A0}^2}$, consistent with fast magnetosonic waves, which travel at these speeds isotropically in the fluid, at least for trans-sonic turbulence \citep{Makwana2020}.} compression ($\delta B_{\parallel})$ waves are excited (shown with dot-dashed, red line), which propagate down $\Bo$. These magnetic compression waves become amplified in the sub-Alfv\'enic regime when shock formation becomes highly oriented along $\Bo$ \citep{Beattie2020c}, favouring the excitation of high-amplitude, fast magnetosonic compression waves compared to perpendicular shear Alfv\'en waves. \citet{Beattie2020c} found that the compression waves preferentially oppose the direction of $\Bo$, reducing the total $\vecB{B}$, and allowing pressure-supported vortices to be maintained in the turbulence through $\sim\nabla B^2$. Away from the shock front the $\delta B_{\parallel}$ gradient becomes very small, $|\partial B / \partial \ell_{\parallel}| \sim 0$. This has repercussions for the magnetic tension. The magnetic tension written in normal $\vecB{\hat{e}}_n$ and tangential $\vecB{\hat{e}}_t$ $\vecB{B}$-field coordinates is,
    \begin{align}
        (\vecB{B}\cdot \nabla) \vecB{B} = B\frac{\partial B}{\partial \lambda}\vecB{\hat{e}}_t - \kappa B^2 \vecB{\hat{e}}_n,
    \end{align}
    where $\lambda$ is the arc length parameterisation variable along $\vecB{B}$, and $\kappa$ is the $\vecB{B}$-field curvature. If gradients along the field are small, as they are everywhere except at the shock interface in the left panel of Figure~\ref{fig:shocks}, then clearly  
    \begin{align}
        (\vecB{B}\cdot \nabla) \vecB{B} \approx - \kappa B^2 \vecB{\hat{e}}_n \implies (\vecB{B}\cdot \nabla) \vecB{B} \perp \Bo,
    \end{align}
    and the tension becomes purely a restoring force for the Alfv\'enic fluctuations. At the shock interface the parallel $\vecB{B}$-field gradient is very large compared to the amplitudes of the shear Alfv\'en waves as we show in the right panel of Figure~\ref{fig:shocks}, and 
    \begin{align}
        (\vecB{B}\cdot \nabla) \vecB{B} \approx B\frac{\partial B}{\partial \lambda}\vecB{\hat{e}}_t \implies (\vecB{B}\cdot \nabla) \vecB{B} \parallel \Bo.
    \end{align}    
    This demonstrates that in the sub-Alfv\'enic mean-field turbulence regime the $(\vecB{B}\cdot \nabla) \vecB{B}$ term becomes anisotropic. This has repercussions for both the energy cascade in MHD turbulence, where $(\vecB{B}\cdot \nabla) \vecB{B}$ is found to play a role in nonlocally suppressing the kinetic energy transfer \citep{Grete2020}, and the nature of solenoidal modes in the velocity field, where $(\vecB{B}\cdot \nabla) \vecB{B}$ is found to be a strong source of vorticity generation when the $\Bo$ is strong \citep{Lim2020}.
    
    \begin{figure*}
        \centering
        \includegraphics[width=\linewidth]{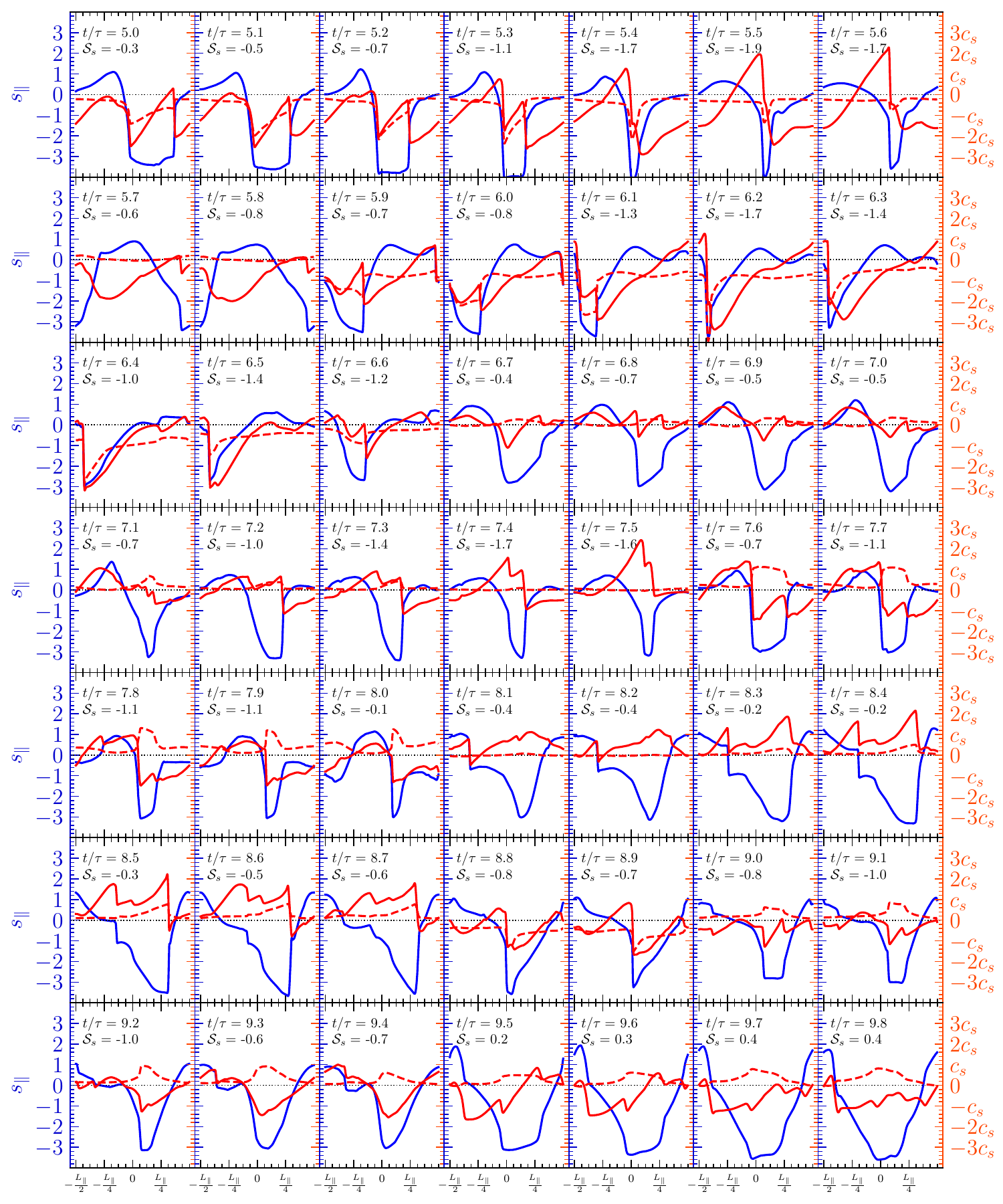}
        \caption{The same as the left panel of Figure~\ref{fig:shocks} but for the events in the turbulence that give rise to the largest under-densities across $5 \leq t/\tau \leq 10$, i.e. a proxy for the events in the turbulence that contribute the most to the \citetalias{Hopkins2013}-like intermittency in the $s$-PDF. The panels are organised such that top-left panel corresponds to the 1D pencil beam at $t/\tau = 5$, and the bottom-right panel is $t/\tau = 9.8$. We find that most of the extreme under-densities ($s_{\parallel}=\ln(\rho_{\parallel}/\rho_0)$, blue, solid lines) correspond to single, strong sawtooth shocks (red, solid lines) that evacuate mass from large-in-volume regions along the direction of $\Bo$. Sometimes these intermittent events are also coupled with large parallel magnetic field fluctuations caused by fast magnetosonic waves (red, dashed lines).}
        \label{fig:intermittent_through_time}
    \end{figure*}    
    
    \begin{figure*}
        \centering
        \includegraphics[width=\linewidth]{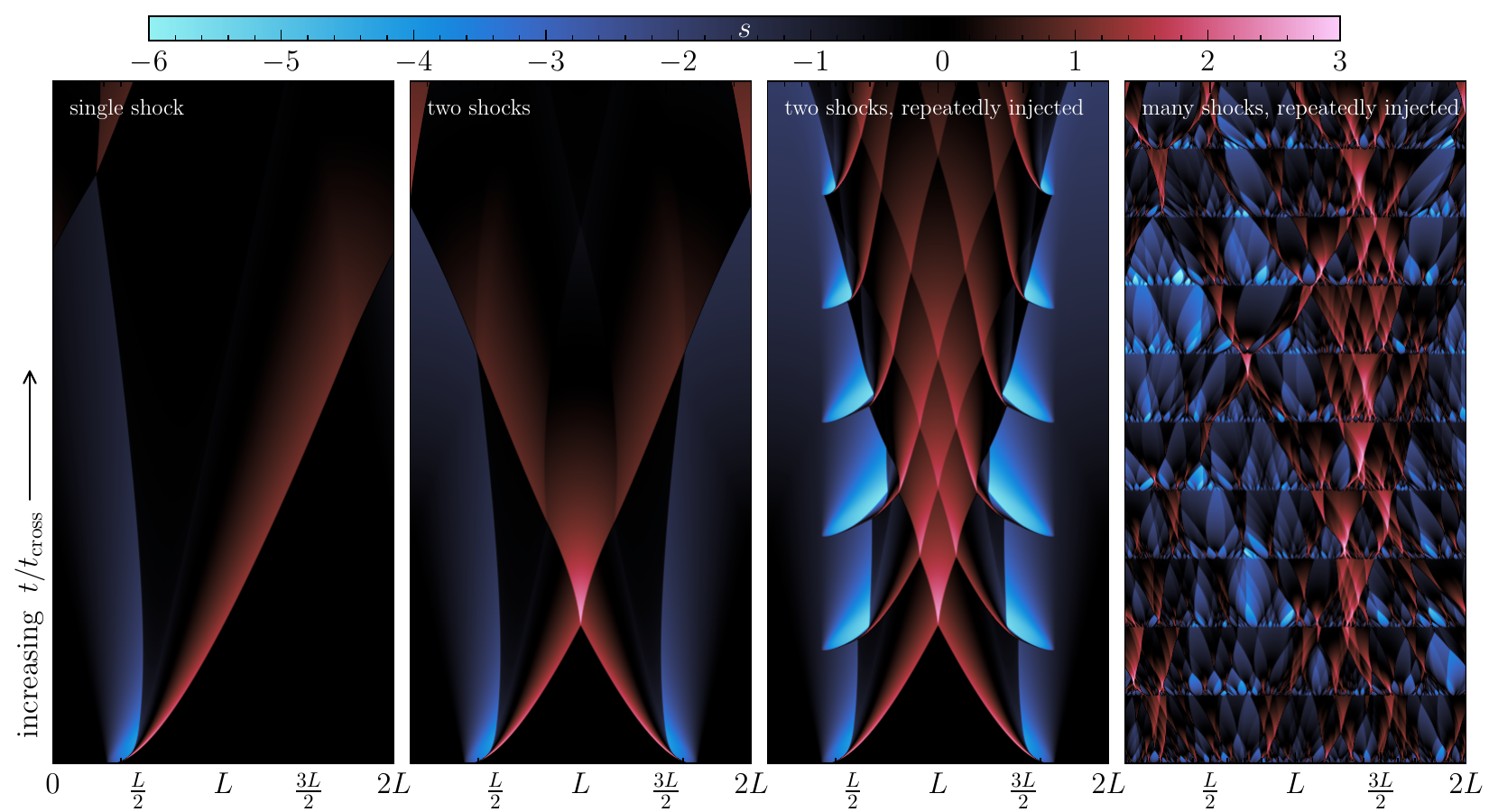}
        \caption{The space-time diagrams of $s$-fluctuations for the four 1D shock experiments: (i) single shock, (ii) two shocks, (iii) two shocks, constant driving, (iv) many shocks, constant driving. We use these experiments to explore what kinds of 1D shock dynamics lead to the largest non-Gaussian components in the $s$-field, analogous to the dynamics that are constrained to along $\Bo$ in the 3D sub-Alfv\'enic mean-field turbulence. $t/t_{\rm cross} = 0$ is at the bottom of the space-times, and $t/t_{\rm cross} = 2$ is at the top.}
        \label{fig:shock_spacetimes}
    \end{figure*}    
    
    \begin{figure}
        \centering
        \includegraphics[width=\linewidth]{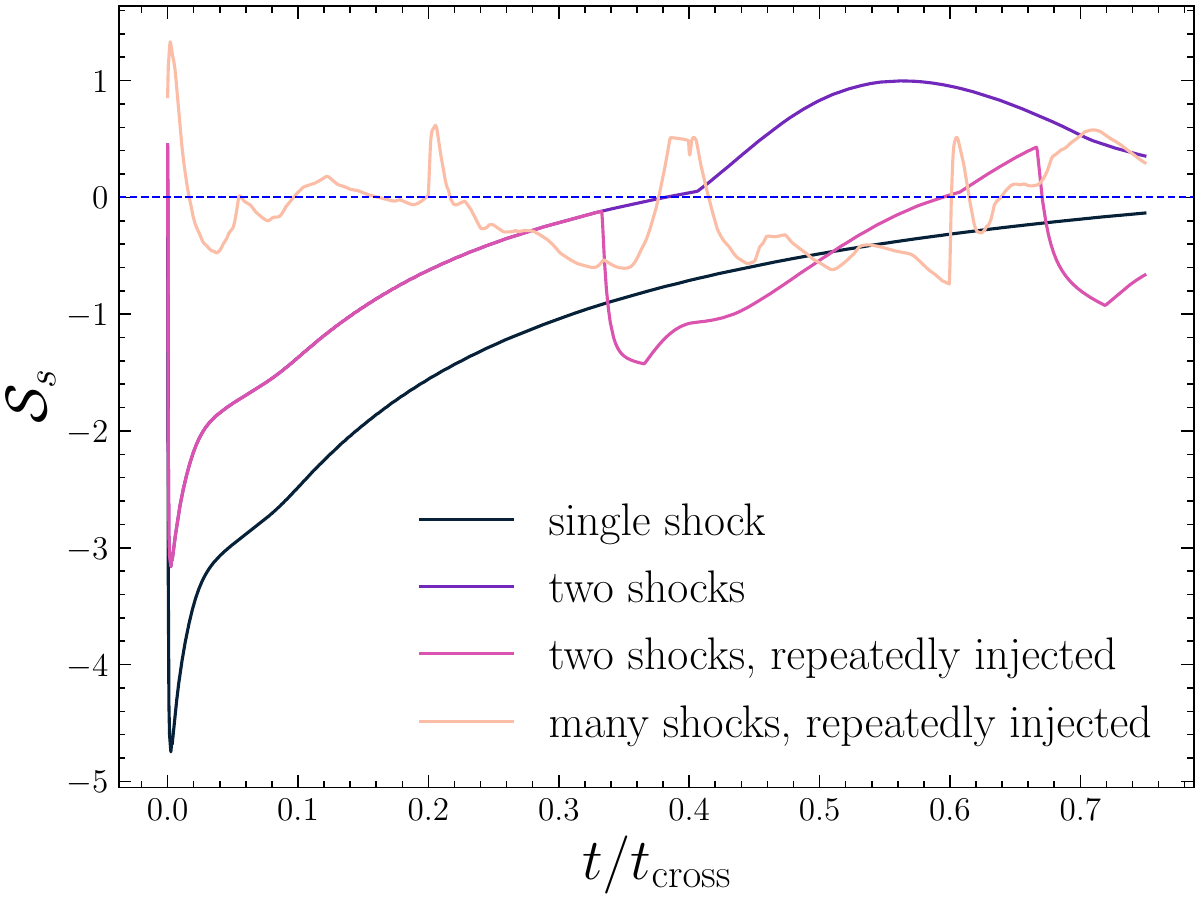}
        \caption{Skewness of $s$ as a function of $t/t_{\rm cross}$ for each of our four 1D shock experiments. The generation of single shocks results in the extreme, negatively skewed $s$-PDFs at the time that the shock is generated. Many-shock simulations give rise to stationary, slightly negatively-skewed distributions that fluctuate through $\mathcal{S}_{s} = 0$.}
        \label{fig:shock_skewness}
    \end{figure}

    \subsection{Across $\Bo$-field pencil beam}
    Turning our attention to the pencil beam perpendicular to $\Bo$ in the right panel of Figure~\ref{fig:shocks}, we find significant amounts of high-frequency density structure, which do not correlate with the perpendicular kinetic velocity (dashed, red line). The perpendicular velocities trace the structure of two large, supersonic, rigid body ($v_{\perp}/c_s \propto \ell/L_{\perp}$) counter-rotating vortices in the plane $\ell_{\perp}$ around $\Bo$. These vortices are visualised in 3D, through the velocity streamlines in Figure~\ref{fig:3d_M2MA01}, and previously in the 3D sub-Alfv\'enic turbulence renderings in Figure~2 in \cite{Beattie2020c}. The large density contrasts, up to roughly $\M^2$, are more strongly correlated with $v_{\parallel}/c_s$ (red, solid line), which we observed were caused from converging flows along $\Bo$. We conclude that the largest over-densities in the strong-$B_0$ MHD regime are caused by convergent flows along $\Bo$, consistent with the anisotropic $\rho/\rho_0$-variance model proposed by \citet{Beattie2021}. The shear Alfv\'en wave fluctuations (dot-dashed, red line) are extremely weak and on small scales compared to both the kinetic velocities and the magnetic compression waves in the left panel, consistent with the ratios between $\delta B_{\parallel}/\delta B_{\perp} > 2$ found for the sub-Alfv\'enic regime in \citet{Beattie2020c}. This suggests that models like \citet{Goldreich1995}, which rely on a critical balance between the turbulent eddies and shear Alfv\'en wave packets may only be appropriate on very small scales in supersonic turbulence, where the flow becomes strongly Alfv\'enic (with respect to the Aflv\'enic fluctuations and not the mean field). This illustrates what is a commonly not considered by some of the astrophysical turbulence community -- sub-Alfv\'enic mean-field turbulence does not mean that the shear Alfv\'en waves are dominating the fluid dynamics. In fact, the perpendicular fluctuating field is clearly super-Alfv\'enic with respect to the kinetic turbulence, and all of the magnetic energy is stored in $\Bo$ \citep{Beattie2022_energy_balance}. 
    
    \subsection{Intermittent events and 1D shock experiments}\label{sec:1d_experiments}
    Now that we understand how to interpret the pencil beams, we turn our attention to using them to seek the events that contribute to the $s$-PDF intermittency. The \citetalias{Hopkins2013}-like intermittency manifests itself in the volume-weighted $s$-PDF by elongating the low-$s$ tail, and truncating the high-$s$ tail. As a proxy for understanding what kind of flow events give rise to the elongated low-$s$ tail we create pencil beams that intersect through the grid element in the simulation that has the minimum value of $s$, $s_{\rm min}$. We construct parallel pencil beams, because as we noted in the previous section, the largest density contrasts are formed by compressive motions along $\Bo$. We show these pencil beams in Figure~\ref{fig:intermittent_through_time} for the density, and kinetic and fast magnetosonic field fluctuations (with the same colouring and line style scheme as Figure~\ref{fig:shocks}), for time realisations between $t/\tau = 5-9.8$, where $t/\tau = 5$ is shown in the top-left corner, and $t/\tau = 9.8$ in the bottom-right corner of the plot. We annotate the skewness, $\mathcal{S}_s$, in the top-left corner of the plot.

    Our first conclusion is that many of the 49 visualised extreme low-$s$ events are formed by single, strong shocks, which is especially apparent in the time series from $t/\tau = 6.2-7.5$. In this time-correlated\footnote{Note that there need not be any time correlation, by construction, but clearly some of the most under-dense events are so extreme that they define $s_{\rm min}$ for a full $\sim \tau = L/(2c_s\M)$.} set of pencil beams we see a $\M \approx 3$ shock (a $\sim 1.5\sigma$ event in the velocity field for the $\texttt{M2MA01}$ simulation) moving down the field, evacuating a large void in the turbulence, where $s_{\rm min}$ resides. The strong shock also facilitates an equally large fast magnetosonic modes. Clearly this hints at density, magnetic and velocity intermittency being linked through these strong-shock events, and explains the physical origins of the fast magnetosonic wave intermittency studied in \citet{Ho2021}, i.e., as a side effect of shock generation in the supersonic plasma. The median, 16$^{\rm th}$ and 84$^{\rm th}$ percentiles for the skewness of all of the intermittent beams is $\mathcal{S}_s = -0.70^{-1.35}_{-0.31}$, almost a factor of 2 more skewed than the average $\mathcal{S}_s$ calculated for the global $s$-field in \texttt{M2MA01} in \S\ref{sec:intermittency_results}. This demonstrates that these events contribute significantly to the non-Gaussian features of the $s$-PDF. The most skewed beam is at $t/\tau = 5.5$, with $\mathcal{S}_s = -1.9$, where a strong, almost $\M \approx 4$ ($2\sigma$ event in the velocity) shock compresses an under-dense region into a thin, low-volume rarefaction. 
    
    A few of the panels show multi-shock interactions along $\Bo$ give rise to $s_{\rm min}$, like, for example, at $t/\tau = 5.0$ and $t/\tau = 8.4$, where two sawtooth shocks travel along the field. To confirm that strong, single shocks, give rise to the most non-Gaussian features we perform four simple numerical experiments. We give details about the numerical experiments in \S\ref{app2:1d_shocks}. To summarise the experiments, we follow the 1D setup in \citet{Mocz2018}, solving the compressible Euler equations on a periodic domain $x \in [0,2L]$ and creating shocks by perturbing the velocity field with a Gaussian pulse that has an amplitude which sets the root-mean-squared $\M$. We choose $\M \approx 2$, mimicking the characteristic shock velocities from the \texttt{M2MA01} simulation. The four experiments are:
    \begin{enumerate}
        \item a single shock travelling across the domain,
        \item two shocks colliding at the centre of the domain,
        \item two shocks repeatedly injected,
        \item multiple shocks repeatedly injected.
    \end{enumerate}
    We show the space-time diagrams for $s$, in each of the experiments in Figure~\ref{fig:shock_spacetimes}, from left-to-right for experiments (i)-(iv), and the $s$-profiles as a function of $t/t_{\rm cross}$ in Figure~\ref{fig:shock_experiments_time evolution}. In Figure~\ref{fig:shock_spacetimes}, the over-densities are coloured in red, which trace the shock fronts and the under-densities in blue, which are almost stationary in the first three experiments. We evolve the experiments for $2t_{\rm cross}$, which is shown by going up the vertical axis as indicated with the $t/t_{\rm cross}$ annotation. In the experiments (iii) and (iv) one can see the that repeated pulses in the $s$-field give rise to intricate, interacting shock networks, analogous to \citet{Burgers1948} turbulence \citep{Mocz2018}. 
    
    We compute the $\mathcal{S}_s$ for each of the experiments as a function of $t/t_{\rm cross}$, shown in Figure~\ref{fig:shock_skewness}. We only take statistics for $0 \leq t/t_{\rm cross} \leq 0.8$, the time for which the periodic boundaries in the first three experiments do not play a role in the flow. We find that the largest $|\mathcal{S}_s|$ values, $\mathcal{S}_s \approx -5$, are associated with the single shock experiment, consistent with our previous intermittent pencil beam analysis. For both the one-shock (i) and two-shock (ii), (iii) experiments, the initial generation of the over-dense regions, where both the under-density and over-density occupy the smallest volumes, have the most negative $\mathcal{S}_s$, similar to the features we saw in the most non-Gaussian event in Figure~\ref{fig:intermittent_through_time} at $t/\tau = 5.5$. The interaction between the two shocks in experiment (ii), which happens at $t/t_{\rm cross} \approx 0.4$, results in $\mathcal{S}_s > 0$ as the shocked gas fills the volume of the domain after the interaction. The two-shock experiment with repeated injections almost matches experiment (ii), but the $\mathcal{S}_s$ reduces after each pulse in the driving, before slowly relaxing back to a more Gaussian field. The many-shock experiment settles into a stationary state quickly, fluctuating around $\mathcal{S}_s = 0$, slightly tending towards a negative value of $\mathcal{S}_s$, similar to our global $\mathcal{S}_s$ statistics we computed in \S\ref{sec:intermittency_results}. Even in this experiment it is probably the rare, single, strong shocks that we can see traced by the red, over-dense gas, that contribute to the non-Gaussian components by creating the strongest under-densities in the $s$-field. 
    
    The key conclusion from this analysis is that we find strong, single shocks in the 3D, low-$\M$, sub-Alfv\'enic mean-field turbulence that give rise to some of the lowest values in $s_{\rm min}$. We show that it is these strong, single shocks that give rise to the greatest non-Gaussian contributions to the $s$-PDF. We have now solidified our understanding of the $s$-intermittency for the $\M \lesssim 4$ sub-Alfv\'enic mean-field turbulence, and now turn focus on the $\M \gtrsim 4$ regime.
    
    \section{Phenomenology of high-$\M$ Gaussianisation}\label{sec:high_mach_gauss}
    In this section we outline a number of reasonable explanations for the high-$\M$ Gaussianity that we observed in the trans- and sub-Alv\'enic mean-field regime in Figures~\ref{fig:skewness_vol}, \ref{fig:variance_compare}~and~\ref{fig:f_mach}. The arguments for this are based upon the self-similarity of $s$ and timescale phenomenologies developed by \citetalias{Hopkins2013} and \citetalias{Mocz2019}, respectively. 
    
    
    \subsection{Self-similarity interpretation}
    Because the \citetalias{Hopkins2013} $s$-PDF model is characterised by the $T$ parameter (Equations~\ref{eq:T_average_jump}~\&~\ref{eq:T_sigma_var}), discussed in \S\ref{sec:intermittent}, which is an average measure of how much the $s$-field in the turbulence deviates from perfect self-similarity at each neighbouring spatial scale, $\Exp{s_{L/\Gamma^n}-s_{L/\Gamma^{n+1}}}_{L/\Gamma^n}$, we call this the self-similarity (SS) phenomenology. The \citetalias{Hopkins2013} SS phenomenology may be the most robust of the phenomenologies we discuss, because the formulation relies upon the \citet{Casting1996} method of probing probabilities for very general steady-state classes of multiplicative-random-relaxation processes within the turbulence cascade. Under this interpretation, the decreasing $T$ with high-$\M$ in the magnetised flows is a simple statement about the $s$-field becoming closer to perfect self-similarity. As \citet{Squire2017} points out, this does not mean that the underlying statistics are Gaussian (we have shown that this cannot be the case if one considers mass conservation or supersonic turbulence, which is not scale-free in \S\ref{sec:problems_with_normal}), but rather, it means that the flow is full of small-in-volume, small-amplitude density contrasts, making $\Exp{s_{L/\Gamma^n}-s_{L/\Gamma^{n+1}}}_{L/\Gamma^n}$ approach zero, and hence decreasing $T$.
    
\begin{figure}
        \centering
        \includegraphics[width=\linewidth]{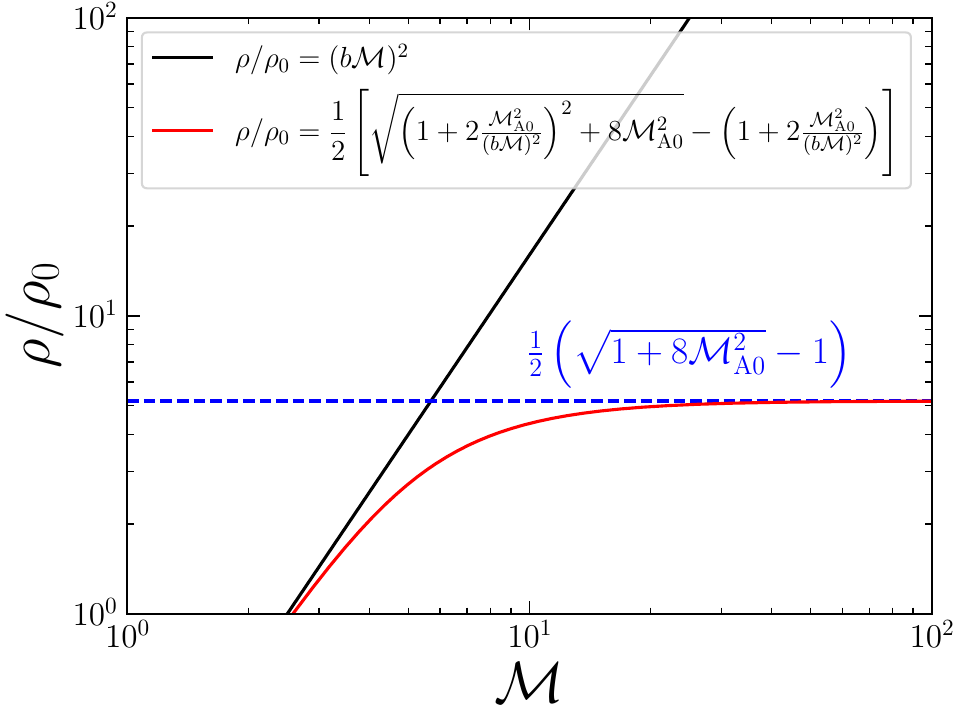}
        \caption{The theoretical density jump for a hydrodynamical (black) and magnetised (red) shocked region of gas as a function of $\M$ derived using the Rankine-Hugoniot jump conditions. The magnetised gas density jump diverges from the hydrodynamical value at high-$\M$ as the small-scale density fluctuations are suppressed by the magnetic field. The high-$\M$ limit, shown in blue, shows that stronger fields suppress these fluctuations more efficiently.}
        \label{fig:dens_jump}
    \end{figure}

    We believe the $T$ trends we find in Figure~\ref{fig:variance_compare} and Figure~\ref{fig:driving_parameter_intermittent} can be at least conceptually understood through (1) the typical shock-jump conditions in MHD turbulence and (2) the isotropic turbulent mixing of the flow as $\Mao$ increases. \citet{Beattie2021} found that the magnetic field preferentially suppresses the $\rho/\rho_0$-variance at high-$\M$ and attributed this to the magnetic field being able to more easily smooth out small-scale fluctuations, which are flux-frozen to the field \citep{Landau1959}, and are introduced into the flow as $\M$ increases \citep{Kim2005}. One can see this immediately by considering a slab of isothermal gas, threaded by magnetic fields. Perpendicular to the magnetic field $B \propto \rho$ \citep{Hennebelle2019}, and by using the regular Rankine-Hugoniot jump conditions \citep{Landau1959} one can show that the density jump for these magnetised fluctuations is  
    \begin{align} \label{eq:mhdJump}
        \frac{\rho}{\rho_0}= \dfrac{1}{2}\left[\sqrt{\left(1 + 2\frac{\MaO{2}}{(b\M)^2}\right)^2 + 8\MaO{2}} -\left(1 + 2\frac{\MaO{2}}{(b\M)^2}\right) \right]
    \end{align}
    \citep{Molina2012,Mocz2019,Beattie2021}. We plot $\rho/\rho_0$ as a function of $\M$ in Figure~\ref{fig:dens_jump} (black), as well as the regular hydrodynamical shock jump conditions $\rho/\rho_0 = (b\M)^2$ (red) for comparison. The magnetised gas density jump diverges from the hydrodynamical behaviour at high-$\M$ because the small-scale density fluctuations are suppressed by the magnetic field. This significantly reduces the amplitude of the fluctuations (both for over- and under-densities; see e.g., Figure~3 in \citealt{Mocz2018}), which means that $\Exp{s_{L/\Gamma^n}-s_{L/\Gamma^{n+1}}}_{L/\Gamma^n}$ is also reduced, giving rise to a more self-similar (small $T$), Gaussian-like set of global $s$-statistics for the high-$\M$ simulations \citep{Squire2017}, regardless of $\Mao$. 
    
    In Figure~\ref{fig:variance_compare} we found that the $T$ parameter depends upon $\Mao$: high-$\Mao$ simulations become lognormal, $T \sim \M^{-3}$ faster than low-$\Mao$, $T \sim \M^{-2}$. We believe this could be due to the global transition from an weakly-isotropic to strongly-anisotropic $s$-field. For the sub-Alfv\'enic mean-field turbulence, the high-energy $\Bo$ provides a domain, $\ell_{\parallel}$, in the turbulence along $\Bo$, where hydrodynamical-like shocks, as we showed in \S\ref{sec:physical_intermittency}, are able to form and increase $|\mathcal{S}_s|$ and hence $T$. But for the trans-Alfv\'enic simulations, $\Mao = 1-2$, $\Bo$ and $\delta\vecB{B}$ are at almost energy equipartition \citep[][see Figure~7]{Beattie2020c}. The transition is qualitatively illustrated in Figure~\ref{fig:16_panel} through the difference between the $\vecB{B}$-field lines in the top row, where the field is highly-coherent across the whole box, dominated by the $\Bo$, and bottom row, where the field is becoming tangled as the $\delta\vecB{B}$ grows. Once there is significant magnetic energy in $\delta\vecB{B}$  there is no longer a region in the turbulence, $\ell_{\parallel}$, where hydrodynamical-like shocks can form (left panel of Figure~\ref{fig:shocks}), since the $\vecB{B}$-fields are tangled into random orientations all through space. This means magnetised $s$-fluctuations are more uniformly distributed through the fluid, and hence, there are many more small amplitude fluctuations in the fluid than in the highly-sub-Alfv\'enic flows. As we discussed above, for the high-$\M$ simulations, having many small amplitude fluctuations leads to lower values of $T$, and more, self-similar, $\Exp{s_{L/\Gamma^n}-s_{L/\Gamma^{n+1}}}_{L/\Gamma^n} \approx 0$, Gaussian-like $s$-statistics. 
    
    Finally, we note that our interpretation relies upon measuring $T$ via the $s$-PDF fits, and interpreting this parameter in the context of SS. One way to directly measure the SS would be to compute the scale-dependent $(s_{L/\Gamma^n}-s_{L/\Gamma^{n+1}})$-PDF, which we do not do in this study, but we think is feasible to do with high-resolution turbulence data that was recently produced in \citet{Federrath2021}. One can compare this with the $T$ computed from the $s$-PDF with the direct measurement, $\Exp{s_{L/\Gamma^n}-s_{L/\Gamma^{n+1}}}_{L/\Gamma^n}$. This would allow us to understand better the exact physics of $T$, and how we can use it to interpret the $s$-PDF intermittency.
    
    \subsection{Timescale interpretation}
    The central tenant of the \citetalias{Mocz2019} $s$-PDF phenomenology is timescales that determine the lifetimes of high-density and low-density structures in the turbulence are different (see \S\ref{sec:intermittent}). High-density objects, $s > s_0$, live on shorter timescales compared to lower-density objects, $s \leq s_0$, reduced by a factor of $1+3f/2$ (see Equation~\ref{eq:advect_timescale}) which amounts to skewness in the $s$-PDF. It is true that the magnetic field can act to shield density fluctuations, prolonging the life of the fluctuations. \citet{Hennebelle2013} found this was the case for $\approx 55 - 65\%$ of the dense, magnetised filamentary structures in a suite of MHD box simulations. The radial Lorentz force was found to mostly point towards the filaments, opposing thermal pressure and preventing the fluctuation from expanding, prolonging the life of the over-density. This could be in part responsible for the Gaussianisation of the statistics, coupled with $\M$ increasing, increasing the number of shocks and getting larger and more complete samples of shock lifetimes. The $\Mao$ dependency could be explained in the same fashion as the previous SS phenomenology, i.e., sub-Alfv\'enic flows allow for hydrodynamical-like shocks to form along $\Bo$, which live on shorter timescales than magnetised shocks, whereas shocks in more trans-Alfv\'enic, isotropic flows, always feel the magnetic field, prolonging the lifetimes for all of the density fluctuations. This is consistent with what we find in Figure~\ref{fig:f_mach}, where we directly compute $1+3f/2$ for our simulations. It shows a spread in $1+3f/2$ for different $\vecB{B}$-field strengths, where $\Mao = 2$ is about 25\% slower than $\Mao = 0.1$, at $\M \approx 4$, before becoming completely homogenised at $\M \gtrsim 4$.  
    
    \citet{Robertson2018} has probably done the most robust measurements of lifetimes for over-dense regions using clustering and filtering techniques on passive tracer particles. The lifetimes of dense shocked regions in the supersonic isothermal turbulence simulation are short, $\mathcal{O}(10^{-3}L/c_s)$. In the frame of turbulent turnover times on the driving scale that is $\sim \mathcal{O}(10^{-2}L/[2\M c_s])$ for $\M\gtrsim 4$. Based on our observations of the shocks in the magnetised flows it is very reasonable to suggest that the shock lifetimes are well beyond that of $\sim \mathcal{O}(10^{-2}L/[2\M c_s])$. For example, we find some of the extreme shock events that we identified in Figure~\ref{fig:intermittent_through_time} persist for $\mathcal{O}(L/[2\M c_s])$. Regardless, a more detailed study of the lifetimes of over-dense objects, such as the one performed for hydrodynamical simulations in \citet{Robertson2018} would be needed to test this directly, not relying on the indirect $f$ parameter to determine how large of an affect the magnetic field has on the lifetimes of the dense regions with different $\M$ and $\Mao$.
    
\section{Summary and key findings}\label{sec:conclusion}
    We provide a detailed analysis of the 1-point logarithmic density ($s \equiv \ln\rho/\rho_0$) statistics for magnetised supersonic, isothermal turbulence in the sub-to-trans-Alfv\'enic large-scale field ($\Mao \lesssim 2$), supersonic ($\M > 1$) regime and driven by different ratios of solenoidal and compressive turbulent modes, relevant to the cool interstellar medium. We list the key results below:
    
    \begin{itemize}
        \item We show that the mass and volume-weighted statistics (Figures~\ref{fig:variances_in_time}~\&~\ref{fig:skewness_vol}) only become stationary in trans-to-sub-Alfv\'enic mean-field turbulence after about five turbulent turnover times, $t = \ell_0/(c_s\M)$, where $\ell_0$ is the driving scale, more than double the time required for hydrodynamical turbulence to reach a stationary state. \\
        
        \item The temporal fluctuations in the volume-weighted (Figure~\ref{fig:3DPDF_v}) and mass-weighted (Figure~\ref{fig:3DPDF_M}) $s$-PDFs show how the low-density, high volume-filling structures, such as rarefactions and voids, fluctuate significantly in volume, and much less in mass. In contrast, the highest-density and lowest volume-filling structures, such as shocked regions, filaments, and over-dense sheets, fluctuate significantly in mass, but not in volume. This paints a picture of magnetised turbulence where the densest regions are gaining and losing mass over short timescales, using neighbouring voids as reservoirs, which are expanding and contracting in volume as mass is exchanged between them.\\
        
        \item We compare the non-Gaussian models, \citetalias{Hopkins2013} (Equation~\ref{eq:Hopkins1}-\ref{eq:Hopkins2}) and \citetalias{Mocz2019} (Equation~\ref{eq:mocz_pdf_model}), and the Gaussian (lognormal in $\rho/\rho_0$) model (Equation~\ref{eq:Vdis}) for the volume-weighted $s$-PDF. We find that unlike hydrodynamical turbulence, which monotonically increases the non-Gaussian features of the $s$-PDF with $\M$, the \citetalias{Hopkins2013} or \citetalias{Mocz2019} are required to capture the PDF morphology at $\M \lesssim 4$, but for $\M \gtrsim 4$ the Gaussian model is empirically (but not theoretically) sufficient.\\
        
        \item Motivated by the complex non-Gaussian behaviour of the $s$-PDF we calculate four independent measures of the global, volume-weighted intermittency (1) the skewness (Equation~\ref{eq:skewness}, Figure~\ref{fig:skewness_vol}), (2) the mass-weighted versus volume-weighted $s$ standard deviation relation (right panel of Figure~\ref{fig:variance_compare}), (3) the \citetalias{Hopkins2013} $T$ parameter (Equation~\ref{eq:T_average_jump}, left panel of Figure~\ref{fig:variance_compare}) and (4) the \citetalias{Mocz2019} $f$ parameter (Equation~\ref{eq:advect_timescale}). All four independent statistics confirm that hydrodynamical-like intermittency is found for low-$\M$ highly-magnetised simulations, but the intermittency monotonically reduces as a function of $\M$, with some dependence upon $\Mao$, completely opposite to the hydrodynamical turbulence studied in \citetalias{Hopkins2013} and \citet{Squire2017}.\\
        
        \item We explore how changing the modes from compressive to solenoidal in the turbulent driving contributes to the non-Gaussian components of the $s$-PDF in Section~\ref{sec:driving_parameter}. We find that the intermittency increases monotonically with shrinking $\zeta$ (Equation~\ref{eq:zeta_equation}; more compressive modes driving the turbulence), and is strongest when $\Mao$ and $\zeta$ are low, resulting in very significant (orders of magnitude in the $s>0$ tails) deviations away from Gaussian $s$ behaviour (Figures~\ref{fig:driving_parameter_pdfs} and \ref{fig:driving_parameter_intermittent}). Because star-formation may be triggered by highly-compressive driving events, we therefore suggest that intermittency must be taken into account when using predictive star-formation rate models based on the density statistics.  \\        
        
        \item Using 1D pencil beams (Figure~\ref{fig:shocks}, \S\ref{sec:physical_intermittency}) we explore the low-$\M$ local parallel and perpendicular dynamics along $\Bo$. Along $\Bo$ large-scale exponential shocks form through convergent flows and sawtooth shocks. We stress, that unlike simple shock-tube experiments, the over-densities have an internal structure, similar to over-densities described in \citet{Robertson2018}. The shocks excite magnetic compression waves at the shock-front and give rise to the largest over-densities in the flow. This also leads to anisotropies in the magnetic tension. Across $\Bo$ we find large-scale, rigid body vortices and very small-amplitude shear Alfv\'en waves (compared to the kinetic and fast magnetosonic compression waves). We comment that in sub-Alfv\'enic mean-field turbulence shear Alfv\'en waves can only possibly play a dominant role in the dynamics on very small scales and the large scales are dominated by sawtooth shocks in the gas velocity (i.e., akin to large-scale \citet{Burgers1948} turbulence).\\
        
        \item Using our pencil beams we search for the low-$s$ intermittent events that are captured by the \citetalias{Hopkins2013}-type intermittency. We look for the most extreme under-dense events through 49 time realisations (Figure~\ref{fig:intermittent_through_time}), and find that the largest contributors to the non-Gaussian $s$-statistics are strong, single shocks ($1.5-2\sigma$ events in the velocity) that form along $\Bo$ and create volume-poor $s$ rarefactions, which are seemingly also coupled to rare velocity and magnetosonic field fluctuation events. We show explicitly that it is single shocks and not multi-shock interactions along $\Bo$ that are the largest contributors to the non-Gaussian components of $s$ using 1D numerical shock experiments that we show in Figures~\ref{fig:shock_spacetimes},\ref{fig:shock_skewness} and \ref{fig:shock_experiments_time evolution}. \\
        
        \item For the high-$\M$ flows, we discuss the two phenomenologies that we use to understand why the $s$-statistics become more Gaussian with increasing $\M$ (\S\ref{sec:high_mach_gauss}). These are the self-similarity (\citetalias{Hopkins2013}, \citealt{Squire2017}) and inhomogeneous timescale \citepalias{Mocz2019} interpretations of the $s$-statistics. We conclude it is most likely that the self-similarity of the flow is maintained through suppression of over- and under-dense $s$-fluctuations by $\vecB{B}$. This effect is stronger when the flows are trans-Alfv\'enic and are becoming dominated by isotropic magnetised $s$-fluctuations. We also conjecture that the magnetic field plays a role in facilitating more uniform dynamical timescales between the low- and high-density structures in the gas.
    \end{itemize}
    
\section*{Acknowledgements}
    We thank the anonymous reviewers who helped enhance the clarity and presentation of our study. J.~R.~B.~thanks Christoph Federrath's and Mark Krumholz's research groups for the many productive discussions, in particular, Shyam H. Menon, and acknowledges financial support from the Australian National University, via the Deakin PhD and Dean's Higher Degree Research (theoretical physics) Scholarships, the Research School of Astronomy and Astrophysics, via the Joan Duffield Research Scholarship, the Australian Government via the Australian Government Research Training Program Fee-Offset Scholarship and the Australian Capital Territory Government funded Fulbright scholarship. 
    P.~M.~acknowledges support for this work provided by NASA through Einstein Postdoctoral Fellowship grant number PF7-180164 awarded by the \textit{Chandra} X-ray Centre, which is operated by the Smithsonian Astrophysical Observatory for NASA under contract NAS8-03060.
    C.~F.~acknowledges funding provided by the Australian Research Council (Future Fellowship FT180100495), and the Australia-Germany Joint Research Cooperation Scheme (UA-DAAD).
    R.~S.~K.~acknowledges financial support from the German Research Foundation (DFG) via the Collaborative Research Center (SFB 881, Project-ID 138713538) 'The Milky Way System' (subprojects A1, B1, B2, and B8). He also thanks for funding from the Heidelberg Cluster of Excellence STRUCTURES in the framework of Germany's Excellence Strategy (grant EXC-2181/1 - 390900948) and for funding from the European Research Council via the ERC Synergy Grant ECOGAL (grant 855130).
    We further acknowledge high-performance computing resources provided by the Australian National Computational Infrastructure (grant~ek9) in the framework of the National Computational Merit Allocation Scheme and the ANU Merit Allocation Scheme, and by the Leibniz Rechenzentrum and the Gauss Centre for Supercomputing (grants~pr32lo, pr48pi, pr74nu, pn73fi).\\[1em]

    The simulation software, \textsc{flash}, was in part developed by the DOE-supported Flash Centre for Computational Science at the University of Chicago. Data analysis and visualisation software used in this study: \textsc{C++} \citep{Stroustrup2013}, \textsc{cython} \citep{Behnel2011}, \textsc{visit} \citep{Childs2012}, \textsc{numpy} \citep{Oliphant2006,numpy2020}, \textsc{matplotlib} \citep{Hunter2007}, \textsc{scipy} \citep{Virtanen2020}, \textsc{scikit-image} \citep{vanderWalts2014}, \textsc{pandas} \citep{pandas2020}.

\section*{Data Availability}

The data underlying this article will be shared on reasonable request to the corresponding author, James R. Beattie.


\bibliographystyle{mnras.bst}
\bibliography{Aug2021.bib} 

\appendix

\section{Intermittency parameters as a function of $\Mao$}\label{app:intermittent_Mao_plots}
\begin{figure}
        \centering
        \includegraphics[width=\linewidth]{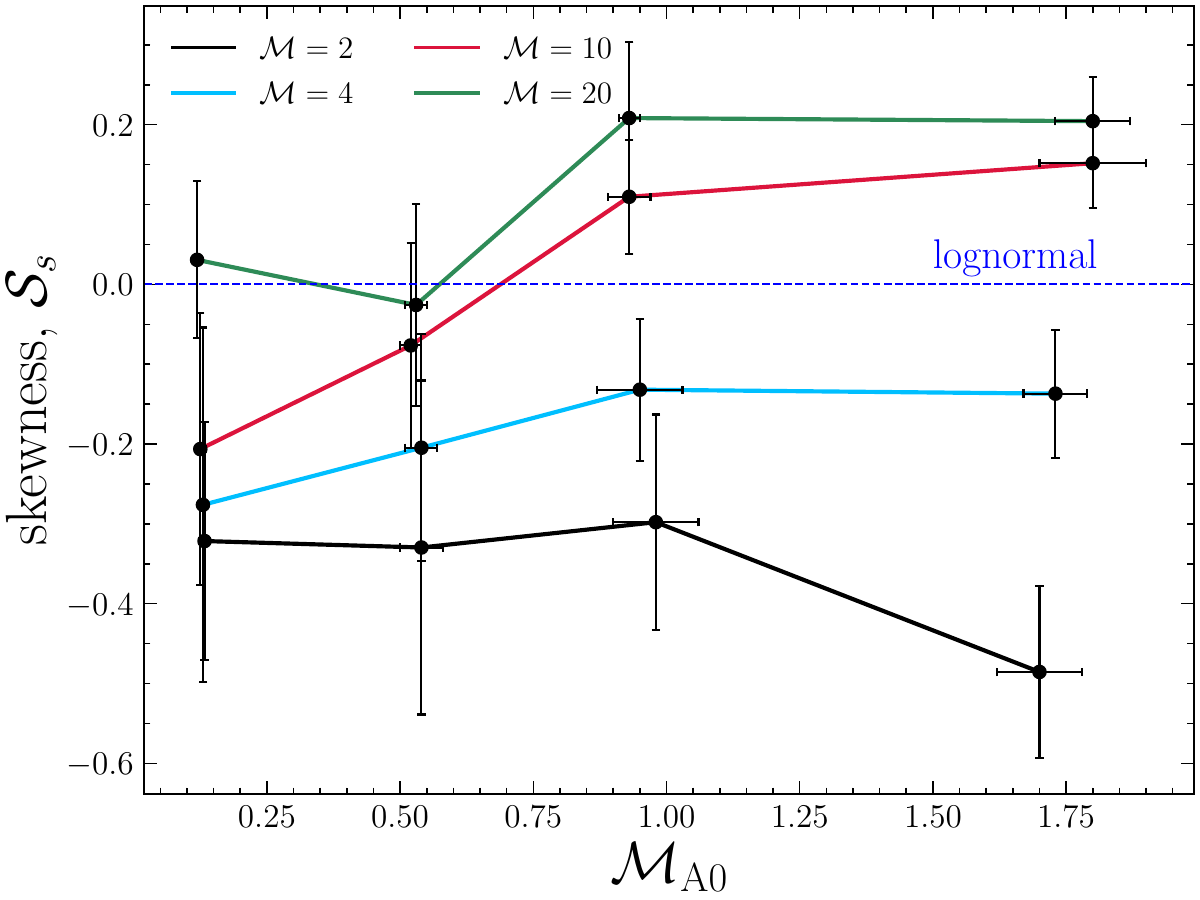}
        \caption{The same as Figure~\ref{fig:skewness_vol} but plotted as a function of $\Mao$.}
        \label{fig:skewness_ma0}
    \end{figure}

\begin{figure}
        \centering
        \includegraphics[width=\linewidth]{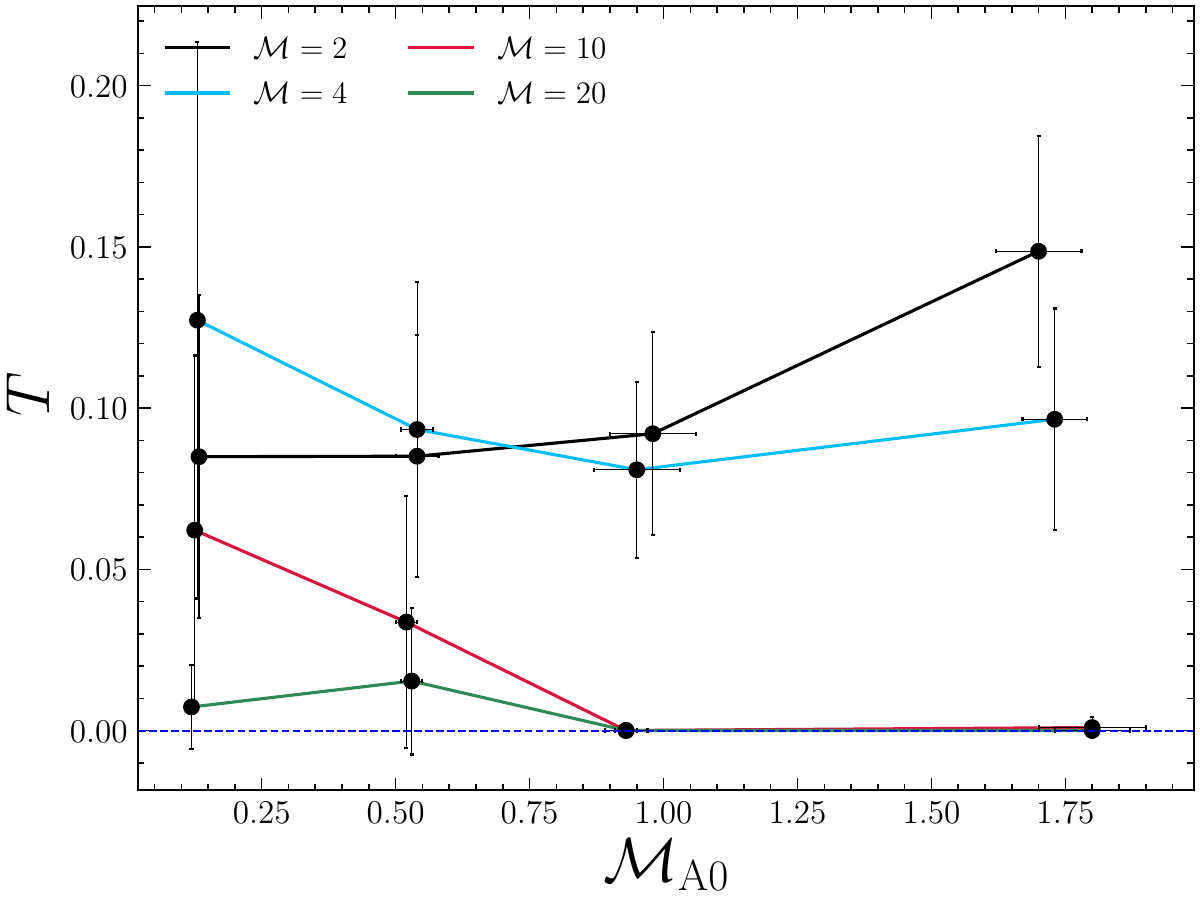}
        \caption{The same as the left panel of Figure~\ref{fig:variance_compare} but plotted as a function of $\Mao$.}
        \label{fig:T_mao}
    \end{figure}

\begin{figure}
        \centering
        \includegraphics[width=\linewidth]{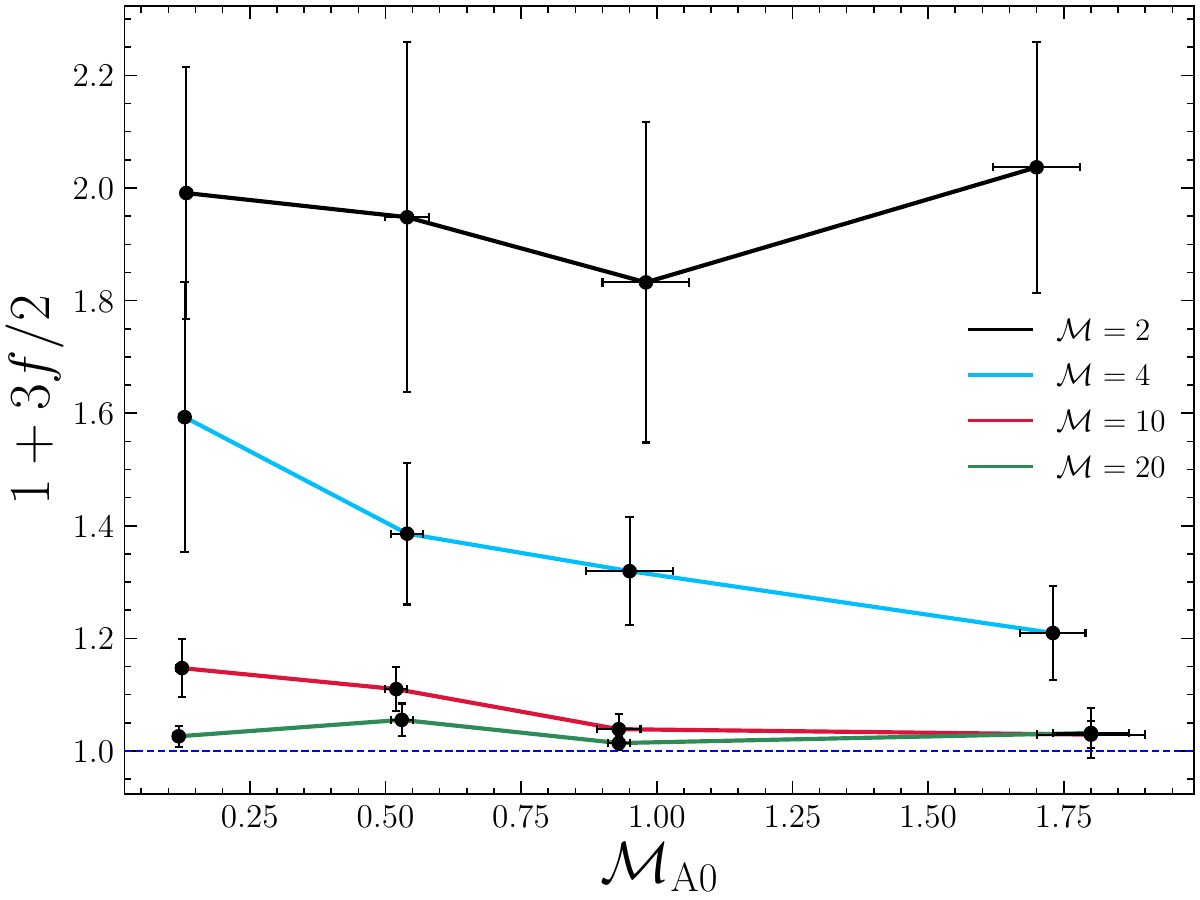}
        \caption{The same as Figure~\ref{fig:f_mach} but plotted as a function of $\Mao$.}
        \label{fig:f_ma0}
    \end{figure}

    In the main text of this study we presented most of our intermittency plots as a function of $\M$, because $\M$ is a strong function of the intermittency in the density statistics. In this section we present the three intermittency parameters in Figures~\ref{fig:skewness_ma0}, \ref{fig:T_mao} and  \ref{fig:f_ma0}, plotted as a function of $\Mao$, providing a different perspective on the functional dependency of the density intermittency. 

\section{Random pencil beam realisations}\label{app:random_pencils}
\begin{figure*}
        \centering
        \includegraphics[width=\linewidth]{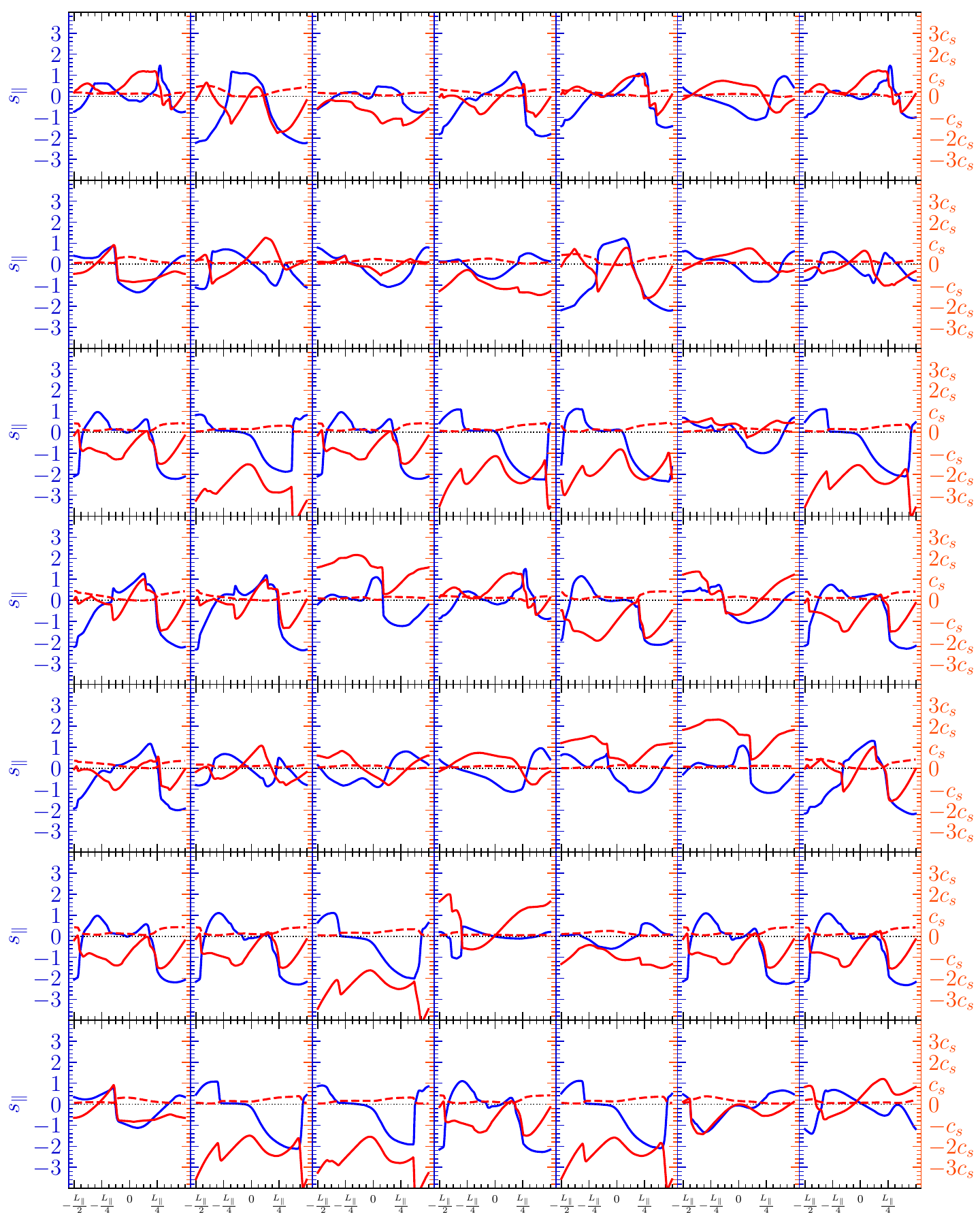}
        \caption{Random pencil beams $\ell\parallel\vecB{B}_0$ in the \texttt{M2MA01} simulation, coloured the same way as Figure~\ref{fig:intermittent_through_time} (blue, $\ln(\rho/\rho_0)$; red-solid, $v$; red-dashed, $v_A = B/\sqrt{4\pi\rho}$).}
        \label{fig:random_parallel}
    \end{figure*}

\begin{figure*}
        \centering
        \includegraphics[width=\linewidth]{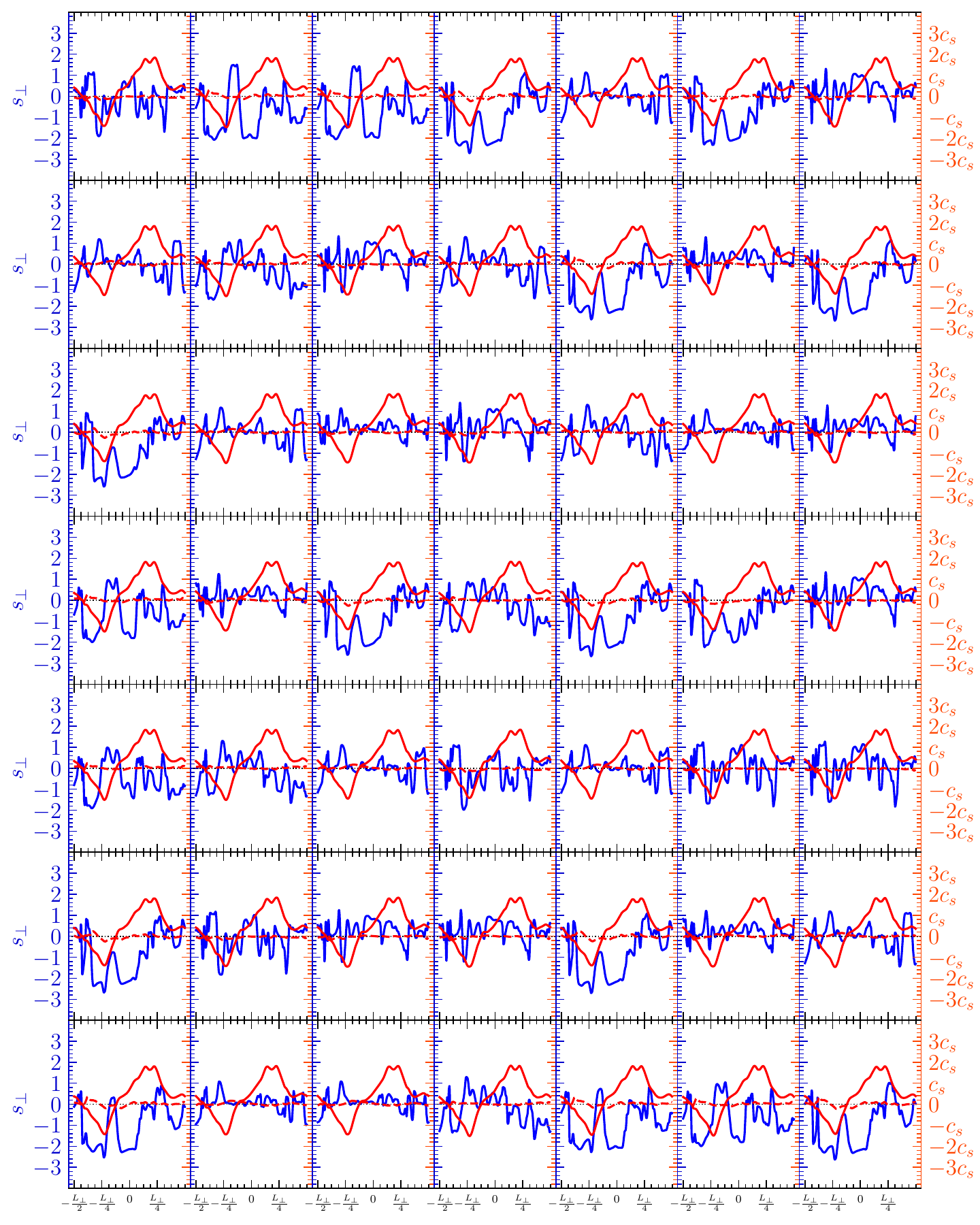}
        \caption{Random pencil beams $\ell\perp\vecB{B}_0$ in the \texttt{M2MA01} simulation, coloured the same way as Figure~\ref{fig:intermittent_through_time} (blue, $\ln(\rho/\rho_0)$; red-solid, $v$; red-dashed, $v_A = B/\sqrt{4\pi\rho}$). Note the two vortices pervade the entire $\ell_{\parallel}$ domain.}
        \label{fig:random_perpendicular}
    \end{figure*}

    To illustrate that Figure~\ref{fig:shocks} was representative of the pencil beams through the sub-Alfv\'enic turbulence, as discussed in \S\ref{sec:physical_intermittency} we plot 49 random further realisations of both the density, velocity and Alfv\'en velocity parallel beams (Figure~\ref{fig:random_parallel}) and perpendicular beams (Figure~\ref{fig:random_perpendicular}), all coloured in the same fashion as Figure~\ref{fig:intermittent_through_time}, using the same simulation as in Figure~\ref{fig:shocks}. As in the left panel of Figure~\ref{fig:shocks}, Figure~\ref{fig:random_parallel} shows sawtooth shocks and converging flows interacting along the magnetic field, exciting parallel magnetic field fluctuations, and developing strong over-densities and deep voids. Likewise, for Figure~\ref{fig:random_perpendicular} we see the strong signature of two rigid body vortices (outer scale on the $k = 2$ mode) that pervades every beam in the perpendicular velocity, coupled with high-frequency density structure from the converging flows along the field lines, and negligible magnetic (Alfv\'en, i.e., those from Alfv\'en modes) fluctuations, similar to the right panel of Figure~\ref{fig:shocks}.

\section{1D shock experiments}\label{app2:1d_shocks}
    \begin{figure}
        \centering
        \includegraphics[width=\linewidth]{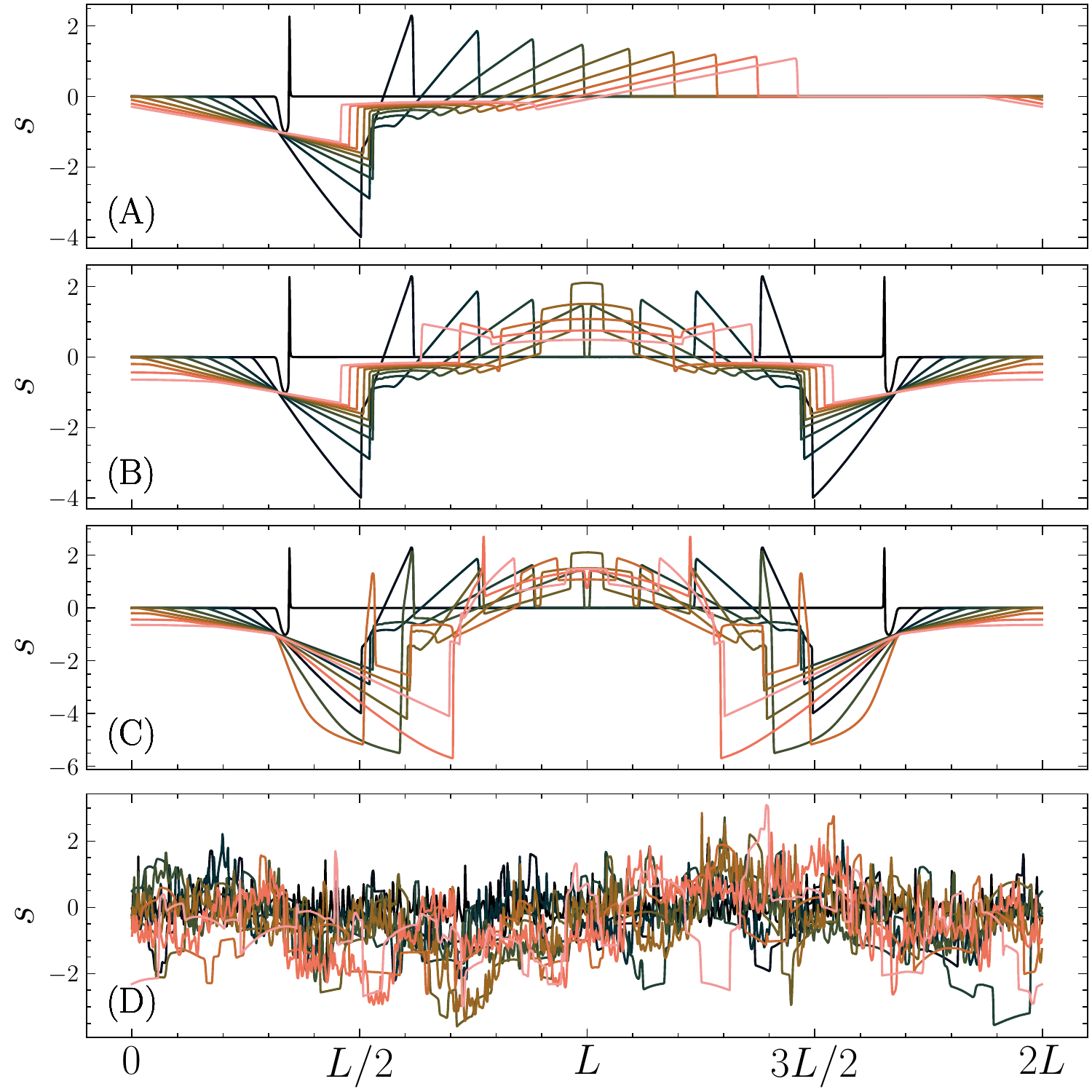}
        \caption{The time-evolution of logarithmic density profiles from the four 1D shock experiments: (A) single shock, (B) two shocks, (C) two shocks, constant driving, (D) many shocks, constant driving, that we use to explore the low-$\M$ intermittency. The profiles are coloured such that black corresponds to $t/t_{\rm cross} = 0$, and pink $t/t_{\rm cross} \gg 0$.}
        \label{fig:shock_experiments_time evolution}
    \end{figure} 

    We perform 1D shock experiments with a similar setup to the one described in \citet{Mocz2018}. We solve the compressible Euler equations in 1D,
    \begin{align}
        \frac{\partial \rho}{\partial t} + \frac{\partial}{\partial x}\rho v_x &= 0,
    \end{align}
    \begin{align}
        \rho \left( \frac{\partial}{\partial t} + v_x\frac{\partial}{\partial x} \right) v_x &= c_s^2 \frac{\partial}{\partial x} \rho, 
    \end{align}
    based on the multi-state Harten–Lax–van Leer approximate Riemann method described in \mbox{\citet{Miyoshi2005}}. We solve the equations on a periodic domain $x \in [0, 2L]$, and with $\rho(x,t=0) = \rho_0 = 1$. Following \mbox{\citet{Mocz2018}}, to create an over-dense region we \mbox{perturb the velocity field with a Gaussian pulse},
    \begin{align}
    v(x,t=0)/c_s = \M\exp\left\{ \frac{-(x - x_0)^2}{2\ell_{0}^2}\right\},
    \end{align}
    hence we are able to tune $\M$ and driving scale, $\ell_{0}$ directly through the pulse. For all of the experiments we set $\M \approx 2.0$, capturing along-$\Bo$-field dynamics of the \texttt{M2MA01} simulation studied in detail, in \S\ref{sec:physical_intermittency}. We run all experiments from $t/t_{\rm cross} = 0-2$, where $t_{\rm cross} = L/c_s$. As outlined in \S\ref{sec:1d_experiments}, we perform four experiments,
    \begin{enumerate}
        \item a single shock travelling across the domain,
        \item two shocks colliding at the centre of the domain,
        \item two shocks repeatedly injected,
        \item multiple shocks repeatedly injected.
    \end{enumerate}
    which we show in terms of the space-time diagrams, Figure~\ref{fig:shock_spacetimes}, and the time-evolving $s$ profiles, Figure~\ref{app2:1d_shocks}. For experiments (i) and (ii) we simply initialise the velocity perturbations and track the over-densities. For (iii) and (iv) we repeatedly add pulses to the velocity. For (iii) we pulse every $(1/3)t_{\rm cross}$ and for (iv) $(1/5)t_{\rm cross}$, for the latter, ensuring that the 1D ``turbulence" becomes stationary. For (i)-(iii) $\ell_{0} = 2.5\times10^{-3}L$, providing ample amounts of space for the shocks to develop and evolve, and for (iv) $\ell_{0} = 1.25\times10^{-3}L$, which allows us to create many small-scale shocks that fill the domain completely.\\

\bsp	
\label{lastpage}
\end{document}

%% file: header.tex
\usepackage[T1]{fontenc}
\usepackage{ae,aecompl}

\DeclareRobustCommand{\VAN}[3]{#2}
\let\VANthebibliography\thebibliography
\def\thebibliography{\DeclareRobustCommand{\VAN}[3]{##3}\VANthebibliography}

\usepackage{graphicx}	
\usepackage{amsmath}	
\usepackage{amssymb}	
\usepackage{color}
\usepackage{algorithm}
\usepackage{booktabs}
\usepackage[noend]{algpseudocode}
\usepackage{pifont}
\usepackage{soul}
\usepackage{float}
\usepackage{multibib}
\newcites{Appendix}{Appendix References}
\usepackage{natbib}
\usepackage[flushleft]{threeparttable}
\usepackage{mathrsfs}
\usepackage{widetext}
\usepackage{textcomp}

\graphicspath{{Figures/}}

\newcommand{\appropto}{\mathrel{\vcenter{
  \offinterlineskip\halign{\hfil$##$\cr
    \propto\cr\noalign{\kern2pt}\sim\cr\noalign{\kern-2pt}}}}}

\newcommand{\vecB}[1]{\mathrm{{\bmath{\mathit{#1}}}}}
\newcommand{\Exp}[1]{\left\langle #1 \right\rangle}

\renewcommand{\d}[1]{\ensuremath{\operatorname{d}\!{#1}}}

\DeclareMathOperator\V{\mathcal{V}}
\DeclareMathOperator\M{\mathcal{M}}
\DeclareMathOperator\Ma{\mathcal{M}_{\text{A}}}
\DeclareMathOperator\Mao{\mathcal{M}_{\text{A0}}}
\newcommand{\MaO}[1]{\mathcal{M}^{#1}_{\text{A0}}}

\DeclareMathOperator\Bo{\vecB{B}_0}

\defcitealias{Brunt2010a}{BFP2010}
\defcitealias{Beattie2020}{BF2020}
\defcitealias{Kolmogorov1941}{K41}
\defcitealias{Menon2020}{MFK2020}
\defcitealias{Hopkins2013}{H2013}
\defcitealias{Mocz2019}{MB2019}

\usepackage{scalerel,tikz}
\usetikzlibrary{svg.path}
\definecolor{orcidlogocol}{HTML}{A6CE39}
\tikzset{orcidlogo/.pic={\fill[orcidlogocol] svg{M256,128c0,70.7-57.3,128-128,128C57.3,256,0,198.7,0,128C0,57.3,57.3,0,128,0C198.7,0,256,57.3,256,128z}; \fill[white] svg{M86.3,186.2H70.9V79.1h15.4v48.4V186.2z} svg{M108.9,79.1h41.6c39.6,0,57,28.3,57,53.6c0,27.5-21.5,53.6-56.8,53.6h-41.8V79.1z M124.3,172.4h24.5c34.9,0,42.9-26.5,42.9-39.7c0-21.5-13.7-39.7-43.7-39.7h-23.7V172.4z} svg{M88.7,56.8c0,5.5-4.5,10.1-10.1,10.1c-5.6,0-10.1-4.6-10.1-10.1c0-5.6,4.5-10.1,10.1-10.1C84.2,46.7,88.7,51.3,88.7,56.8z};}}
\newcommand\orcidicon[1]{\href{https://orcid.org/#1}{\mbox{\scalerel*{
\begin{tikzpicture}[yscale=-1,transform shape]\pic{orcidlogo};
\end{tikzpicture}}{|}}}}

%% file: table.tex
    \begin{table*}
        \caption{Main simulation and derived parameters.}
        \centering
        \resizebox{\textwidth}{!}{\begin{tabular}{l r@{}l r@{}l c r@{}l r@{}l r@{}l r@{}l r@{}l l}
            \hline
            \hline
            Sim. ID & \multicolumn{2}{c}{$\M\,(\pm1\sigma)$} & \multicolumn{2}{c}{$\Mao\,(\pm1\sigma)$} & $\zeta$ & \multicolumn{2}{c}{$\mathcal{S}_s\,(\pm1\sigma)$} & \multicolumn{2}{c}{$T\,(\pm1\sigma)$} & \multicolumn{2}{c}{$\sigma_{s,V}\,(\pm1\sigma)$} & \multicolumn{2}{c}{$\sigma_{s,M}\,(\pm1\sigma)$} & \multicolumn{2}{c}{$1+3f/2\,(\pm1\sigma)$} & $N^3$ \\
            \multicolumn{1}{c}{(1)} & \multicolumn{2}{c}{(2)} & \multicolumn{2}{c}{(3)} & (4) & \multicolumn{2}{c}{(5)} & \multicolumn{2}{c}{(6)} & \multicolumn{2}{c}{(7)} & \multicolumn{2}{c}{(8)} & \multicolumn{2}{c}{(9)} & (10) \\
            \hline
            \multicolumn{17}{c}{Main Simulations $(\zeta = 0.5)$}\\
            \hline
            \texttt{M2Ma01}       & 2.6\,    & $ \pm$ 0.2      & 0.131\, & $ \pm$ 0.008 & 0.5  & $-$0.32\,  & $ \pm$ 0.15   & (8.49\, & $ \pm$ 5.01$)\times 10^{-2}$ & 0.51\, & $ \pm$ 0.04 & 0.40\, & $ \pm$ 0.03 & 1.99\, & $ \pm$ 0.22  & $512^3$   \\
            \texttt{M4Ma01}       & 5.2\,    & $ \pm$ 0.4      & 0.13\,  & $ \pm$ 0.01  & 0.5  & $-$0.28\,  & $ \pm$ 0.22   & (1.27\, & $ \pm$ 0.86$)\times 10^{-1}$ & 0.52\, & $ \pm$ 0.04 & 0.41\, & $ \pm$ 0.04 & 1.59\, & $ \pm$ 0.24  & $512^3$   \\
            \texttt{M10Ma01}      & 12\,     & $ \pm$ 1        & 0.125\, & $ \pm$ 0.006 & 0.5  & $-$0.21\,  & $ \pm$ 0.17   & (6.21\, & $ \pm$ 5.42$)\times 10^{-2}$ & 0.55\, & $ \pm$ 0.03 & 0.43\, & $ \pm$ 0.03 & 1.15\, & $ \pm$ 0.05  & $512^3$   \\
            \texttt{M20Ma01}      & 24\,     & $ \pm$ 1        & 0.119\, & $ \pm$ 0.003 & 0.5  & $-$0.03\,  & $ \pm$ 0.10   & (0.73\, & $ \pm$ 1.30$)\times 10^{-2}$ & 0.66\, & $ \pm$ 0.05 & 0.45\, & $ \pm$ 0.02 & 1.03\, & $ \pm$ 0.02  & $512^3$   \\
            \texttt{M2Ma05}       & 2.2\,    & $ \pm$ 0.2      & 0.54\,  & $ \pm$ 0.04  & 0.5  & $-$0.33\,  & $ \pm$ 0.21   & (8.51\, & $ \pm$ 3.76$)\times 10^{-2}$ & 1.38\, & $ \pm$ 0.10 & 0.96\, & $ \pm$ 0.07 & 1.99\, & $ \pm$ 0.31  & $512^3$   \\
            \texttt{M4Ma05}       & 4.4 \,   & $ \pm$ 0.2      & 0.54\,  & $ \pm$ 0.03  & 0.5  & $-$0.20\,  & $ \pm$ 0.14   & (9.34\, & $ \pm$ 4.57$)\times 10^{-2}$ & 1.32\, & $ \pm$ 0.07 & 1.00\, & $ \pm$ 0.06 & 1.39\, & $ \pm$ 0.13  & $512^3$   \\
            \texttt{M10Ma05}      & 10.5\,   & $ \pm$ 0.5      & 0.52\,  & $ \pm$ 0.02  & 0.5  & $-$0.08\,  & $ \pm$ 0.13   & (3.36\, & $ \pm$ 3.91$)\times 10^{-2}$ & 1.46\, & $ \pm$ 0.02 & 1.10\, & $ \pm$ 0.05 & 1.11\, & $ \pm$ 0.04  & $512^3$   \\
            \texttt{M20Ma05}      & 21 \,    & $ \pm$ 1        & 0.53\,  & $ \pm$ 0.02  & 0.5  & $-$0.03\,  & $ \pm$ 0.13   & (1.54\, & $ \pm$ 2.28$)\times 10^{-2}$ & 1.60\, & $ \pm$ 0.02 & 1.19\, & $ \pm$ 0.03 & 1.06\, & $ \pm$ 0.03  & $512^3$   \\
            \texttt{M2Ma1}       & 2.0\,     & $ \pm$ 0.1      & 0.98\,  & $ \pm$ 0.07  & 0.5  & $-$0.30\,  & $ \pm$ 0.13   & (9.21\, & $ \pm$ 3.15$)\times 10^{-2}$ & 2.16\, & $ \pm$ 0.08 & 1.79\, & $ \pm$ 0.06 & 1.83\, & $ \pm$ 0.29  & $512^3$   \\
            \texttt{M4Ma1}       & 3.8\,     & $ \pm$ 0.3      & 0.95\,  & $ \pm$ 0.08  & 0.5  & $-$0.13\,  & $ \pm$ 0.09   & (8.08\, & $ \pm$ 2.73$)\times 10^{-2}$ & 2.02\, & $ \pm$ 0.03 & 1.83\, & $ \pm$ 0.05 & 1.32\, & $ \pm$ 0.10  & $512^3$   \\
            \texttt{M10Ma1}      & 9.3\,     & $ \pm$ 0.5      & 0.93\,  & $ \pm$ 0.05  & 0.5  & 0.11\,     & $ \pm$ 0.07   & (2.49\, & $ \pm$ 4.41$)\times 10^{-5}$ & 2.30\, & $ \pm$ 0.06 & 2.21\, & $ \pm$ 0.05 & 1.04\, & $ \pm$ 0.03  & $512^3$   \\
            \texttt{M20Ma1}      & 18.8\,    & $ \pm$ 0.7      & 0.93\,  & $ \pm$ 0.03  & 0.5  & 0.21\,     & $ \pm$ 0.10   & (3.54\, & $ \pm$ 5.28$)\times 10^{-5}$ & 2.40\, & $ \pm$ 0.04 & 2.24\, & $ \pm$ 0.05 & 1.01\, & $ \pm$ 0.01  & $512^3$   \\
            \texttt{M2Ma2}       & 1.7\,     & $ \pm$ 0.1      & 1.7\,   & $ \pm$ 0.1   & 0.5  & $-$0.49\,  & $ \pm$ 0.11   & (1.49\, & $ \pm$ 0.36$)\times 10^{-1}$ & 1.96\, & $ \pm$ 0.05 & 2.12\, & $ \pm$ 0.06 & 2.04\, & $ \pm$ 0.22  & $512^3$   \\
            \texttt{M4Ma2}       & 3.5\,     & $ \pm$ 0.1      & 1.73\,  & $ \pm$ 0.07  & 0.5  & $-$0.14\,  & $ \pm$ 0.08   & (9.65\, & $ \pm$ 3.44$)\times 10^{-2}$ & 2.29\, & $ \pm$ 0.07 & 2.25\, & $ \pm$ 0.07 & 1.21\, & $ \pm$ 0.08  & $512^3$   \\
            \texttt{M10Ma2}      & 9.0\,     & $ \pm$ 0.4      & 1.8\,   & $ \pm$ 0.1   & 0.5  & 0.15\,     & $ \pm$ 0.06   & (0.09\, & $ \pm$ 3.24$)\times 10^{-2}$ & 2.47\, & $ \pm$ 0.08 & 2.63\, & $ \pm$ 0.06 & 1.03\, & $ \pm$ 0.02  & $512^3$   \\
            \texttt{M20Ma2}      & 18\,      & $ \pm$ 1        & 1.8\,   & $ \pm$ 0.1   & 0.5  & 0.20\,     & $ \pm$ 0.06   & (3.17\, & $ \pm$ 4.82$)\times 10^{-5}$ & 2.79\, & $ \pm$ 0.05 & 2.83\, & $ \pm$ 0.05 & 1.03\, & $ \pm$ 0.04  & $512^3$   \\
            \texttt{M2Ma}$\infty$       & 2.15\,& $ \pm$ 0.07  & \,      & $ \infty$    & 0.5  &            & -             & (4.89\, & $ \pm$ 3.53$)\times 10^{-2}$ & 0.65\, & $ \pm$ 0.03 & 0.58\, & $ \pm$ 0.02 &        & -            & $576^3$   \\
            \texttt{M4Ma}$\infty$       & 4.1\, & $ \pm$ 0.1   & \,      & $ \infty$    & 0.5  &            & -             & (5.94\, & $ \pm$ 4.10$)\times 10^{-2}$ & 1.59\, & $ \pm$ 0.05 & 1.33\, & $ \pm$ 0.03 &        & -            & $576^3$   \\
            \texttt{M10Ma}$\infty$      & 10.1\,& $ \pm$ 0.5   & \,      & $ \infty$    & 0.5  &            & -             & (1.18\, & $ \pm$ 1.03$)\times 10^{-1}$ & 3.31\, & $ \pm$ 0.12 & 2.33\, & $ \pm$ 0.08 &        & -            & $576^3$   \\
            \texttt{M20Ma}$\infty$      & 20.2\,& $ \pm$ 0.7   & \,      & $ \infty$    & 0.5  &            & -             & (2.03\, & $ \pm$ 0.73$)\times 10^{-1}$ & 4.84\, & $ \pm$ 0.07 & 2.74\, & $ \pm$ 0.07 &        & -            & $576^3$   \\
            \hline 
            \multicolumn{17}{c}{Driving Parameter Simulations}\\
            \hline
            \texttt{M2Ma01$\zeta$0}  & 2.0\,   & $ \pm$ 0.1     & 0.100\, & $ \pm$ 0.001 & 0.00  & -0.52\,  & $ \pm$ 0.21  & (5.44\,  & $ \pm$ 1.34$)\times 10^{-1}$  &  1.93\, & $ \pm$ 0.12    & 1.06\, & $ \pm$ 0.05  & 4.50\, & $ \pm$ 0.34  & $288^3$   \\
            \texttt{M2Ma01$\zeta$025} & 2.01\, & $ \pm$ 0.08    & 0.100\, & $ \pm$ 0.004 & 0.25  & -0.46\,  & $ \pm$ 0.14  & (2.02\,  & $ \pm$ 0.14$)\times 10^{-1}$  &  1.04\, & $ \pm$ 0.04    & 0.81\, & $ \pm$ 0.03  & 2.36\, & $ \pm$ 0.44  & $288^3$   \\             
            \texttt{M2Ma01$\zeta$05} & 1.86\,  & $ \pm$ 0.06    & 0.092\, & $ \pm$ 0.001 & 0.50  & -0.37\,  & $ \pm$ 0.15  & (6.24\,  & $ \pm$ 1.20$)\times 10^{-2}$  &  0.51\, & $ \pm$ 0.04    & 0.47\, & $ \pm$ 0.03  & 2.16\, & $ \pm$ 0.91  & $288^3$   \\
            \texttt{M2Ma01$\zeta$075} & 1.88\, & $ \pm$ 0.19    & 0.094\, & $ \pm$ 0.009 & 0.75  & -0.25\,  & $ \pm$ 0.11  & (2.42\,  & $ \pm$ 1.63$)\times 10^{-2}$  &  0.37\, & $ \pm$ 0.02    & 0.36\, & $ \pm$ 0.02  & 2.45\, & $ \pm$ 1.07  & $288^3$   \\            
            \texttt{M2Ma01$\zeta$1}  & 2.5\,   & $ \pm$ 0.3     & 0.124\, & $ \pm$ 0.002 & 1.00  & -0.30\,  & $ \pm$ 0.21  & (2.94\,  & $ \pm$ 2.98$)\times 10^{-2}$  &  0.42\, & $ \pm$ 0.03    & 0.40\, & $ \pm$ 0.03  & 2.50\, & $ \pm$ 1.12  & $288^3$   \\ 
            \texttt{M2Ma1$\zeta$0}   & 2.03\,  & $ \pm$ 0.09    & 1.02\,  & $ \pm$ 0.04  & 0.00  & -0.29\,  & $ \pm$ 0.17  & (3.41\,  & $ \pm$ 0.97$)\times 10^{-1}$  &  1.77\, & $ \pm$ 0.07    & 1.23\, & $ \pm$ 0.08  & 3.20\, & $ \pm$ 0.14  & $288^3$   \\
            \texttt{M2Ma1$\zeta$025} & 2.11\,  & $ \pm$ 0.08    & 1.05\,  & $ \pm$ 0.04   & 0.25  & -0.41\,  & $ \pm$ 0.13  & (2.64\, & $ \pm$ 0.34$)\times 10^{-1}$  &  1.33\, & $ \pm$ 0.09    & 0.99\, & $ \pm$ 0.05  & 1.96\, & $ \pm$ 0.43  & $288^3$   \\             
            \texttt{M2Ma1$\zeta$05}  & 2.0\,   & $ \pm$ 0.1     & 0.99\,  & $ \pm$ 0.05   & 0.50  & -0.36\,  & $ \pm$ 0.06  & (1.14\, & $ \pm$ 0.06$)\times 10^{-1}$  &  0.84\, & $ \pm$ 0.04    & 0.72\, & $ \pm$ 0.03  & 1.95\, & $ \pm$ 1.10  & $288^3$   \\
            \texttt{M2Ma1$\zeta$075} & 1.99\,  & $ \pm$ 0.09    & 0.99\,   & $ \pm$ 0.05  & 0.75  & -0.25\,  & $ \pm$ 0.10  & (7.06\, & $ \pm$ 0.72$)\times 10^{-2}$  &  0.68\, & $ \pm$ 0.03    & 0.63\, & $ \pm$ 0.03  & 1.60\, & $ \pm$ 0.83  & $288^3$   \\ 
            \texttt{M2Ma1$\zeta$1}   & 1.95\,  & $ \pm$ 0.09    & 0.97\,  & $ \pm$ 0.04  & 1.00  & -0.25\,  & $ \pm$ 0.13  & (4.88\,  & $ \pm$ 0.67$)\times 10^{-2}$  &  0.62\, & $ \pm$ 0.03    & 0.58\, & $ \pm$ 0.01  & 1.67\, & $ \pm$ 1.53  & $288^3$   \\    
            \texttt{M10Ma01$\zeta$0} & 11.0\,  & $ \pm$ 0.4     & 0.110\, & $ \pm$ 0.001 & 0.00  &  0.26\,  & $ \pm$ 0.10  & (4.65\,  & $ \pm$ 1.50$)\times 10^{-2}$  &  2.65\, & $ \pm$ 0.06    & 1.61\, & $ \pm$ 0.05  & 2.55\, & $ \pm$ 0.20  & $288^3$   \\
            \texttt{M10Ma01$\zeta$025} & 11.1\, & $ \pm$ 0.5    & 0.111\, & $ \pm$ 0.005 & 0.25  & -0.25\,  & $ \pm$ 0.10  & (2.70\,   & $ \pm$ 0.47$)\times 10^{-1}$  &  1.93\, & $ \pm$ 0.05    & 1.45\, & $ \pm$ 0.05  & 1.32\, & $ \pm$ 0.03  & $288^3$   \\             
            \texttt{M10Ma01$\zeta$05}& 11.3\,  & $ \pm$ 0.3     & 0.113\, & $ \pm$ 0.001 & 0.50  & -0.18\,  & $ \pm$ 0.19  & (4.32\,  & $ \pm$ 2.82$)\times 10^{-2}$  &  1.18\, & $ \pm$ 0.07    & 1.13\, & $ \pm$ 0.02  & 1.04\, & $ \pm$ 0.02  & $288^3$   \\
            \texttt{M10Ma01$\zeta$075} & 10.9\, & $ \pm$ 0.7    & 0.109\, & $ \pm$ 0.007  & 0.75  &  0.01\,  & $ \pm$ 0.15  & (1.09\,   & $ \pm$ 2.17$)\times 10^{-4}$  &  0.91\, & $ \pm$ 0.04   & 0.97\, & $ \pm$ 0.07  & 1.01\, & $ \pm$ 0.03  & $288^3$   \\            
            \texttt{M10Ma01$\zeta$1} & 11.1\,  & $ \pm$ 0.8     & 0.110\, & $ \pm$ 0.001 & 1.00  & -0.10\,  & $ \pm$ 0.15  & (2.00\,  & $ \pm$ 1.80$)\times 10^{-4}$  &  0.97\, & $ \pm$ 0.02    & 1.01\, & $ \pm$ 0.02  & 1.01\, & $ \pm$ 0.01  & $288^3$   \\ 
            \texttt{M10Ma1$\zeta$0}  & 9.8\,   & $ \pm$ 0.5     & 0.98\,  & $ \pm$ 0.05   & 0.00  &  0.09\,  & $ \pm$ 0.14  & (1.81\, & $ \pm$ 8.24$)\times 10^{-2}$  &  2.53\, & $ \pm$ 0.08    & 1.88\, & $ \pm$ 0.09  & 1.13\, & $ \pm$ 0.04  & $288^3$   \\
            \texttt{M10Ma1$\zeta$025} & 10.6\,  & $ \pm$ 0.5    & 1.06\,  & $ \pm$ 0.05   & 0.25  &  0.02\,  & $ \pm$ 0.18  & (6.11\,  & $ \pm$ 2.58$)\times 10^{-2}$  &  2.27\, & $ \pm$ 0.13    & 1.87\, & $ \pm$ 0.11  & 1.10\, & $ \pm$ 0.02  & $288^3$   \\            
            \texttt{M10Ma1$\zeta$05} & 9.7\,   & $ \pm$ 0.9     & 0.97\,  & $ \pm$ 0.09   & 0.50  & -0.01\,  & $ \pm$ 0.17  & (1.79\, & $ \pm$ 3.90$)\times 10^{-2}$  &  1.50\, & $ \pm$ 0.04    & 1.35\, & $ \pm$ 0.04  & 1.10\, & $ \pm$ 0.02  & $288^3$   \\
            \texttt{M10Ma1$\zeta$075} & 10.9\,  & $ \pm$ 0.6    & 1.09\,  & $ \pm$ 0.06   & 0.75  &  0.11\,  & $ \pm$ 0.09  & (3.20\,  & $ \pm$ 5.12$)\times 10^{-5}$  &  1.38\, & $ \pm$ 0.04    & 1.41\, & $ \pm$ 0.03  & 1.02\, & $ \pm$ 0.01  & $288^3$   \\            
            \texttt{M10Ma1$\zeta$1}  & 9.7\,   & $ \pm$ 0.3     & 0.96\,  & $ \pm$ 0.03   & 1.00  &  0.12\,  & $ \pm$ 0.10  & (2.00\, & $ \pm$ 9.00$)\times 10^{-5}$  &  1.30\, & $ \pm$ 0.04    & 1.33\, & $ \pm$ 0.04  & 1.00\, & $ \pm$ 0.01  & $288^3$   \\
            \hline 
            \hline
        \end{tabular}
        } \\
        \begin{tablenotes}
        \item{\textit{\textbf{Notes:}} For each simulation we extract 51 realisations at 0.1$\, \tau$ intervals, where $\tau = L/(2c_s\M)$ is the correlation time of the turbulent driving source, between $5$-$10 \, \tau$. All $1\sigma$ fluctuations listed in the table and study are from time-averaging quantities over the $5\, \tau$. Column (1): the simulation ID. Column (2): the rms turbulent Mach number, $\M = \sigma_V / c_s$. Column (3): the Alfv\'en Mach number for the mean-$\vecB{B}$ component, $\vecB{B}_0$, $\Mao = (2c_s\M\sqrt{\pi\rho_0}) / |\vecB{B}_0|$, where $\rho_0$ is the mean density and $c_s$ is the sound speed. Column (4): $\zeta$ (Equation~\ref{eq:zeta_equation}) the driving parameter controlling the amount of $\nabla\times\vecB{F}$ and $\nabla\cdot\vecB{F}$ modes in the driving source. Column (5): the skewness, Equation~\ref{eq:skewness}, of the logarithmic density. Column (6): the \citet{Hopkins2013} intermittency parameter, $T$ Equation~\ref{eq:T_average_jump} or \ref{eq:T_sigma_var}, for the logarithmic density. Column (7): the volume-weighted variance of the logarithmic density. Column (8): the mass-weighted variance of the logarthimic density. Column (9): the \citet{Mocz2019} intermittency parameter $f$, scaled so that it defined the dynamical time-scale of the voids versus over-density gas structures in turbulence. Column (10): the number of grid cells in the discretisation of the spatial domain, $\mathcal{V}=L^3$. }
        \end{tablenotes}
        \label{tb:simtab}
    \end{table*}